\documentclass[12pt]{article}

\usepackage{natbib}
\usepackage[nottoc,notlof,notlot]{tocbibind} 

\bibpunct{(}{)}{;}{a}{}{;}
\usepackage{breakcites}
\usepackage[breaklinks=true]{hyperref}
\usepackage{graphicx}
\graphicspath{ {images/} }

\usepackage{amsmath}
\usepackage{bbm}
\usepackage{amsfonts}%
\usepackage{amssymb}%

\usepackage{geometry} 
\usepackage{graphicx} 
\usepackage{framed}
\usepackage{caption}
\usepackage{subfigure}

\usepackage{booktabs} 
\usepackage{array} 
\usepackage{paralist} 
\usepackage{verbatim} 

\usepackage{lipsum}
\usepackage{setspace}

\usepackage{lipsum}
\usepackage[singlelinecheck=false]{caption}
\usepackage{tabularx}

\usepackage{multicol}
\usepackage{multirow}

\title{\setstretch{1} Heterogeneous Treatment Effects for Networks, Panels, and other Outcome Matrices}

\author{Eric Auerbach\footnote{Department of Economics, Northwestern University. E-mail: eric.auerbach@northwestern.edu.} \and Yong Cai \footnote{Department of Economics, Northwestern University. E-mail: yongcai2023@u.northwestern.edu.\newline We thank Jon Auerbach, Lori Beaman, Vivek Bhattacharya, Stephane Bonhomme, Federico Bugni, Ivan Canay, Ben Golub, Bryan Graham, Joel Horowitz, Gaston Illanes, Chuck Manski, Roger Moon, Rob Porter, Guillaume Pouliot, Chris Udry and Martin Weidner  for helpful feedback. }}
 \date{\parbox{\linewidth}{\centering%
  \today\endgraf}} 

\begin{document}
\maketitle
%
\begin{abstract} \setstretch{1}\noindent
We are interested in the distribution of treatment effects for an experiment where units are randomized to a treatment but outcomes are measured for pairs of units. For example, we might measure risk sharing links between households enrolled in a microfinance program, employment relationships between workers and firms exposed to a trade shock, or bids from bidders to items assigned to an auction format. Such a double randomized experimental design may be appropriate when there are social interactions, market externalities, or other spillovers across units assigned to the same treatment. Or it may describe a natural or quasi experiment given to the researcher. In this paper, we propose a new empirical strategy that compares the eigenvalues of the outcome matrices associated with each treatment. Our proposal is based on a new matrix analog of the Fr\'echet-Hoeffding bounds that play a key role in the standard theory. We first use this result to bound the distribution of treatment effects. We then propose a new matrix analog of quantile treatment effects that is given by a difference in the eigenvalues. We call this analog spectral treatment effects.   \looseness=-1 
\end{abstract}

\section{Introduction}
Consider a market designer tasked with learning how an intervention alters the transactions between buyers and sellers in a marketplace. For example, the designer may be an online platform such as Amazon or AirBnB and the intervention a change in a search algorithm or website layout, see recently \cite{bajari2021multiple,johari2022experimental}. The designer conducts an experiment where they randomly assign buyers and sellers to two groups. They implement the intervention in one group, maintain the status quo in the other group, and measure outcome matrices describing how much each buyer buys from each seller in each group.\looseness=-1 

How can the market designer characterize the impact of the intervention on the buyer and seller transactions using such a double randomized experiment? To address this question, we propose a new empirical strategy for identifying heterogeneous treatment effects that compares the eigenvalues of the outcome matrices associated with each treatment. \looseness=-1

To motivate our proposal, Section 2 reviews two approaches standard in the conventional setting of a single randomized experiment. The first approach bounds the distribution of treatment effects using arguments of \cite{frechet1951tableaux,hoeffding1940masstabinvariante,makarov1982estimates} \citep[see for example][]{manski1997mixing,manski2003partial,heckman1997making,fan2010sharp,abadie2018econometric,firpo2019partial,masten2020inference,molinari2020microeconometrics,frandsen2021partial}. The second approach makes a rank invariance assumption and computes the difference in the quantiles of the outcomes associated with each treatment. This is often called quantile treatment effects \citep[see for example][]{abadie2002instrumental,chernozhukov2005iv,bitler2006mean,firpo2007efficient,imbens2009identification,masten2018identification}. While comparing the average outcome across treatments only characterizes an expected treatment effect, the above two approaches can identify treatment effect heterogeneity because they reveal information about the entire distribution of treatment effects.      \looseness=-1

Section 3 contains our main results: new analogs of the Frech\'et-Hoeffding bounds and quantile treatment effects for the double randomized experiment with outcome matrices. A key complication is that two dimensions of randomization make the relevant optimization problem quadratic rather than linear. Exact solutions are not generally computable. \looseness=-1 

Our main idea is to instead consider relaxations of the quadratic problem solved by rearranging the eigenvalues of the outcome matrices associated with each treatment. Section 4 sketches the logic behind this solution. We first use it to bound the distribution of treatment effects building on arguments of \cite{whitt1976bivariate,finke1987quadratic,lovasz2012large}. We then show that under a matrix generalization of rank invariance, the distribution of treatment effects is point identified and characterized by a difference in eigenvalues. We call this matrix analog of quantile treatment effects, spectral treatment effects.\looseness=-1 

Section 5 discusses extensions including covariates, spillovers, and estimation. Section 6 shows results from two empirical demonstrations and Section 7 concludes. Proof of our main claims are collected in Appendix A, supplementary material can be found in an online appendix, and an R package can be found at  \url{https://github.com/yong-cai/MatrixHTE}. \looseness=-1

\subsection{Motivating examples}
We describe four examples of double randomized experiments or quasi experiments with outcome matrices. They are used to motivate our framework and results below. \looseness=-1 

\subsubsection{Example 1: risk sharing}
\cite{banerjee2021changes} study the impact of a microfinance program in a sample of Indian villages. They argue that the program decreases informal risk sharing between some households. \cite{comola2021treatment} study the impact of savings accounts in a sample of Nepalese villages.  They argue that the program increases informal risk sharing between some households. In this example, the units are households, the treatment is program participation, and the outcomes are surveyed risk sharing links between pairs of households. We revisit this example in the first empirical demonstration of Section 6 below.     \looseness=-1 

\subsubsection{Example 2: superstar extinction}
\cite{azoulay2010superstar} study the impact of a superstar researcher's death in a sample of research groups in the life sciences. They argue that the death of a superstar decreases the quality of research conducted by researchers nearby in the coauthorship network. In this example, the units are researchers, the treatment is the death of a superstar, and the outcomes are the amount of research conducted between coauthors.  \looseness=-1 

\subsubsection{Example 3: auction format}
\cite{athey2011comparing} study the impact of a sealed versus open bid design in a sample of US timber auctions. They argue that the sealed bid design incentivizes some firms to participate who otherwise would not in the open bid design. In this example, the units are firms and tracts of land, the treatment is the auction format, and the outcomes are the bids made by firms on the tracts.  We revisit this example in the second empirical demonstration of Section 6 below.      \looseness=-1 

\subsubsection{Example 4: buyer-seller experiment}
\cite{bajari2021multiple} model the impact of an information policy on the likelihood that a buyer buys an item from a seller. They consider a multiple randomization experimental design where the researcher independently randomizes buyers and sellers to groups and then assigns policies to pairs of buyers and sellers depending on their group memberships, see for example their Definitions 7 and 8. In this example, the units are buyers and sellers, the treatment is the information policy, and the outcomes are transactions between buyers and sellers. We revisit this example in our discussion of treatment spillovers in Section 5.4 below. \looseness=-1

\section{Review of the single randomized experiment}
We review a standard framework and results for the conventional single randomized experiment following \cite{whitt1976bivariate}. This review is used to motivate our framework and results for the double randomized experiment with outcome matrices in Section 3. \looseness=-1 

\subsection{Model and econometric problem}

\subsubsection{Model}
A population of agents is randomized to a binary treatment $t \in \{0,1\}$. The population may be finite or infinite.  Potential outcomes are defined for each agent in the population and may be fixed or random. The realized potential outcomes of an agent selected uniformly at random from the population are described by a joint distribution function $F$ on $\mathbb{R}^{2}$. 

We define the measurable function $(Y^{*}_{1},Y^{*}_{0}):[0,1] \to \mathbb{R}^{2}$ so that $(Y^{*}_{1}(U),Y^{*}_{0}(U))$ has distribution $F$ when $U$ is standard uniform, see Lemma 2.7 of \cite{whitt1976bivariate}. We sometimes interpret $(Y^{*}_{1},Y^{*}_{0})$ as the fixed potential outcomes of a continuum of agent types indexed by $[0,1]$, although this function representation is valid for both finite and infinite populations. \looseness=-1

As an example, consider an experiment where the researcher randomizes $N$ workers to participate $(t = 1)$ or not participate $(t=0)$ in a training program. Let $\{Y^{*}_{i,1},Y^{*}_{i,0}\}_{i \in [N]}$ describe the fixed potential wages of the $N$ workers and define $Y^{*}_{t}(u) = \sum_{i=1}^{N}Y_{i,t}^{*}\mathbbm{1}\{u \in \tau_{i}\}$ where $\tau_{i} = \{u \in [0,1]: \lceil Nu\rceil = i\}$. Then $Y^{*}_{t}(u)$ describes the potential wage of the $\lceil Nu \rceil$th worker under treatment $t$ and the distribution of $(Y_{1}^{*}(U),Y_{0}^{*}(U))$ is the empirical distribution of the potential wages of the $N$ workers, i.e. $F(y_{1},y_{0}) = \frac{1}{N}\sum_{i=1}^{N}\mathbbm{1}\{Y^{*}_{i,1} \leq y_{1},Y^{*}_{i,0} \leq y_{0}\}$.  \looseness=-1

\subsubsection{Parameters of interest}
We focus on the joint distribution of potential outcomes (DPO) and distribution of treatment effects (DTE). The DPO is
\begin{align}\label{joint_simple}
F(y_{1},y_{0}) 
:= P\left(Y^{*}_{1}(U) \leq y_{1}, Y^{*}_{0}(U) \leq y_{0} \right)
= \int\prod_{t \in \{0,1\}}\mathbbm{1}\{Y^{*}_{t}(u)\leq y_{t}\}du
\end{align}
where $y_{1},y_{0} \in \mathbb{R}$ are arbitrary and $U$ is a standard uniform random variable. In words, the DPO is the mass of agent types with potential outcome less than $y_{1}$ under treatment $1$ and less than $y_{0}$ under treatment $0$.   \looseness=-1 

The DTE is
\begin{align}\label{te_simple}
\Delta(y) 
:= P\left(Y^{*}_{1}(U) - Y^{*}_{0}(U) \leq y\right) 
= \int\mathbbm{1}\{Y^{*}_{1}(u) - Y^{*}_{0}(u) \leq y\} du.
\end{align}
In words, $Y^{*}_{1}(u) - Y^{*}_{0}(u)$ is the change in outcome associated with switching the treatment status of an agent with type $u$ from $0$ to $1$.  The DTE is the mass of agents for which this individual treatment effect is less than $y$.   \looseness=-1

\subsubsection{Econometric problem}
Our task is to identify the DPO and DTE. The problem is that the researcher does not observe both $Y_{1}^{*}$ and $Y_{0}^{*}$ for the same population of agents. Agents are assigned to treatment $1$ or treatment $0$ but not both.  \looseness=-1 

For example, the researcher may assign workers to participate or not participate in a training program. If a worker participates in the program, then the researcher observes the potential wage associated with program participation. They do not observe the wage that the participating worker would have received had they not participated in the program. To infer this missing potential outcome, the researcher must use the wages of the workers that did not participate in the training program.  \looseness=-1 

Formally, the standard assumption is that the researcher observes the marginal distributions of the potential outcomes given by $F_{1}(\cdot) := F(\cdot,\infty)$ and $F_{0}(\cdot) := F(\infty,\cdot)$ on $\mathbb{R}$, but not any other feature of their joint distribution $F(\cdot.\cdot)$ on $\mathbb{R}^{2}$. The econometric problem is then to identify the DPO and DTE using only $F_{1}$ and $F_{0}$. \looseness=-1 

To motivate our results for the double randomized experiment in Section 3, we restate this formulation of the econometric problem using measure preserving transformations. Specifically, we assume the the researcher observes not $(Y_{1}^{*},Y_{0}^{*})$ but  $(Y_{1},Y_{0}) : [0,1] \to \mathbb{R}^{2}$ where $Y_{t}$ is equivalent to $Y_{t}^{*}$ up to an unknown measure preserving transformation $\varphi_{t}$. That is, 
 \begin{align}\label{perm_simple}
 Y_{t}(\varphi_{t}(u)) = Y_{t}^{*}(u)
 \end{align}
  for some unknown $\varphi_{t} \in \mathcal{M} := \{\phi: [0,1] \to [0,1]\text{ with } |\phi^{-1}(A)| = |A| \text{ for any measurable } A \subseteq [0,1]\}$ where $|A|$ refers to the Lebesgue measure of $A$. Intuitively, $Y_{t}$ is a rearranged version of $Y_{t}^{*}$ so that the two have the same marginal distribution, but no other feature of the joint distribution of $Y^{*}_{0}$ and $Y^{*}_{1}$ can be learned from $Y_{0}$ and $Y_{1}$.  \looseness=-1 
  
The restated econometric problem is then to identify the DPO and DTE using only $Y_{1}$ and $Y_{0}$. The two versions of the econometric problem are equivalent because $Y_{t}$ contains exactly the same information as the marginal distribution or quantile function associated with $Y_{t}^{*}$, see Theorem 5.1 of \cite{whitt1976bivariate}. However, we use measure preserving transformations and not marginal or quantile functions in our formulation of the econometric problem because there is no natural analog of the latter for outcome matrices under double randomization.   \looseness=-1

\subsection{Some standard results for the single randomized experiment}
We first bound the DPO and DTE following \cite{frechet1951tableaux,hoeffding1940masstabinvariante,makarov1982estimates}. We then show that under a rank invariance assumption the DPO and DTE are point identified and can be written as functionals of the quantiles of the outcomes associated with each treatment following \cite{doksum1974empirical,lehmann1975nonparametrics,whitt1976bivariate}. These results are known to the econometrics literature. We state them to motivate our results for the double randomized experiment with outcome matrices in Section 3.  \looseness=-1 

\subsubsection{Bounds on the DPO and DTE}
Plugging (\ref{perm_simple}) into (\ref{joint_simple}) gives sharp bounds on the DPO
\begin{align}\label{lap}
\min_{\varphi_{0},\varphi_{1} \in \mathcal{M}}\int \prod_{t \in \{0,1\}} \mathbbm{1}\{Y_{t}(\varphi_{t}(u)) \leq y_{t}\}du
\leq F(y_{1},y_{0}) \nonumber  \\
\leq \max_{\varphi_{0},\varphi_{1} \in \mathcal{M}}\int \prod_{t \in \{0,1\}} \mathbbm{1}\{Y_{t}(\varphi_{t}(u)) \leq y_{t}\}du.
\end{align} 
These bounds have a simple analytical solution. 
\begin{flushleft}
\textbf{Standard result 1:} For any $(y_{1},y_{0}) \in \mathbb{R}^{2}$
\begin{align*}
\max\left(F_{1}(y_{1})+F_{0}(y_{0}) - 1,0\right)
\leq F(y_{1},y_{0}) 
\leq \min\left(F_{1}(y_{1}),F_{0}(y_{0})\right).
\end{align*} 
\end{flushleft}  \looseness=-1 

Standard result 1 is often attributed to \cite{frechet1951tableaux,hoeffding1940masstabinvariante}, although
our proof sketch in Section 4.2 follows \cite{whitt1976bivariate}. The bounds are straightforward to compute or estimate (in cases of sampled, mismeasured, or missing outcomes) using standard tools.  \looseness=-1

The bounds on the DPO imply bounds on the DTE.
\begin{flushleft}
\textbf{Standard result 2:} For any $y \in \mathbb{R}$
\begin{align*}
\sup_{\substack{(y_{1},y_{0}) \in \mathbb{R}^{2}:\\ y_{1}-y_{0} = y}}\max\left(F_{1}(y_{1})-F_{0}(y_{0}) ,0\right) \leq \Delta(y) 
\leq 1 + \inf_{\substack{(y_{1},y_{0}) \in \mathbb{R}^{2}:\\ y_{1}-y_{0} = y}}\min\left(F_{1}(y_{1})-F_{0}(y_{0}),0\right).
\end{align*} 
\end{flushleft} \looseness=-1 

Standard result 2 is often attributed to \cite{makarov1982estimates}. These bounds are also straightforward to compute or estimate using standard tools.  \looseness=-1 

 \subsubsection{Point identification of the DTE under rank invariance}
The Quantile Treatment Effects parameter (QTE) refers to the difference in the quantile functions of $Y_{1}$ and $Y_{0}$. Specifically, $QTE(u) := Q_{1}(u)-Q_{0}(u)$ where $Q_{t}(u) := \inf\{y \in \mathbb{R}: u \leq F_{t}(y)\}$ is the inverse marginal distribution (quantile) function associated with $Y_{t}^{*}$, or equivalently, $Y_{t}$. Although $Q_{t}$ and $Y_{t}^{*}$ generally have the same marginal distribution for $t\in \{0,1\}$,  $Q_{1}-Q_{0}$ and $Y_{1}^{*}-Y_{0}^{*}$ do not. In fact, the difference in quantiles is a more conservative notion of the effect of treatment as measured by mean squared error. That is, \looseness=-1 
\begin{flushleft}
\textbf{Standard result 3:} $\int \left(Q_{1}(u) - Q_{0}(u)\right)^{2}du \leq \int \left(Y_{1}^{*}(u)-Y_{0}^{*}(u)\right)^{2}du$.
\end{flushleft}
See Corollary 2.9 of \cite{whitt1976bivariate}. However, under a rank invariance assumption, $Q_{1}-Q_{0}$ and $Y_{1}^{*}-Y_{0}^{*}$ do have the same distribution. We say that a treatment effect is rank invariant if $Y_{1}^{*} = g(Y_{0}^{*})$ for some nondecreasing $g: \mathbb{R} \to \mathbb{R}$. \looseness=-1 

\begin{flushleft}
\textbf{Standard result 4}: $\Delta(y) = \int \mathbbm{1}\{Q_{1}(u)-Q_{0}(u) \leq y\}du$ under rank invariance.  
\end{flushleft}
See Theorem 2.5 of \cite{whitt1976bivariate}. The DPO is similarly identified under rank invariance with $F(y_{1},y_{0}) = \int \prod_{t \in \{0,1\}}\mathbbm{1}\{Q_{t}(u) \leq y_{t}\}du$. In words, rank invariance says that the rank of an agent's outcome in the population is the same under both treatments. That is, $\int \mathbbm{1}\{Y_{0}^{*}(s) \leq Y_{0}^{*}(u)\}ds = \int \mathbbm{1}\{Y_{1}^{*}(s) \leq Y_{1}^{*}(u)\}ds$ for every $u \in [0,1]$.

\section{The double randomized experiment}
We propose analogs of the Section 2 framework and results for a double randomized experiment with outcome matrices. Our focus is on symmetric matrices indexed by one population as in Examples 1 and 2 of Section 1.1. Asymmetric matrices or matrices indexed by two different populations as in Examples 3 and 4 are handled by symmetrization in Section 5.1.1. \looseness=-1 

\subsection{Model and econometric problem}

\subsubsection{Model}
A population of agents is randomized to two groups. The population may be finite or infinite. Pairs of agents are assigned a binary treatment $t \in \{0,1\}$ depending on the individual group assignments. \cite{bajari2021multiple} call this a simple multiple randomization design, see their Definition 8. For other examples of double randomization in the literature see \cite{graham2008identifying,graham2011econometric,graham2014complementarity,johari2022experimental}.

To simplify our exposition, we suppose that a pair of agents is assigned to treatment $1$ if both agents belong to the first group and assigned to treatment $0$ if both agents belong to the second group, ignoring any outcomes between agents assigned to different groups. However our main arguments below are not specific to this particular comparison, see Section 5.4 below. Potential outcomes are bounded (this can be relaxed) and defined for each pair of agents in the population. They may be fixed or random.  \looseness=-1 

Recall that the potential outcomes in the single randomized setting are represented by $(Y^{*}_{1},Y^{*}_{0}):[0,1] \to \mathbb{R}^{2}$ where $Y_{t}^{*}(u)$ describes the fixed potential outcome associated with treatment $t \in \{0,1\}$ and agent type $u \in [0,1]$. We consider an analogous representation for the double randomized setting with outcome matrices where the potential outcomes are represented by a symmetric measurable function $(Y^{*}_{1},Y^{*}_{0}):[0,1]^{2} \to \mathbb{R}^{2}$. $Y_{t}^{*}(u,v)$ describes the fixed potential outcome associated with treatment $t \in \{0,1\}$ and agent types $u, v \in [0,1]$. \looseness=-1 

Following \cite{lovasz2012large}, we sometimes interpret $Y_{t}^{*}$ as an infinite dimensional population matrix, although this representation is valid for both finite and infinite populations. Let $U$ and $V$ be independent standard uniform random variables. Then the random vector $(Y_{1}^{*}(U,V),Y_{0}^{*}(U,V))$ describes the distribution of potential outcomes between a pair of agent types each drawn independently and uniformly at random from $[0,1]$. Both the choice of state space $[0,1]$ and the assumption that agents are selected uniformly at random are standard normalizations also made in the conventional single randomized setting. \looseness=-1 

As an example, consider Example 1 from Section 1.1 where the researcher randomizes $N$ households to participate or not participate in a microfinance program. Let $\{Y_{ij,1}^{*},Y_{ij,0}^{*}\}_{i,j \in [N]}$ describe the fixed potential risk sharing links between every pair of households when both enroll ($t=1$) or do not enroll ($t=0$) in the program. Define $Y_{t}(u,v) = \sum_{i=1}^{N}\sum_{j=1}^{N}Y_{ij,t}^{*}\mathbbm{1}\{u \in \tau_{i}, v \in \tau_{j}\}$ where  $\tau_{i} = \{u \in [0,1]: \lceil Nu\rceil = i\}$. Then $Y_{t}^{*}(u,v)$ describes the potential risk sharing link between the $\lceil Nu\rceil$th and $\lceil Nv \rceil$th households under treatment $t$ and the distribution of $(Y_{1}^{*}(U,V),Y_{0}^{*}(U,V))$ is the empirical distribution of the potential risk sharing links of the $N^{2}$ household pairs, i.e. $F(y_{1},y_{0}) = \frac{1}{N^{2}}\sum_{i=1}^{N}\sum_{j=1}^{N}\mathbbm{1}\{Y_{ij,1}^{*} \leq y_{1}, Y_{ij,0}^{*} \leq y_{0}\}$.  \looseness=-1 

\subsubsection{Interpreting the model}
Our model explicitly describes one source of randomness: sampling agents uniformly at random from a fixed population. This implies a specific notion of treatment effect heterogeneity that defines the distribution of treatment effects, following exactly the logic of the conventional single randomized experiment from Section 2. A consequence of the model is that the sampled potential outcomes are independent across pairs of agents that do not have an agent in common. This dependence structure is frequently used in the network econometrics literature, see Section 6 of \cite{graham2020network} and Section 3 of \cite{de2016econometrics} for many examples. \looseness=-1 

We do not intend for this model to necessarily describe any data generating process. In particular, the researcher may believe that the outcomes they actually observe are also influenced by measurement error, missing data, spillovers, strategic interactions, etc. While such variation is not used to explicitly define treatment effect heterogeneity in our  framework, as in the setting of the conventional single randomized experiment, it can be incorporated into the data generating process and still play a role in identification, estimation, and inference. We discuss some extensions along these lines in Sections 5.3 and 5.4 below. \looseness=-1 

As an example, consider the two-way effects model $Y_{ij,t} = f_{t}(\alpha_{i,t},\alpha_{j,t},\varepsilon_{ij,t})$ where the agent effects $\{\alpha_{i,t}\}_{i \in [N]}$ have distribution $F_{\alpha,t}$ and the idiosyncratic errors $\{\varepsilon_{ij,t}\}_{i,j \in [N]}$ have distribution $F_{\varepsilon,t}$. One might define the potential outcome function of interest to be $Y^{*}_{t}(u,v) = \int f_{t}(F_{\alpha,t}^{-1}(u),F_{\alpha,t}^{-1}(v),x)dF_{\varepsilon,t}(x)$. Intuitively, this function indexes a collection of expected outcomes (with respect to $\varepsilon_{ij,t}$) generated by sampling agents uniformly at random from the population (with respect to $\alpha_{i,t}$). Inferring $Y^{*}_{t}$ from the data may be complicated if some entries of the matrix are missing, the idiosyncratic errors are dependent across agent pairs or correlated with the individual effects, etc. However, as in the setting of the conventional single randomized experiment, we do not use this variation in our characterization the heterogeneous impact of the treatment as given by the parameters of interest below.  \looseness=-1

\subsubsection{Parameters of interest}
We define the joint distribution of potential outcomes (DPO) and the distribution of treatment effects (DTE) as in Section 2. The DPO is 
\begin{align}\label{joint}
F(y_{1},y_{0}) 
:= P\left(Y^{*}_{1}(U,V) \leq y_{1}, Y^{*}_{0}(U,V) \leq y_{0} \right)
= \int\int \prod_{t \in \{0,1\}}\mathbbm{1}\{Y^{*}_{t}(u,v)\leq y_{t}\}dudv
\end{align}
where $y_{1},y_{0} \in \mathbb{R}$ and $U$ and $V$ are independent standard uniform random variables. In words, the DPO is the mass of agent type pairs with potential outcome less than $y_{1}$ under treatment $1$ and less than $y_{0}$ under treatment $0$. \looseness=-1

Similarly, the DTE is
\begin{align}\label{te}
\Delta(y) 
:= P\left(Y^{*}_{1}(U,V) - Y^{*}_{0}(U,V) \leq y\right) 
= \int\int\mathbbm{1}\{Y^{*}_{1}(u,v) - Y^{*}_{0}(u,v) \leq y\} dudv.
\end{align}
In words,  $Y^{*}_{1}(u,v) - Y^{*}_{0}(u,v)$ is the change in outcome associated with switching the treatment status of a pair of agents with types $u$ and $v$ from $0$ to $1$. The DTE is the mass of agent type pairs for which this treatment effect is less than $y$. Under the treatment assignment rule described in Section 3.1.1, it is the distributional analog of what \cite{bajari2021multiple} call the average effect for the treated pairs. Heterogeneous analogs of other parameters such as their spillover or direct effects can be similarly constructed, see for example Section 5.4. \looseness=-1

\subsubsection{Econometric problem}
As before, our task is to identify the DPO and the DTE. The problem is also that the researcher observes at most one potential outcome for any pair of agents.  \looseness=-1

For example, the researcher may assign households to participate or not participate in a microfinance program. If both households participate in the program, then the researcher observes the potential risk sharing link associated with joint program participation. They do not observe whether these households would have formed a link under the counterfactual treatment that neither household participates. To infer this missing potential outcome, the researcher must use the risk sharing links between the nonparticipating households.  \looseness=-1

Following the second econometric problem formulation of Section 2.1.3, we suppose that the researcher observes not $(Y_{1}^{*},Y_{0}^{*})$ but $(Y_{1},Y_{0}) : [0,1]^{2} \to \mathbb{R}^{2}$ where $Y_{t}$ is equivalent to $Y_{t}^{*}$ up to an unknown measure preserving transformation. That is,
 \begin{align}\label{perm}
 Y_{t}(\varphi_{t}(u),\varphi_{t}(v)) = Y_{t}^{*}(u,v)
 \end{align}
  for some unknown $\varphi_{t} \in \mathcal{M}$. Like the conventional single randomized setting, (\ref{perm}) says that $Y_{t}$ and $Y_{t}^{*}$ represent the same random object. However, $Y_{0}$ and $Y_{1}$ do not reveal any additional information about how the entries of $Y_{0}^{*}$ and $Y_{1}^{*}$ are related. Unlike the single randomized setting, there is no canonical $Y_{t}$ that  serves the role of the marginal distribution or quantile function in the double randomized setting with outcome matrices. The ``marginal distribution of $Y_{t}^{*}$'' is instead represented by an equivalence class of functions $Y_{t}$ that satisfy (\ref{perm}). \cite{lovasz2012large} calls such functions weakly isomorphic, see generally his Sections 7.3, 10.7, and 13.2.  \looseness=-1

\subsection{Some new results for the double randomized experiment}
We first bound the DPO and DTE. We then propose a new matrix generalization of rank invariance under which the DPO and DTE are point identified and can be written as functionals of the eigenvalues of the potential outcome functions associated with each treatment. Eigenvalues of functions are defined a bit differently than their matrix counterparts, see our Appendix Section A.1 or \cite{lovasz2012large}, Section 7.5 for a review. Proof of our main propositions can be found in Appendix Sections A.2-4.   \looseness=-1

\subsubsection{Bounds on the DPO and DTE}
As in the single randomized setting, plugging (\ref{perm}) into (\ref{joint}) gives sharp bounds on the DPO
\begin{align}\label{qap}
\min_{\varphi_{0},\varphi_{1} \in \mathcal{M}}\int\int \prod_{t \in \{0,1\}} \mathbbm{1}\{Y_{t}(\varphi_{t}(u),\varphi_{t}(v)) \leq y_{t}\}dudv 
\leq F(y_{1},y_{0}) \nonumber \\
\leq \max_{\varphi_{0},\varphi_{1} \in \mathcal{M}}\int\int \prod_{t \in \{0,1\}} \mathbbm{1}\{Y_{t}(\varphi_{t}(u),\varphi_{t}(v)) \leq y_{t}\}dudv.
\end{align} 
We do not consider these bounds, however, because their quadratic structure makes them analytically and computationally intractable in general. See \cite{cela2013quadratic}, Section 1.5. \looseness=-1 

We instead propose bounds that are not generally sharp but are tractable. Let $\lambda_{1t}(y_{1}) \geq  \lambda_{2t}(y_{1}) \geq ... \geq \lambda_{Rt}(y_{t})$ be the $R$ largest (in absolute value) eigenvalues of $\mathbbm{1}\{Y_{t}(\cdot,\cdot) \leq y_{t}\}$ ordered to be decreasing and $s_{R}(r) = R - r + 1$. For any $t, t' \in \{0,1\}$, let $\sum_{r}\lambda_{rt}\lambda_{rt'} := \lim_{R \to \infty}\sum_{r=1}^{R}\lambda_{rt}(y_{t})\lambda_{rt'}(y_{t'})$, $\sum_{r}\lambda_{rt}\lambda_{s(r)t'} :=    \lim_{R \to \infty}\sum_{r=1}^{R}\lambda_{rt}(y_{t})\lambda_{s_{R}(r)t'}(y_{t'})$ and $\sum_{r}\lambda_{rt}^{2} := \sum_{r}\lambda_{rt}\lambda_{rt}$. When the population is finite and equal to $N$ we drop the limit and take $R = N$.  \looseness=-1 

Our first result is 
\begin{flushleft}
\textbf{Proposition 1:} For any $(y_{1},y_{0}) \in \mathbb{R}^{2}$
\begin{align}\label{SFH}
\max\left(\sum_{r}\left(\lambda_{r1}^{2} + \lambda_{r0}^{2}\right) - 1,\sum_{r}\lambda_{r1}\lambda_{s(r)0},0\right)
\leq F(y_{1},y_{0})  \nonumber \\
\leq \min\left(\sum_{r}\lambda_{r1}^{2},\sum_{r}\lambda_{r0}^{2},\sum_{r}\lambda_{r1}\lambda_{r0}\right).
\end{align} 
\end{flushleft}
We defer a discussion of Proposition 1 to Section 4.3, only remarking here that unlike the infeasible bounds in  (\ref{qap}), those in  (\ref{SFH}) are straightforward to compute because they only depend on the eigenvalues of $\mathbbm{1}\{Y_{t}^{*}(\cdot,\cdot) \leq y_{t}\}$, or equivalently, $\mathbbm{1}\{Y_{t}(\cdot,\cdot) \leq y_{t}\}$. They can be computed or estimated (in cases of sampled, mismeasured, or missing outcomes) using standard tools, see Section 5.3.  \looseness=-1 

As in Section 2, bounds on the DPO imply bounds on the DTE. Our second result is
\begin{flushleft}
\textbf{Proposition 2:} For any $y \in \mathbb{R}$
\begin{align}\label{SM}
\sup_{\substack{(y_{1},y_{0}) \in \mathbb{R}^{2}:\\ y_{1}-y_{0} = y}}\max\left(\sum_{r}\left(\lambda_{r1}^{2} - \lambda_{r0}^{2}\right),\sum_{r}\left(\lambda_{r1}^{2}-\lambda_{r1}\lambda_{r0}\right),0\right)
\leq \Delta(y) \nonumber \\
\leq 1 + \inf_{\substack{(y_{1},y_{0}) \in \mathbb{R}^{2}:\\ y_{1}-y_{0} = y}}\min\left(\sum_{r}\left(\lambda_{r1}^{2} - \lambda_{r0}^{2}\right),\sum_{r}\left(\lambda_{r1}\lambda_{r0}- \lambda_{r0}^{2}\right),0\right)
\end{align}
\end{flushleft}
where the eigenvalue $\lambda_{rt}$ is implicitly a function of $y_{t}$. In finite data, these bounds only require the researcher to compute eigenvalues for at most $N(N+1)$ values of $y_{1}$ and $y_{0}$ where $N$ is the number of  agents. Optimizing over a smaller set also gives valid but potentially wider bounds.  \looseness=-1 

\subsubsection{Definition of spectral treatment effects}
 We propose a matrix analog of the QTE. Let $\{\sigma_{rt}\}_{r=1}^{R}$ be the $R$ largest (in absolute value) eigenvalues of $Y_{t}$ ordered to be decreasing and $\{\phi_{r}\}_{r=1}^{\infty}$ be any orthogonal basis of $L^{2}([0,1])$.  \looseness=-1 
 
\begin{flushleft}
\textbf{Definition 1:} The Spectral Treatment Effects parameter (STE) is 
\begin{align}\label{ste}
STE(u,v;\phi) := \lim_{R \to \infty}\sum_{r=1}^{R}(\sigma_{r1}-\sigma_{r0})\phi_{r}(u)\phi_{r}(v).
\end{align}
\end{flushleft}
 The STE is similar to the diagonalized difference in the eigenvalues of $Y_{1}$ and $Y_{0}$, but its exact values depend on a choice of basis. Two natural choices are the eigenfunctions of $Y_{1}$ and $Y_{0}$, denoted $\{\phi_{r1}\}_{r=1}^{\infty}$ and $\{\phi_{r0}\}_{r=1}^{\infty}$ respectively, see Appendix Section A.1. We call $STE(\phi_{1}) $ and $STE(\phi_{0})$ the Spectral Treatment Effects on the Treated (STT) and Untreated (STU). In words, the STT takes the observed matrix $Y_{1}$ and subtracts a counterfactual formed by keeping the eigenfunctions of $Y_{1}$ and inserting the eigenvalues of $Y_{0}$. That is, 
 \begin{align*}
STT(u,v) &=  Y_{1}(u,v) - \lim_{R \to \infty}\sum_{r=1}^{R}\sigma_{r0}\phi_{r1}(u)\phi_{r1}(v). 
 \\&= Y_{1}(u,v)  -  \int\int Y_{0}(s,t)W(u,s)W(v,t)dsdt
 \end{align*}
where $W(u,s) = \lim_{R \to \infty}\sum_{r=1}^{R}\phi_{r1}(u)\phi_{r0}(s)$. The second line suggests an alternative interpretation of the STT where the counterfactual outcome for a pair of agents assigned to treatment $1$ is formed by a weighted average of the outcomes of agent pairs assigned to treatment $0$. Without additional assumptions, the weights are potentially extrapolative in that they may be negative and not necessarily integrate to $1$. In some cases the researcher may wish to explicitly restrict the weights so that they satisfy these properties. We describe one way to do this in Online Appendix Section D.4. The weights will however necessarily be nonnegative and integrate to $1$ under the rank invariance condition that we introduce in the next section. \looseness=-1 

The STT is analogous to the QTE which imputes a counterfactual for an agent assigned to treatment $1$ by using the outcome of a similarly ranked agent assigned to treatment $0$. In this analogy, the eigenfunctions serve the role of the agent ranks and the eigenvalues serve the role of the quantiles associated with each rank. As in the case of the conventional single randomized experiment, this parameter may also be justified by a rank invariance assumption. \looseness=-1

 \subsubsection{Point identification of the DTE under rank invariance}
 Like the QTE, our STE is also a more conservative notion of the effect of treatment than $Y_{1}^{*}-Y_{0}^{*}$ as measured by mean squared error. Our third result is 
 \begin{flushleft}
\textbf{Proposition 3:} For any orthogonal basis $\{\phi_{r}\}_{r=1}^{\infty}$ of $L^{2}([0,1])$
\begin{align}\label{HWineq}  
\int\int STE(u,v;\phi)^{2}dudv \leq \int\int\left(Y_{1}^{*}(u,v)-Y_{0}^{*}(u,v)\right)^{2}dudv.
\end{align}
\end{flushleft}  \looseness=-1

In addition, under a rank invariance assumption, the STT, STU, and $Y_{1}^{*}-Y_{0}^{*}$ all have the same distribution. To extend rank invariance to matrices, we use the notion of a matrix function from \cite{horn1991topics}, Chapter 6.1. For any $f: \mathbb{R} \to \mathbb{R}$ that admits the representation $f(x) = \sum_{r=1}^{\infty}c_{r}x^{r}$ and square matrix $A$ (or function $A: [0,1]^{2}\to\mathbb{R}$), the matrix lift of $f$ is $f(A) = \sum_{r=1}^{\infty}c_{r}A^{r}$ where $A^{r}$ is the $r$th matrix (or operator) power of $A$, i.e. $A^{r}(u,v) = \int\int...\int A(u,t_{1})A(t_{1},t_{2})...A(t_{r-1},v)dt_{1}dt_{2}...dt_{r-1}$. \looseness=-1 

\begin{flushleft}
\textbf{Definition 2:} A treatment effect is rank invariant if $Y^{*}_{1} =  g(Y^{*}_{0})$ where $g$ is the matrix lift of some nondecreasing $g : \mathbb{R} \to \mathbb{R}$. 
\end{flushleft}  \looseness=-1 

We call Definition 2 a matrix generalization of rank invariance because it is equivalent to the definition from Section 2 when $Y^{*}_{1}$ and $Y^{*}_{0}$ are scalars. Our fourth result is  \looseness=-1 

 \begin{flushleft}
\textbf{Proposition 4:} Under rank invariance, 
\begin{align}
\Delta(y) = \int\int\mathbbm{1}\{STT(u,v) \leq y\}dudv =  \int\int\mathbbm{1}\{STU(u,v) \leq y\}dudv.
\end{align}
\end{flushleft}

Intuitively, if we think of the treatment working by taking in $Y_{0}^{*}$ and producing $Y_{1}^{*} = g(Y_{0}^{*})$, then rank invariance implies that the treatment affects the eigenvalues but not the eigenfunctions of $Y_{0}^{*}$. This is analogous to rank invariance in the conventional single randomization setting, where the treatment affects the quantiles but not the ranks of the outcomes. As in that setting, rank invariance is a strong assumption. But there are also many settings where it can be justified by economic theory. We provide four concrete examples from the literature on information diffusion, factor models, social interaction, and network formation in Online Appendix Section C.1.  \looseness=-1

\section{Sketch and discussion of the proof of Proposition 1}
We demonstrate some of the main technical ideas behind our results by sketching a proof of Proposition 1. To simplify arguments we consider a finite approximation as in \cite{whitt1976bivariate,heckman1997making}. A full proof can be found in Appendix Section A.2. \looseness=-1 

\subsection{Finite approximation}
For the single randomized experiment, we assume that $Y_{t}^{*}$ is an $N\times 1$ vector, the DPO is $\frac{1}{N}\sum_{i=1}^{N}\prod_{t \in \{0,1\}}\mathbbm{1}\{Y^{*}_{i,t} \leq y_{t}\}$, and $Y_{i,t} = \sum_{j = 1}^{N}Y_{j,t}^{*}P_{ij,t}$ is observed where $P_{t}$ is an unknown $N\times N$ permutation matrix. Intuitively, there are $N$ types of agents. One agent of each type is assigned to treatment $1$ and one agent of each type is assigned to treatment $0$.  Our task is to compare the outcomes of agents with the same type and different treatment assignments, but we do not know which agent is of which type. Bounds on the DPO are given by maximizing and minimizing $\frac{1}{N}\sum_{i=1}^{N}\sum_{j=1}^{N}\prod_{t \in \{0,1\}}\mathbbm{1}\{Y_{j,t} \leq y_{t}\}P_{ij,t}$ over all $N \times N$ permutation matrices $P_{0}$ and $P_{1}$. That this discrete problem is a good approximation to the continuous (\ref{lap}) is demonstrated in Section 2 of \cite{whitt1976bivariate}. See also Section 3 of \cite{heckman1997making}. \looseness=-1 

For the double randomized experiment, we similarly assume that $Y_{t}^{*}$ is an $N\times N$ matrix, the DPO is $\frac{1}{N^{2}}\sum_{i=1}^{N}\sum_{j=1}^{N}\prod_{t \in \{0,1\}}\mathbbm{1}\{Y^{*}_{ij,t} \leq y_{t}\}$, and $Y_{ij,t} = \sum_{k=1}^{N}\sum_{l = 1}^{N}Y_{kl,t}^{*}P_{ik,t}P_{jl,t}$ is observed where $P_{t}$ is an unknown permutation matrix. The intuition is the same as in the single randomized experiment. There are $N$ types of agents, one agent of each type is assigned to each treatment, and though we want to compare the outcomes of agents with the same type but different treatment assignments, we do not know which agent is of which type. Tight bounds on the DPO are given by maximizing and minimizing $\frac{1}{N^2}\sum_{i=1}^{N}\sum_{j=1}^{N}\sum_{k=1}^{N}\sum_{l=1}^{N}\prod_{t \in \{0,1\}}\mathbbm{1}\{Y_{kl,t} \leq y_{t}\}P_{ik,t}P_{jl,t}$ over $P_{0}$ and $P_{1}$, which we show is a good approximation to the continuous problem (\ref{qap}) in Appendix Section A.2. Since this discrete problem is an intractable ``Quadratic Assignment Problem'' or QAP, see generally \cite{cela2013quadratic}, our bounds are instead based on a conservative but tractable relaxation.   \looseness=-1 

\subsection{Standard Result 1 from Section 2}
The DPO for the finite approximation to the single randomized experiment is $F_{N}(y_{1},y_{0}) = \frac{1}{N}\sum_{i=1}^{N}\sum_{j=1}^{N}\prod_{t \in \{0,1\}}\mathbbm{1}\{Y_{j,t} \leq y_{t}\}P_{ij,t}$. We show that 
\begin{align*}
\max\left(F_{N1}(y_{1})+F_{N0}(y_{0}) - 1,0\right)
\leq F_{N}(y_{1},y_{0}) 
\leq \min\left(F_{N1}(y_{1}),F_{N0}(y_{0})\right)
\end{align*} 
where $F_{Nt}(y_{t}) = \frac{1}{N}\sum_{i=1}^{N}\mathbbm{1}\{Y_{i,t} \leq y_{t}\}$ following \cite{whitt1976bivariate}, Theorem 2.1. The proof relies on the following rearrangement inequality often attributed to \cite{hardy1952inequalities}.
 \begin{flushleft}
\textbf{Theorem 368 (Hardy-Littlewood-P\'olya):} For any $m\in \mathbb{N}$ and $g,h \in \mathbb{R}^{m}$ we have $\sum_{r} g_{(r)}h_{(m-r+1)} \leq  \sum_{r} g_{r}h_{r}  \leq  \sum_{r} g_{(r)}h_{(r)}$ where $g_{(r)}$ is the $r$th order statistic of $g$. 
\end{flushleft}  \looseness=-1

\subsubsection{Sketch of proof of Standard Result 1}
Theorem 368 implies that 
\begin{align*}
\sum_{i=1}^{N}\mathbbm{1}\{Y_{(N-i+1),0} \leq y_{0}\}\mathbbm{1}\{Y_{(i),1} \leq y_{1}\} \leq 
NF_{N}(y_{1},y_{0}) 
\leq \sum_{i=1}^{N}\mathbbm{1}\{Y_{(i),0} \leq y_{0}\}\mathbbm{1}\{Y_{(i),1} \leq y_{1}\}
\end{align*}
where $Y_{(i),t}$ is the $i$th order statistic of $Y_{t}$. The upper bound follows 
\begin{align*}
\sum_{i=1}^{N}\mathbbm{1}\{Y_{(i),0} \leq y_{0}\}\mathbbm{1}\{Y_{(i),1} \leq y_{1}\} 
\leq \min_{t \in \{0,1\}}\sum_{i=1}^{N}\mathbbm{1}\{Y_{i,t} \leq y_{t}\}.
\end{align*} 
The lower bound follows 
\begin{align*}
\sum_{i=1}^{N}\mathbbm{1}\{Y_{(N-i+1),0} \leq y_{0}\}\mathbbm{1}\{Y_{(i),1} \leq y_{1}\} 
= \sum_{i=1}^{N}\left( 1 -\mathbbm{1}\{Y_{(N-i+1),0} > y_{0}\} \right)\mathbbm{1}\{Y_{(i),1} \leq y_{1}\} \\
\geq \sum_{i=1}^{N}\mathbbm{1}\{Y_{i,1} \leq y_{1}\} - \min\left(\sum_{i=1}^{N}\mathbbm{1}\{Y_{i,1} \leq y_{1}\}, \sum_{i=1}^{N}\mathbbm{1}\{Y_{i,0} > y_{0}\}\right) \\
= \max\left(\sum_{i=1}^{N}\mathbbm{1}\{Y_{i,1} \leq y_{1}\}  + \sum_{i=1}^{N}\mathbbm{1}\{Y_{i,0} \leq y_{0}\} - N,0\right). 
\end{align*}  \looseness=-1 

\subsection{Proposition 1 from Section 3}
The DPO for the finite approximation to the double randomized experiment is 
\begin{align*}
F_{N}(y_{1},y_{0}) = \frac{1}{N^2}\sum_{i=1}^{N}\sum_{j=1}^{N}\sum_{k=1}^{N}\sum_{l=1}^{N}\prod_{t \in \{0,1\}}\mathbbm{1}\{Y_{kl,t} \leq y_{t}\}P_{ik,t}P_{jl,t}.
\end{align*}
We show that 
\begin{align*}
\max\left(\sum_{r=1}^{N}\left(\lambda_{r1}^{2}+\lambda_{r0}^{2}\right) - N^2,\sum_{r=1}^{N}  \lambda_{r1}\lambda_{s_{N}(r)0},0\right) \leq N^2 F_{N}(y_{1},y_{2}) \\
\leq \min\left(\sum_{r=1}^{N}\lambda_{r1}^{2},\sum_{r=1}^{N}\lambda_{r0}^{2},\sum_{r=1}^{N}\lambda_{r1}\lambda_{r0}\right)
\end{align*}
 where $\lambda_{rt}$ is the $r$th largest eigenvalue of the matrix $\mathbbm{1}\{Y_{t} \leq y_{t}\}$ and $s_{N}(r) = N-r+1$. Our sketch has two parts. The second part is based on work by \cite{finke1987quadratic} and relies on the following result due to  \cite{birkhoff1946three}. We say that a square matrix is doubly stochastic if its entries are nonnegative and if every row and column sum to $1$.   \looseness=-1 
 
 \begin{flushleft}
\textbf{Theorem (Birkhoff):} If $M$ is doubly stochastic then there exist an $m \in \mathbb{N}$, $\alpha_{1},...,\alpha_{m} > 0$, and permutation matrices $P_{1},...,P_{m}$ such that $\sum_{t=1}^{m}\alpha_{t} = 1$ and $M_{ij} = \sum_{t=1}^{m}\alpha_{t}P_{ij,t}$.
\end{flushleft}  \looseness=-1 
 
 \subsubsection{Sketch of proof of Proposition 1, part 1}
We first show $\max\left(\sum_{r=1}^{N}\left(\lambda_{r1}^{2}+\lambda_{r0}^{2}\right) - N^2,0\right)
\leq N^2 F_{N}(y_{1},y_{0}) \leq \min\left(\sum_{r=1}^{N}\lambda_{r1}^{2},\sum_{r=1}^{N}\lambda_{r0}^{2}\right)$. Write $N^{2}F_{N}(y_{1},y_{0}) = \sum_{r=1}^{N^{2}}\sum_{s=1}^{N{^2}}\prod_{t \in \{0,1\}}\mathbbm{1}\{\tilde{Y}_{r,t} \leq y_{t}\}\tilde{P}_{rs,t}$ where $i_{r} = \lfloor \frac{r-1}{N}\rfloor + 1$, $j_{r} = r-N\lfloor \frac{r-1}{N}\rfloor$, $\tilde{Y}_{r,t} = Y_{i_{r}j_{r},t}$, and $\tilde{P}_{rs,t} = P_{i_{r}i_{s},t}P_{j_{r}j_{s},t}$. In words, $\tilde{Y}_{t}$ and $\tilde{P}_{t}$ are vectorized versions of $Y_{t}$ and $P_{t}\otimes P_{t}$ formed by iteratively appending their rows. Theorem 368 implies that 
\begin{align*}
\sum_{r=1}^{N^{2}}\mathbbm{1}\{\tilde{Y}_{(N^2 - r + 1),0} \leq y_{0}\}\mathbbm{1}\{\tilde{Y}_{(r),1} \leq y_{1}\} \leq 
N^{2}F_{N}(y_{1},y_{0}) 
\leq \sum_{r=1}^{N^{2}}\mathbbm{1}\{\tilde{Y}_{(r),0} \leq y_{0}\}\mathbbm{1}\{\tilde{Y}_{(r),1} \leq y_{1}\}
\end{align*}
and so following the arguments of Section 4.2.1, we have 
\begin{align*}
\max\left(\sum_{r=1}^{N^{2}}\mathbbm{1}\{\tilde{Y}_{r,1} \leq y_{1}\} + \sum_{r=1}^{N^{2}}\mathbbm{1}\{\tilde{Y}_{r,0} \leq y_{0}\} - N^{2},0\right)
\leq N^{2}F_{N}(y_{1},y_{0}) 
\leq \min_{t \in \{0,1\}}\sum_{r=1}^{N^{2}}\mathbbm{1}\{\tilde{Y}_{r,t} \leq y_{t}\}.
\end{align*}
The bounds follow since
\begin{align*}
\sum_{r = 1}^{N^{2}}\mathbbm{1}\{\tilde{Y}_{r,t} \leq y_{t}\}  
= \sum_{i = 1}^{N}\sum_{j=1}^{N}\mathbbm{1}\{Y_{ij,t} \leq y_{t}\}  
= \sum_{r=1}^{N}\lambda_{rt}^{2}.
\end{align*}  \looseness=-1

 \subsubsection{Sketch of proof of Proposition 1, part 2}
We now show $\sum_{r=1}^{N}  \lambda_{r1}\lambda_{s_{N}(r)0} \leq N^2 F_{N}(y_{1},y_{2}) \leq \sum_{r=1}^{N}\lambda_{r1}\lambda_{r0}$. These bounds follow
  \begin{align*}
\sum_{i,j=1}^{N}\sum_{k,l=1}^{N}\prod_{t \in \{0,1\}}\mathbbm{1}\{Y_{kl,t} \leq y_{t}\}P_{ik,t}P_{jl,t}
= \sum_{i,j = 1}^{N}\left[\sum_{r=1}^{N}\lambda_{r1}\phi_{ir,1}\phi_{jr,1}\right]\left[\sum_{s=1}^{N}\lambda_{s0}\phi_{is,0}\phi_{js,0}  \right]
= \sum_{r,s = 1}^{N} \lambda_{r1}\lambda_{s0}W^{\phi}_{rs}
\end{align*}
where $(\lambda_{rt},\phi_{rt})$ is the $r$th eigenvalue and eigenvector pair of $\sum_{k,l=1}^{N}\mathbbm{1}\{Y_{kl,t} \leq y_{t}\}P_{ik,t}P_{jl,t}$ and $W^{\phi}$, a matrix with $rs$th entry $W_{rs}^{\phi} = \left[\sum_{i=1}^{N}\phi_{ir,1}\phi_{is,0}\right]^{2}$, is the Hadamard square of an orthogonal matrix and so is doubly stochastic. Birkhoff's Theorem implies 
\begin{align*}
\sum_{i,j=1}^{N}\sum_{k,l=1}^{N}\prod_{t \in \{0,1\}}\mathbbm{1}\{Y_{kl,t} \leq y_{t}\}P_{ik,t}P_{jl,t} 
 = \sum_{r,s = 1}^{N} \lambda_{r1}\lambda_{s0}W_{rs}^{\phi} = \sum_{k=1}^{K}\alpha^{\phi}_{k}\sum_{r,s=1}^{N}  \lambda_{r1}\lambda_{s0}P^{\phi}_{rs,k}.
\end{align*}
where  $\alpha^{\phi}_{1},...,\alpha^{\phi}_{k} > 0$, $\sum_{k=1}^{K}\alpha^{\phi}_{k} = 1$, $P^{\phi}_{1},...,P^{\phi}_{K}$ are permutation matrices, and $W_{rs}^{\phi} = \sum_{k=1}^{K}\alpha^{\phi}_{k}P^{\phi}_{rs,k}$. Theorem 368 implies that for any permutation matrix $P$
\begin{align*}
\sum_{r=1}^{N}  \lambda_{r1}\lambda_{s_{N}(r)0} \leq \sum_{r,s=1}^{N}  \lambda_{r1}\lambda_{s0}P_{rs} \leq \sum_{r=1}^{N}  \lambda_{r1}\lambda_{r0}.
\end{align*} 
The claim follows.  \looseness=-1

\subsubsection{Discussion}
Our bounds on the DPO follow by intersecting those from parts 1 and 2. Each part describes a different relaxation of the intractable QAP. Take for instance the upper bound
\begin{align}\label{QAP_upper}
\max_{P \in \mathcal{P}_{N}}\frac{1}{N^{2}}\sum_{i=1}^{N}\sum_{j=1}^{N}\sum_{k=1}^{N}\sum_{l=1}^{N}\prod_{t \in \{0,1\}}\mathbbm{1}\{Y_{kl,t} \leq y_{t}\}P_{ik,t}P_{jl,t}.
\end{align}
where $\mathcal{P}_{N}$ is the set of all $N \times N$ permutation matrices. Part 1 bounds it from above with 
\begin{align}\label{relax1}
\max_{P \in \mathcal{P}_{N^{2}}}\frac{1}{N^{2}}\sum_{i,j=1}^{N}\sum_{k,l=1}^{N}\prod_{t \in \{0,1\}}\mathbbm{1}\{Y_{kl,t} \leq y_{t}\}P_{r(i,j)r(k,l),t}
\end{align}
where $\mathcal{P}_{N^{2}}$ describes permutations of pairs of agents and $r(i,j) = N(i-1)+j$. Intuitively, this relaxation treats the $N \times N$ outcome matrices as vectors of length $N^2$ and uses the fact that $\{P_{ij}\}_{i,j = 1}^{N} \in \mathcal{P}_{N}$ implies that $\{P_{ik}P_{jl}\}_{i,j,k,l = 1}^{N} \in \mathcal{P}_{N^{2}}$. Whereas  (\ref{QAP_upper}) is an intractable QAP, (\ref{relax1}) is linear and can be bounded using Theorem 368.  \looseness=-1 

Part 2 bounds the QAP from above with 
\begin{align}\label{relax2}
\max_{O \in \mathcal{O}_{N}}\frac{1}{N^{2}}\sum_{i,j=1}^{N}\sum_{k,l=1}^{N}\prod_{t \in \{0,1\}}\mathbbm{1}\{Y_{kl,t} \leq y_{t}\}O_{ik,t}O_{jl,t}
\end{align}
 where $\mathcal{O}_{N}$ is the set of orthogonal $N\times N$ matrices. This is an upper bound because $\mathcal{P}_{N} \subset \mathcal{O}_{N}$. The insight, see \cite{finke1987quadratic}, is to use the fact that $\{O_{ij}\}_{i,j=1}^{N} \in \mathcal{O}_{N}$ implies that $\{O_{ij}^{2}\}_{i,j =1}^{N} \in \mathcal{D}^{+}_{N}$ where $\mathcal{D}_{N}^{+}$ is the set of doubly stochastic $N \times N$ matrices. This allows us to rewrite  (\ref{relax2}) as $\max_{W \in \mathcal{D}_{N}^{+}}\sum_{r,s}\lambda_{r1}\lambda_{rs}W_{rs}$ which is linear in $W$ and can be bounded using  Birkhoff's Theorem and  Theorem 368.  \looseness=-1
 
 Our full proof of Proposition 1 as stated in Section 3 is complicated by the fact that the infinite dimensional analog of $W^{\phi}$ is not generally doubly stochastic and so Birkhoff's Theorem cannot be directly applied. We address this problem by first approximating the function $\mathbbm{1}\{Y_{t} \leq y_{t}\}$ with a finite dimensional matrix, applying the logic of Part 2, and then showing convergence as the dimensions of the matrix are taken to infinity. \cite{whitt1976bivariate} uses a similar strategy to demonstrate his Theorem 2.1 (our Standard Result 1) in the setting of Section 2.  \looseness=-1 

These bounds describe two of many possible relaxations of the intractable QAP, see broadly \cite{cela2013quadratic}, Section 2. We chose these relaxations because they are straightforward to compute, characterize, and appear to work well in practice. Intersecting our bounds with others may lead to smaller identified sets for the DPO and DTE, but potentially at the cost of greater computational complexity or statistical uncertainty. We leave this to future work.  \looseness=-1 

\section{Extensions}
We describe some extensions to the Section 3 framework. Additional details can be found in Online Appendix Sections C and D. \looseness=-1 

\subsection{Asymmetric outcome matrices}
Asymmetric matrices or matrices indexed by two different populations can be handled in the following way. A population of workers and firms are randomized (or as good as randomized in the case of a quasi experiment) to two groups. Worker and firm pairs are then assigned a binary treatment $t \in \{0,1\}$ depending on the individual group assignments. For example, the groups may correspond to economic regions where one region is exposed to a trade shock ($t=1$) and the other is not ($t=0$). 

Potential outcomes are defined for each worker and firm pair. We index the workers with latent types in $[0,1]$ and firms with latent types in $[2,3]$. The potential outcomes are then represented by  $(Y_{0}^{*},Y_{1}^{*}):S \to \mathbb{R}^{2}$ where $S = [0,1]\times[2,3]$. For example, $Y_{t}^{*}(u,v)$ may describe the potential wage that a worker of type $u$ would earn at a firm of type $v$ when exposed or not exposed to the trade shock. Following Section 3.1.4, we assume that the researcher observes $Y_{1}$ and $Y_{0}$ where $Y_{t}(\phi_{t}(u),\psi_{t}(v)) =Y_{t}^{*}(u,v)$ for unknown measure preserving functions $\phi_{t}$ and $\psi_{t}$. The DPO is $F(y_{1},y_{0}) = \int\int\prod_{t \in \{0,1\}}\mathbbm{1}\{Y_{t}(\phi_{t}(u),\psi_{t}(v)) \leq y_{t}\} dudv$ and the DTE is $\Delta(y) = \int\int\mathbbm{1}\{Y_{1}(\phi_{1}(u),\psi_{1}(v))- Y_{0}(\phi_{0}(u),\psi_{0}(v)) \leq y\} dudv$. \looseness=-1 

We symmetrize the potential outcome matrices along the lines of \cite{auerbach2022testing}. Let  $S^{2} =  \left([0,1]\cup[2,3]\right) \times \left([0,1]\cup[2,3]\right)$ and define $(Y_{0}^{\dagger},Y_{1}^{\dagger}): S^{2} \to \mathbb{R}^{2}$ so that
\begin{equation*}
  Y_{t}^{\dagger}(u,v) :=
    \begin{cases}
      Y_{t}(u,v) & \text{if $(u,v) \in [0,1]\times [2,3]$} \\
      Y_{t}(v,u) &  \text{if $(u,v) \in [2,3]\times [0,1]$}\\
      0 & \text{otherwise}
    \end{cases}       
\end{equation*}
and  $\varphi_{t}(u) := \phi_{t}(u)\mathbbm{1}\{u \in [0,1]\} + \psi_{t}(u)\mathbbm{1}\{u \in [2,3]\}$ is measure preserving. Then the DPO is equal to $\frac{1}{2}\int\int\prod_{t \in \{0,1\}}\mathbbm{1}\{Y^{\dagger}_{t}(\varphi_{t}(u),\varphi_{t}(v)) \leq y_{t}\} dudv$. Since $Y^{\dagger}$ is symmetric and defined on one population (the population of workers and firms), the logic of Section 3 can be applied to bound the DPO and DTE. One can similarly define the STE using the eigenvalues of $Y_{t}^{\dagger}$. \looseness=-1 


\subsection{Row and column heterogeneity}
Spectral methods may perform poorly when there is nontrivial heterogeneity in the row and column variances of the outcome matrices, see also \cite{auerbach2022testing}. To address this issue we adapt arguments of \cite{finke1987quadratic}. We decompose $\mathbbm{1}\{Y_{t}^{*}(u,v) \leq y_{t}\} = \alpha_{t}(u) + \alpha_{t}(v) + \epsilon_{t}(u,v)$ where $\int \epsilon_{t}(s,v)ds =  \int \epsilon_{t}(u,s)ds = 0$ for every $u,v \in [0,1]$. The DPO becomes 
\begin{align*}
F(y_{1},y_{0}) = \int\int \prod_{t \in \{0,1\}}\left(\alpha_{t}(\varphi_{t}(u)) + \alpha_{t}(\varphi_{t}(v)) + \epsilon_{t}(\varphi_{t}(u),\varphi_{t}(v))\right) dudv \\
= \int\int \prod_{t \in \{0,1\}}\left(\alpha_{t}(\varphi_{t}(u)) + \alpha_{t}(\varphi_{t}(v))\right) dudv + \int\int \prod_{t \in \{0,1\}}\epsilon_{t}(\varphi_{t}(u),\varphi_{t}(v))dudv.  
\end{align*}
The summand $\int\int \prod_{t \in \{0,1\}}\left(\alpha_{t}(\varphi_{t}(u)) + \alpha_{t}(\varphi_{t}(v))\right) dudv$ can be bounded along the lines of Theorem 368 in Section 4.2. The summand $ \int\int \prod_{t \in \{0,1\}}\epsilon_{t}(\varphi_{t}(u),\varphi_{t}(v))dudv$ can be bounded using the arguments from Section 4.3. One can similarly decompose $Y_{t}^{*}(u,v)$ and redefine the STE using the quantiles of $\alpha_{t}$, $\alpha_{t}$ and the eigenvalues of $\epsilon_{t}$.  See Online Appendix Section D.1 for details. \looseness=-1 

\subsection{Estimation and inference}
In many settings the researcher can exploit a symmetry in the experimental design to conduct randomization-based inference. We illustrate three ways of doing this using the motivating examples from Section 1.1 in Online Appendix Section D.2. Alternatively, if the researcher observes only noisy signals of $Y^{*}_{1}$ and $Y^{*}_{0}$ due to random sampling, missing data, measurement error, etc. then they can estimate the STE or bounds on the DPO and DTE by replacing the eigenvalues of $Y_{t}$ with empirical analogs. We formalize this strategy, give sufficient conditions for consistency, and sketch a strategy for statistical inference in Online Appendix Section D.3. \looseness=-1

\subsection{Spillovers}
One motivation for implementing a double randomized experimental design is to characterize social interactions, market externalities, or other spillovers between agents. Our framework and results can be directly applied to characterize distributional analogs of such spillover effects in many settings. The kinds of spillovers that are identified generally depend on the experimental design and assumptions about the agent interactions, see for instance \cite{bajari2021multiple}. The following example, where we consider heterogeneous spillover effects under the assumption of strong no-interference (\cite{bajari2021multiple}'s Assumption 5.1), is related to their Section 6. We also provide three additional concrete examples concerning spillover effects under local interference (\cite{bajari2021multiple}'s Assumption 5.4), market externlaities, and social interactions in Online Appendix Sections C.1 and C.2.  \looseness=-1 

Consider the setting of the buyer-seller experiment in Example 4 of Section 1.1. Suppose the researcher is interested in how an information treatment assigned to buyers affects their transactions with untreated sellers. To do this, they independently randomize the buyers and sellers to two groups. Only pairs of buyers and sellers that are both assigned to the first group are treated, but transactions may occur between any buyer-seller pair.  \looseness=-1 

\cite{bajari2021multiple} call this a conjunctive simple multiple randomization design in their Definition 8. They define the average buyer spillover effect to be the average difference in the potential transactions between the event that the buyer but not the seller  is assigned to the treated group and the event that both the buyer and the seller are assigned to the untreated group. To characterize this spillover effect using our notation, let $(Y_{1}^{*},Y_{0}^{*}) : [0,1] \to \mathbb{R}^{2}$ record the potential transactions for pairs of buyers and sellers under the two events. \cite{bajari2021multiple}'s average buyer spillover effect is $\int \int \left(Y^{*}_{1}(u,v)-Y^{*}_{0}(u,v) \right)dudv$, or equivalently, $\int \int \left(Y_{1}(u,v)-Y_{0}(u,v) \right)dudv$ because $Y_{t}$ and $Y_{t}^{*}$ are equivalent up to a measure preserving transformation (see also their Lemma 1). After symmetrization as in Section 5.1, the arguments of Section 3 characterize distributional analogs of the average buyer spillover effect. That is, the joint distribution of potential transactions $\int\int\Pi_{t \in \{0,1\}}\mathbbm{1}\{Y^{*}_{t}(u,v) \leq y_{t}\}dudv$ and the distribution of buyer spillover effects $\int\int\mathbbm{1}\{Y_{1}^{*}(u,v) - Y_{0}^{*}(u,v) \leq y\}dudv$.  \looseness=-1

\subsection{Covariates and instruments}
\cite{abadie2002instrumental,chernozhukov2005iv,firpo2007efficient} use covariates or instruments to allow for endogeneity or characterize various conditional treatment effects. Their parameters can be written as solutions to an extremum estimation problem building on the framework of  \cite{koenker1978regression}. We have results for an analogous approach to incorporate covariates and instruments into the framework of this paper, but they are sufficiently complicated to warrant a separate, forthcoming paper.  \looseness=-1

\section{Two empirical demonstrations}
We revisit Examples 1 and 3 from Section 1.1 and find policy relevant heterogeneity in the effect of treatment that might otherwise be missed by focusing exclusively on average effects. An R package can be found at \url{https://github.com/yong-cai/MatrixHTE}. \looseness=-1 

\subsection{Example 1: risk sharing}
Our first demonstration follows \cite{comola2021treatment}.\footnote{The data can be found on the Review of Economics and Statistics data repository: \url{https://doi.org/10.7910/DVN/K6QU2J}.} Households in nineteen villages are randomly provided with a savings account.  A main finding of the authors is that  ``the intervention increased the transfers towards others and the overall informal financial activity in the villages, suggesting that there might be complementarities between formal savings and informal financial networks.'' Our methodology suggests a more complicated relationship between formal savings and informal financial networks. In particular, we find that the treatment also decreased transfers between a nontrivial fraction of household pairs. \looseness=-1 

Table 1 reports our bounds on the joint distribution of risk sharing links across all $19$ villages. Treatment $1$ is the event that both households are provided with a savings account, treatment $0$ is the event that neither household is provided with a savings account, and $Y_{ij,t}$ indicates whether household pair $ij$ reports a risk sharing link under treatment $t$. To construct Table 1, we first compute bounds on the distribution of potential outcomes for each village allowing for row and column heterogeneity as in Section 5.2. We then average the bounds over the $19$ villages, weighting by the number of households in each village. Table 1 also gives bounds on the distribution of treatment effects since $P(Y_{ij,1}-Y_{ij,0} = 1) = P(Y_{ij,1} = 1, Y_{ij,0} = 0)$, $P(Y_{ij,1}-Y_{ij,0} = -1) = P(Y_{ij,1} = 0, Y_{ij,0} = 1)$, and $P(Y_{ij,1}-Y_{ij,0} = 0) = P(Y_{ij,1} = 1, Y_{ij,0} = 1)+P(Y_{ij,1} = 0, Y_{ij,0} = 0)$. \looseness=-1 

\begin{table}[htbp]
	\vspace{5mm}
	\centering
	\title{Table 1: Bounds on the joint distribution of risk sharing links} \\ \vspace{5mm}
	\begin{tabular}{lccc}\hline\hline
	      &       & Lower & Upper \\
		\midrule
		$P(Y_{ij,1} = 1, Y_{ij,0} = 1)$ &       & 0.000 & 0.010 \\
		$P(Y_{ij,1} = 1, Y_{ij,0} = 0)$ &       & 0.017 & 0.027  \\
		$P(Y_{ij,1} = 0, Y_{ij,0} = 1)$ &       & 0.010 & 0.021  \\
		$P(Y_{ij,1} = 0, Y_{ij,0} = 0)$ &       & 0.970 & 0.981 \\
		\hline\hline
	\end{tabular}%
\begin{flushleft} \footnotesize \linespread{.75} Table 1 reports bounds on the distribution of potential outcomes using data from \cite{comola2021treatment}. \normalsize \end{flushleft}
\end{table}%

We find a positive lower bound on $P(Y_{ij,1} = 1, Y_{ij,0} = 0)$ which implies that the savings accounts create links. This is consistent with the main finding of \cite{comola2021treatment}. We also find a positive lower bound on $P(Y_{ij,1} = 0, Y_{ij,0} = 1)$ which implies that the savings accounts also destroys links. This lower bound is at least a third of the upper bound on $P(Y_{ij,1} = 1, Y_{ij,0} = 0)$ and so is nontrivial in magnitude. However, the aggregate effect of the savings accounts on link creation is unlikely to be large and negative. This is because $P(Y_{ij,1} = 1, Y_{ij,0} = 0) - P(Y_{ij,1} = 0, Y_{ij,0} = 1)$ is not less than $-0.004$. The total change in the number of links is either positive or not substantially different from zero.   \looseness=-1

Figure 1 shows a smoothed density plot of the spectral treatment effects on the treated (STT) for households in all 19 villages.  For reference, we also show a smoothed density plot for a collection of conditional average treatment effects (CATE). To construct the CATE, we first bin the households by size and the number of children. Then for every pair of bins, we compute the difference in the fraction of links between households under both treatments. The CATE plot is then the smoothed density of the differences across treatments for every bin pair, weighting by the number of households in each bin. Figure 1 shows that the STT and CATE are similarly distributed, even though the STT is constructed without the use of any covariate information.  \looseness=-1

\begin{figure}
	\vspace{5mm}
	\centering
	\title{Figure 1: Two Characterizations of the Distribution of Treatment Effects} \\ \vspace{5mm}
	\includegraphics[width=0.47\linewidth]{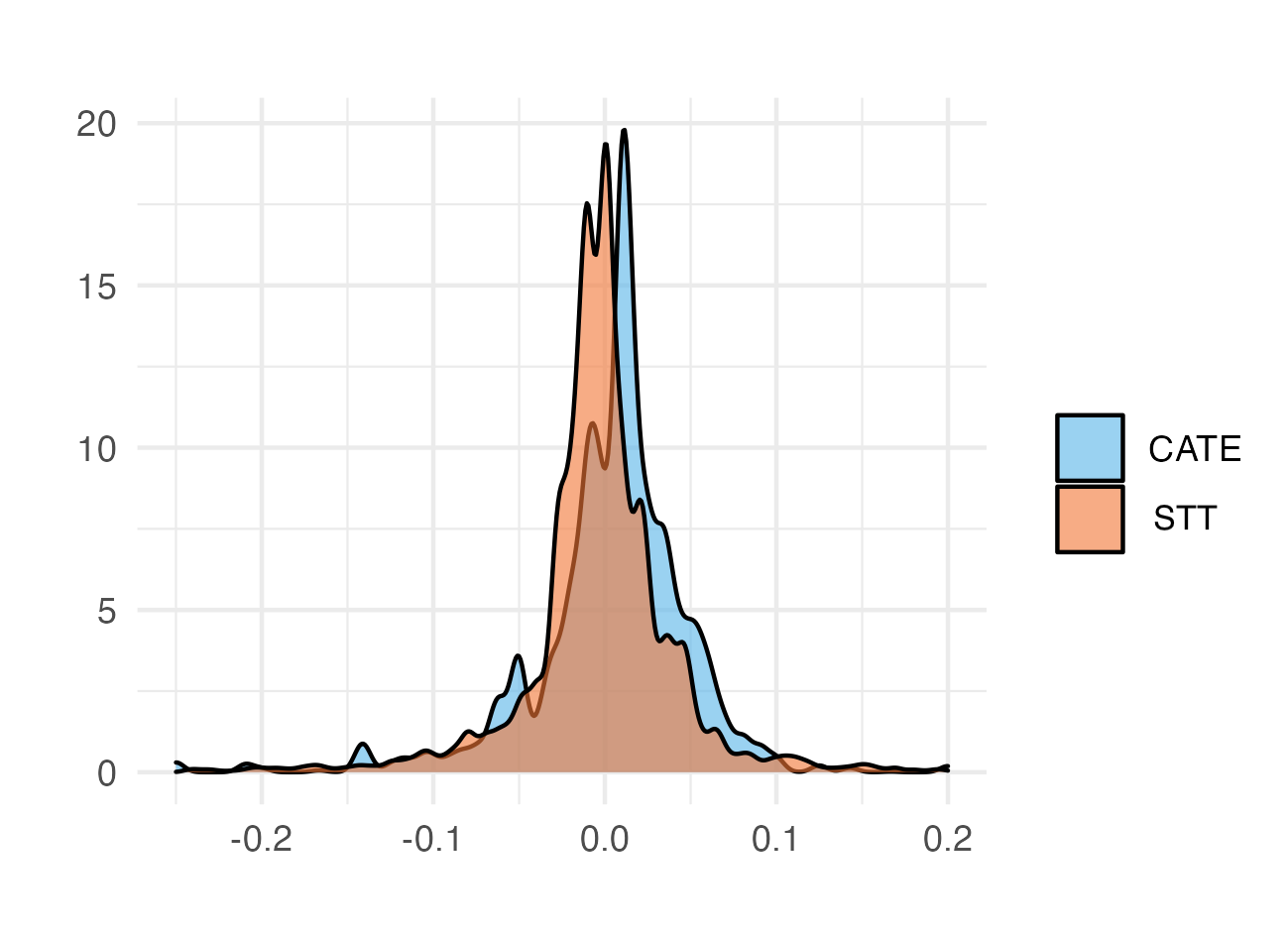}
	\begin{flushleft} \footnotesize \linespread{.75} Figure 1 shows two characterizations of the distribution of treatment effects using data from \cite{comola2021treatment}.  The distribution of spectral treatment effect on the treated (STT) is plotted in orange. The distribution of average treatment effects conditional on household size and the number of children (CATE) is plotted in blue. \normalsize \end{flushleft}
\end{figure}

\subsection{Example 3: auction format}
Our second demonstration follows \cite{athey2011comparing}.\footnote{The data can be found on Phil Haile's website: \url{http://www.econ.yale.edu/~pah29/timber/timber.htm}. We restrict attention to a subsample of auctions proposed by \cite{schuster1994sealed} in which the auction format is randomly assigned.} Tracts of forest land are sold by either an open or sealed bid auction format. A main finding of the authors is that ``sealed bid auctions attract more small bidders [and] shift the allocation towards these bidders.'' Our methodology suggests a more complicated relationship between auction format and firm entry. In particular, we find that the sealed bid format also discourages large firms from entry.    \looseness=-1  

Table 2 reports our bounds on the joint distribution of entry decisions. Treatment $1$ is the sealed bid format, treatment $0$ is the open format, and $Y_{ij,t}$ indicates whether firm $i$ bids on tract $j$ under format $t$. To construct Table 2, we symmetrize the outcome matrices as in Section 5.1 and allow for row and column heterogeneity as in Section 5.2. We report results for the full sample of firms, as well as large and small firms seperately. \looseness=-1  

\begin{table}[htbp]
	\vspace{5mm}
	\centering
	\title{Table 2: Bounds on the joint distribution of entry} \\ \vspace{5mm}
	    \begin{tabular}{clcc} \hline\hline
          &       & Lower & Upper \\
    \midrule
    \multicolumn{1}{c}{\multirow{4}[0]{*}{Full Sample}} 
    	  & P($Y_{ij, 1} = 1$, $Y_{ij, 0} = 1$) & 0.000 & 0.016 \\
          & P($Y_{ij, 1} = 1$, $Y_{ij, 0} = 0$) & 0.017 & 0.033 \\
          & P($Y_{ij, 1} = 0$, $Y_{ij, 0} = 1$) & 0.015 & 0.031 \\
          & P($Y_{ij, 1} = 0$, $Y_{ij, 0} = 0$) & 0.965 & 0.981 \\
    \midrule
    \multicolumn{1}{c}{\multirow{4}[0]{*}{Small Firms}} 
     	  & P($Y_{ij, 1} = 1$, $Y_{ij, 0} = 1$) & 0.000 & 0.023 \\
          & P($Y_{ij, 1} = 1$, $Y_{ij, 0} = 0$) & 0.019 & 0.042 \\
          & P($Y_{ij, 1} = 0$, $Y_{ij, 0} = 1$) & 0.024 & 0.047 \\
          & P($Y_{ij, 1} = 0$, $Y_{ij, 0} = 0$) & 0.950 & 0.973 \\
    \midrule
    \multicolumn{1}{c}{\multirow{4}[0]{*}{Large Firms}} 
          & P($Y_{ij, 1} = 1$, $Y_{ij, 0} = 1$) & 0.000 & 0.098 \\
          & P($Y_{ij, 1} = 1$, $Y_{ij, 0} = 0$) & 0.056 & 0.160 \\
          & P($Y_{ij, 1} = 0$, $Y_{ij, 0} = 1$) & 0.103 & 0.207 \\
          & P($Y_{ij, 1} = 0$, $Y_{ij, 0} = 0$) & 0.766 & 0.870 \\
          \hline\hline
    \end{tabular}
\begin{flushleft} \footnotesize \linespread{.75} Table 2 reports bounds on the joint distribution of entry using data from \cite{schuster1994sealed}. \normalsize \end{flushleft}
\end{table}%

For the full sample, we find a strictly positive lower bound on $P(Y_{ij, 1} = 1, Y_{ij, 0} = 0)$ which implies that the sealed bid design encourages some firms to enter. However, we also find evidence that it discourages the entry of other firms. For the population of small and large firms separately, we find that while the sealed bid design induces entry and exit for both types of firms, there is at least twice as much exit of large firms than exit and entry of small firms. This is consistent with the main finding of  \cite{athey2011comparing}, that sealed bid auctions shift the allocation towards small firms. However, our results suggest that the exit of large firms as well as the entry of small firms drives this outcome.   \looseness=-1

Figure 2 shows smoothed density plots of the STT and the CATE using firm size and tract location as covariates. The two distributions have similar centers with the bulk of the treatment effect above $0$. They also both have large left tails suggesting large negative treatment effects for a small number of firms and tracts. However, there are some noticable differences between the two plots. For example, the CATE plot concentrates at a few discrete spikes, which is not a feature of the STT.    \looseness=-1

\begin{figure}[htbp]
	\vspace{5mm}
	\centering
	\title{Figure 2: Two Characterizations of the Distribution of Treatment Effects} \\ \vspace{5mm}
	\includegraphics[width=0.47\linewidth]{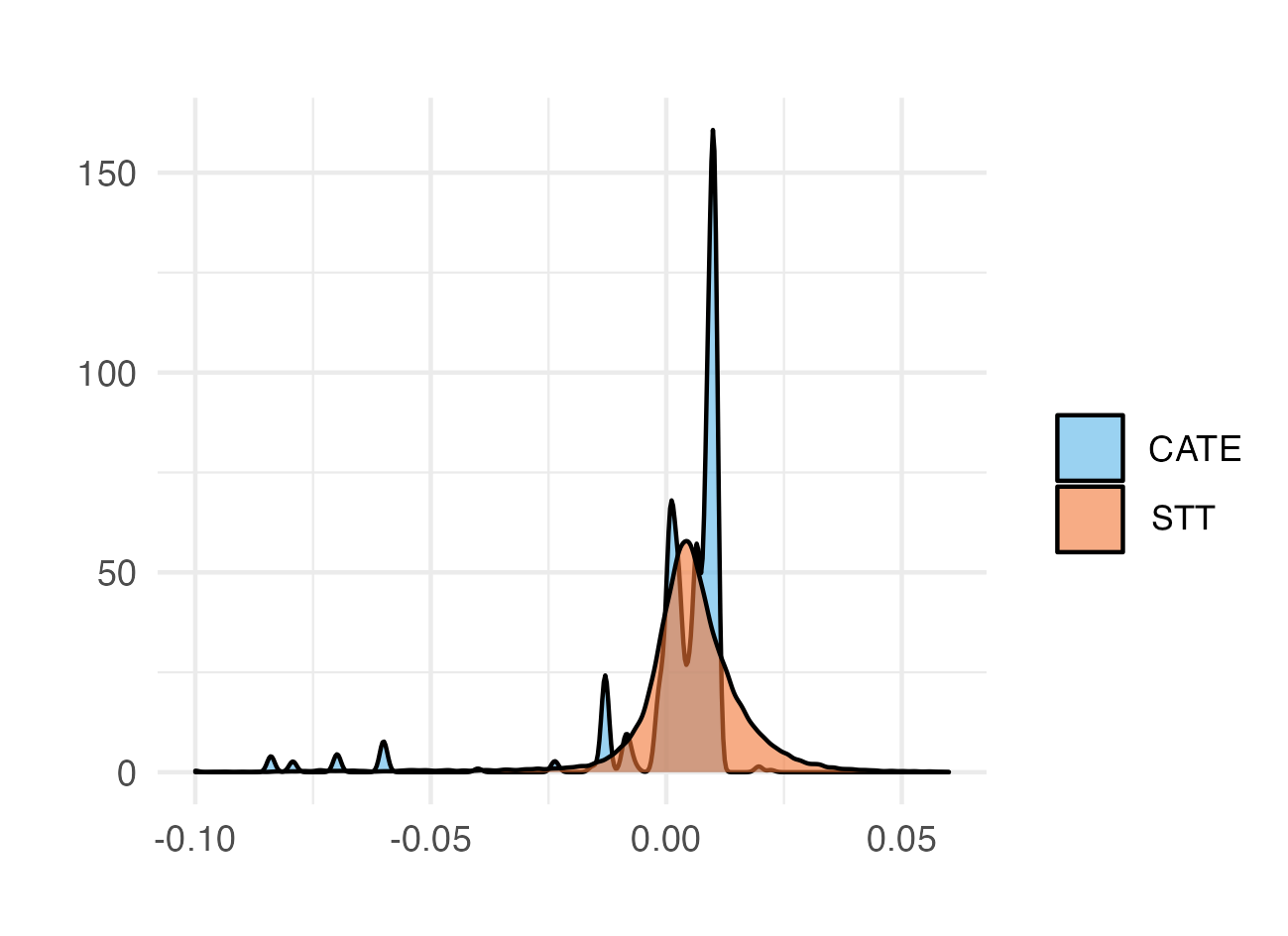}
	\begin{flushleft} \footnotesize \linespread{.75} This figure shows two characterizations of the distribution of treatment effects using data from \cite{schuster1994sealed}.  The distribution of spectral treatment effects on the treated (STT) is plotted in orange. The distribution of average treatment effects conditional on tract location and firm size (CATE) is plotted in blue. \normalsize \end{flushleft}
\end{figure}

\section{Conclusion}
This paper characterizes the distribution of treatment effects in a double randomized experiment where a matrix of outcomes is associated with each treatment. We propose bounds on the distribution of treatment effects and a matrix analog of quantile treatment effects. Our results are based on a new matrix analog of the Fr\'echet-Hoeffding bounds that play a key role in the standard theory. We illustrate our methodology with two empirical demonstrations and find policy relevant heterogeneity that might be missed by focusing exclusively on averages.  \looseness=-1 

\bibliographystyle{aer}
\bibliography{literature}

\appendix

\section{Appendix: proof of Propositions 1-4}
\subsection{Definitions and lemmas}

\subsubsection{Hilbert-Schmidt integral operators and function embeddings}
Our  Section 3 model of a double randomized experiment with outcome matrices uses bounded symmetric measurable functions to describe the potential outcomes associated with pairs of agents in the population. We describe some relevant properties of such functions here.\looseness=-1 

Any bounded symmetric measurable function $f: [0,1]^{2}\to \mathbb{R}$ defines a compact symmetric Hilbert-Schmidt integral operator $T_{f}: L_{2}[0,1] \to L_{2}[0,1]$ where $(T_{f} g)(u) = \int f(u,\tau)g(\tau)d\tau$. It has a bounded countable multiset of real eigenvalues $\{\lambda_{r}\}_{r \in \mathbb{N}}$ with $0$ as the only limit point. It also admits the spectral decomposition $\sum_{r}\lambda_{r}\phi_{r}(u)\phi_{r}(v)$ where $\phi_{r}: [0,1] \to \mathbb{R}$ is the eigenfunction associated with eigenvalue $\lambda_{r}$, i.e. $\int f(u,\tau)\phi_{r}(\tau)d\tau = \lambda_{r}\phi_{r}(u)$. The functions $\{\phi_{r}\}_{r\in \mathbb{N}}$ can be chosen to be orthogonal, i.e. $\int \phi_{r}(u)^{2}du = 1$ and $\int(\phi_{r}(u)- \phi_{s}(u))^{2}du = 2$ if $r \neq s$, and form a basis of $L_{2}[0,1]$. It follows that $\sum_{r}\lambda_{r}^{2} = \int\int f(u,v)^{2}dudv <  \infty$. See Lemma B.2 in Online Appendix Section B.1,  Section 7.5 of \cite{lovasz2012large}, or Chapter 9 of \cite{birman2012spectral}.  \looseness=-1 

Any square symmetric matrix can be represented by a bounded symmetric measurable function, sometimes called its function embedding. Let $F$ be an arbitrary $n\times n$ square symmetric matrix with $ij$th entry $F_{ij}$. The function embedding $f: [0,1]^{2}\to\mathbb{R}$ of $F$  is $f(u,v) = F_{\lceil nu\rceil \lceil nv\rceil}$ for $u,v \in [0,1]$. Intuitively, $f$ assigns the mass of types in the region $S_{i}^{n} := \left(\frac{i-1}{n},\frac{i}{n}\right]$ to observation $i$.  Similarly, any $n \times n$ permutation matrix $\Pi_{t}$ can be represented as a measure preserving transformation $\varphi_{t}(u) = \lceil nu \rceil - nu + \Pi_{t}(\lceil nu \rceil)$ where $\Pi_{t}(k) = \{l \in [n]: \Pi_{kl} = 1\}$. Intuitively, if $\Pi_{kl} = 1$, $\varphi_{t}$ maps the interval $\left(\frac{k-1}{n},\frac{k}{n}\right]$ monotonically to $\left(\frac{l-1}{n},\frac{l}{n}\right]$. See also Section 7.1 of \cite{lovasz2012large}.  \looseness=-1 
 
The eigenvalues of matrices and their function embeddings are scaled differently. Specifically, if $(\lambda^{F}_{r},\phi^{F}_{r})$ is a eigenvalue and eigenvector pair of $F$ then $(\lambda^{F}_{r}/n,\sqrt{n}\phi^{F}_{r}(\lceil n\cdot\rceil))$ is a eigenvalue and eigenfunction pair of $f$ where $\phi^{F}_{r}(i)$ is the $i$th entry of the vector $\phi^{F}_{r}$. \looseness=-1 

As introduced in Section 3, we take inner products of eigenvalues in a very specific way. That is, if $\{\lambda_{r1}\}$ and $\{\lambda_{r0}\}$ are the eigenvalues of functions $f_{1}$ and $f_{0}$ respectively, then $\sum_{r}\lambda_{r1}\lambda_{r0}$ refers to $\lim_{R \to \infty}\sum_{r \in [R]}\lambda_{r1}\lambda_{r0}$ where $\{\lambda_{r1}\}_{r \in [R]}$ and $\{\lambda_{r1}\}_{r \in [R]}$ are the $R$ largest (in absolute value) elements of $\{\lambda_{r1}\}$ and $\{\lambda_{r0}\}$ respectively (counting multiplicities) ordered to be decreasing.  

\subsubsection{Sets}
We use $\mathbb{N}$ for the set of positive integers, $\mathbb{R}$ for the set of real numbers, $[n]$ for the set $\{1,2,...,n\}$, $\mathcal{P}_{n}$ for the set of $n\times n$ permutation matrices (square matrices with $\{0,1\}$ valued entries and row and column sums equal to $1$), $\mathcal{D}_{n}^{+}$ for the set of $n\times n$ doubly stochastic matrices (square matrices with positive entries and row and column sums equal to $1$), $\mathcal{O}_{n}$ for the set of $n\times n$ orthogonal matrices (square matrices where any two rows or any two columns have inner product $1$ if they are the same or $0$ otherwise), and $\mathcal{M} := \{\phi: [0,1] \to [0,1]\text{ with } |\phi^{-1}(A)| = |A| \text{ for any measurable } A \subseteq [0,1]\}$ for the set of all measure preserving transformations on $[0,1]$ where $|A|$ refers to the Lebesgue measure of $A$. \looseness=-1

\subsubsection{Lemmas}
For the following Lemmas, let $f_{t}(u,v)$ refer to either $Y_{t}(\varphi_{t}(u),\varphi_{t}(v))$ or $\mathbbm{1}\{Y_{t}(\varphi_{t}(u),\varphi_{t}(v)) \leq y_{t}\}$ for an arbitrary $y_{t} \in \mathbb{R}$ and $\varphi_{t} \in \mathcal{M}$. For any $n\in \mathbb{N}$ let $S_{i}^{n} := \left(\frac{i-1}{n},\frac{i}{n}\right]$, $F_{t}^{n}$ be an $n\times n$ matrix with $F_{ij,t}^{n} \in \mathbb{R}$ as its $ij$th entry, and $f_{t}^{n}(u,v) = \sum_{ij}F_{ij,t}^{n}\mathbbm{1}\{u \in S_{i}^{n}, v \in S_{j}^{n}\}$ such that $\int\int\left(f_{t}(u,v)-f_{t}^{n}(u,v)\right)^{2}dudv \to 0$ as $n\to\infty$. In words, $F_{t}^{n}$ is an $n\times n$ matrix approximation of $f_{t}$ and $f_{t}^{n}$ is its function embedding. Intuitively, $f^{n}_{t}$ is a histogram approximation to the function $f$. The existence of such a sequence of matrices $F_{t}^{n}$ follows Lemma 1 below. Let $\{\lambda_{rt}\}$ denote the eigenvalues of  $f_{t}$ and $\{\lambda^{n}_{rt}\}$ the eigenvalues of  $f_{t}^{n}$.  \looseness=-1 

\begin{flushleft}
\textbf{Lemma 1:} For every bounded measurable $g: [0,1]^{2} \to \mathbb{R}$ there exists sequences $\{G^{n}\}_{n \in \mathbb{N}}$ and $\{g^{n}\}_{n \in \mathbb{N}}$ where $G^{n}$ is an $n\times n$ matrix with $ij$th entry $G^{n}_{ij}$ and $g^{n}:[0,1]^{2}\to\mathbb{R}$ with $g^{n}(u,v) = \sum_{ij}G_{ij}^{n}\mathbbm{1}\{u \in S_{i}^{n}, v \in S_{j}^{n}\}$ and $S_{i}^{n} := \left(\frac{i-1}{n},\frac{i}{n}\right]$ such that for every $\varepsilon > 0$ there exists an $m \in \mathbb{N}$ such that $\int\int\left(g(u,v)-g^{n}(u,v)\right)^{2}dudv \leq \varepsilon$ for every $n > m$.  
\end{flushleft}

\begin{flushleft}
\textbf{Proof of Lemma 1:} Fix an arbitrary $\varepsilon > 0$. Lusin's Theorem (see Lemma B1 Online Appendix Section B.1) implies that for any measurable $g: [0,1]^{2} \to \mathbb{R}$ and $\epsilon > 0$, there exists a compact $E^{\epsilon}_{g} \subseteq [0,1]^{2}$ of measure at least $1-\epsilon$ such that $g$ is continuous when restricted to $E^{\epsilon}_{g}$.  \newline

For any $N \in \mathbb{N}$, define the $N\times N$ matrix  $G^{N\epsilon}$ with $ij$th entry $G_{ij}^{N\epsilon} = \frac{\int\int_{(u,v) \in E^{\epsilon}_{g}} g_{t}(u,v)\mathbbm{1}\{u \in S_{i}^{N}, v \in S_{j}^{N}\}dudv}{\int\int_{(u,v) \in E^{\epsilon}_{g}}\mathbbm{1}\{u \in S_{i}^{N}, v \in S_{j}^{N}\}dudv}$  if $\int\int_{(u,v) \in E^{\epsilon}_{g}}\mathbbm{1}\{u \in S_{i}^{N}, v \in S_{j}^{N}\}dudv > 0$ and $G_{ij}^{N\epsilon} = 0$ otherwise. Let $g^{N\epsilon}$ be the function embedding of $G^{N\epsilon}$ so that for $u,v \in [0,1]$, $g^{N\epsilon}(u,v) = \sum_{ij}G_{ij}^{N\epsilon}\mathbbm{1}\{u \in S_{i}^{N}, v \in S_{j}^{N}\}$. Also let  $\bar{g} := \sup_{(u,v) \in [0,1]^{2}}|g(u,v)|^{2} < \infty$. \newline 

Since $g$ is continuous when restricted to $E^{\epsilon}_{g}$ there exists an $m(\epsilon) \in \mathbb{N}$ such that $\int\int_{(u,v) \in E^{ \epsilon}_{g}}\left(g(u,v) - g^{N\epsilon}(u,v)\right)^{2}dudv \leq \epsilon$ for every $N > m(\epsilon)$. In addition, $\int\int_{(u,v) \not\in E^{\epsilon}_{g}}\left(g(u,v) - g^{N\epsilon}(u,v)\right)^{2}dudv \leq 4\bar{g}\epsilon$  for every $N$. It follows that $\int\int_{(u,v) \in [0,1]^{2}}\left(g(u,v) - g^{N\epsilon}(u,v)\right)^{2}dudv \leq \left(1+4\bar{g}\right)\epsilon$  for every $N > m(\epsilon)$.  \newline

Let $e^{\dagger}(N) := \inf\{e > 0: m(e) \leq N\}$ where $e^{\dagger}(N) \to 0$ as $N\to \infty$ because $m(\epsilon) \in \mathbb{N}$ for every $\epsilon > 0$. For every $n \in \mathbb{N}$, define $G^{n} = G^{n e^{\dagger}(n)}$ and  $g^{n} = g^{n e^{\dagger}(n)}$. Then $\int\int_{(u,v) \in [0,1]^{2}}\left(g(u,v) - g^{n}(u,v)\right)^{2}dudv \leq \left(1+4\bar{g}\right)e^{\dagger}(N)$ for all $n > m(e^{\dagger}(N))$ and $N \in \mathbb{N}$. The claim follows by taking $N$ sufficiently large so that $\left(1+4\bar{g}\right)e^{\dagger}(N) < \varepsilon$.  $\square$
\end{flushleft}

\begin{flushleft}
\textbf{Lemma 2:} $ \sum_{r \in [n]}\lambda^{n}_{s_{n}(r)0}\lambda^{n}_{r1} \leq \int\int f^{n}_{0}(u,v)f^{n}_{1}(u,v)dudv \leq \sum_{r \in [n]}\lambda^{n}_{r0}\lambda^{n}_{r1}$ where $s_{n}(r) = n-r+1$. 
\end{flushleft}

\begin{flushleft}
\textbf{Proof of Lemma 2:} By construction $\int\int f^{n}_{1}(u,v)f^{n}_{0}(u,v)dudv = \frac{1}{n^{2}}\sum_{ij}F_{ij,1}^{n}F_{ij,0}^{n}$ so it is sufficient to show that $n^2\sum_{r \in [n]}\lambda^{n}_{s_{n}(r)0}\lambda^{n}_{r1} \leq \sum_{ij} F^{n}_{ij,1}F^{n}_{ij,0} \leq n^2\sum_{r \in [n]}\lambda^{n}_{r0}\lambda^{n}_{r1}$. Also if  $\{\lambda_{rt}^{n}\}_{r \in [n]}$ are the $n$ largest (in absolute value) eigenvalues of $f_{t}^{n}$  then $\{n\lambda_{rt}^{n}\}_{r \in [n]}$ are the eigenvalues of $F_{t}^{n}$.  \newline

Since $F^{n}_{t}$ is square and symmetric, the spectral theorem (see Lemma B2 in Online Appendix Section B.1) implies that $F_{ij,t}^{n} = n\sum_{r \in [n]}\lambda_{rt}^{n}\phi_{ir,t}^{n}\phi_{jr,t}^{n}$ where $\phi_{ir,t}^{n}$ is the eigenvector of $F_{ij,t}^{n}$ associated with eigenvalue $n\lambda_{rt}^{n}$. As a result $\sum_{ij}F_{ij,1}^{n}F_{ij,0}^{n} = n^2\sum_{r,s \in [n]}\lambda_{r1}^{n}\lambda_{s0}^{n}\left[\sum_{i}\phi_{ir,1}^{n}\phi_{is,0}^{n}\right]^{2}$. \newline

The matrix $\left[\sum_{i}\phi_{ir,1}^{n}\phi_{is,0}^{n}\right]^{2}$ is doubly stochastic and so Birkhoff's Theorem (see Lemma B4 in Online Appendix Section B.1)  implies that 
\begin{align*}
\sum_{r,s \in [n]}\lambda_{r1}^{n}\lambda_{s0}^{n}\left[\sum_{i}\phi_{ir,1}^{n}\phi_{is,0}^{n}\right]^{2} 
 =  \sum_{r,s \in [n]}\lambda_{r1}^{n}\lambda_{s0}^{n}\sum_{t\in[m]}\alpha_{t}P_{ij,t} 
 =  \sum_{t\in[m]}\alpha_{t}\sum_{r,s \in [n]}\lambda_{r1}^{n}\lambda_{s0}^{n}P_{ij,t}
 \end{align*}
 for some $m \in \mathbb{N}$, $\alpha_{1},...,\alpha_{m} > 0$ with $\sum_{t\in[m]}\alpha_{t} = 1$, and $P_{1},...,P_{m} \in \mathcal{P}_{n}$.  \newline
 
 Hardy-Littlewood-Polya's Theorem 368 (see Lemma B5 in Online Appendix Section B.1) implies that
 \begin{align*}
  \sum_{r \in [n]}\lambda_{r1}^{n}\lambda_{s_{n}(r)0}^{n} 
  \leq \sum_{r,s \in [n]}\lambda_{r1}^{n}\lambda_{s0}^{n}P_{ij} 
  \leq \sum_{r \in [n]}\lambda_{r1}^{n}\lambda_{r0}^{n}
  \end{align*}
   for any $P \in \mathcal{P}_{n}$ and so 
    \begin{align*}
  \sum_{r \in [n]}\lambda_{r1}^{n}\lambda_{s_{n}(r)0}^{n} 
  \leq \sum_{t\in[m]}\alpha_{t}\sum_{r,s \in [n]}\lambda_{r1}^{n}\lambda_{s0}^{n}P_{ij,t}
  \leq \sum_{r \in [n]}\lambda_{r1}^{n}\lambda_{r0}^{n}
  \end{align*}
  because $\sum_{t \in [m]}\alpha_{t} = 1$. The claim follows. $\square$
\end{flushleft}

\begin{flushleft}
\textbf{Lemma 3:} For every $\varepsilon > 0$ there exists an $m \in \mathbb{N}$ such that  
\begin{enumerate}
\item[i.] $\left|\int\int f^{n}_{1}(u,v)f^{n}_{0}(u,v)dudv -  \int\int f_{1}(u,v)f_{0}(u,v)dudv\right| \leq \varepsilon$, and
\item[ii.] $\left|\sum_{r \in [n]}\lambda^{n}_{\sigma_{n}(r)0}\lambda^{n}_{r1} - \sum_{r}\lambda_{\sigma(r)0}\lambda_{r1}\right| \leq \varepsilon$, 
\end{enumerate}
for every $n > m$ where $\sum_{r}\lambda_{\sigma(r)0}\lambda_{r1}$ refers to $\lim_{R \to \infty}\sum_{r\in [R]}\lambda_{\sigma_{R}(r)0}\lambda_{r1}$ where $\{\lambda_{rt}\}_{r \in [R]}$ is ordered to be decreasing and $\sigma_{R}(r)$ refers to either $R$ or $s_{R}(r) := R-r+1$. 
\end{flushleft}

\begin{flushleft}
\textbf{Proof of Lemma 3:} Fix an arbitrary $\varepsilon > 0$. Part i. follows from
\begin{align*}
&\left|\int\int f^{n}_{1}(u,v)f^{n}_{0}(u,v)dudv -  \int\int f_{1}(u,v)f_{0}(u,v)dudv\right| \\
&= \left|\int\int \left(f^{n}_{1}(u,v)-f_{1}(u,v)\right)f^{n}_{0}(u,v)dudv + \int\int \left(f^{n}_{0}(u,v)-f_{0}(u,v)\right)f_{1}(u,v)dudv\right| \\
&\leq \left(\int\int \left(f^{n}_{1}(u,v)-f_{1}(u,v)\right)^{2}dudv\right)^{1/2}\bar{f}^{n}_{0} + \left(\int\int \left(f^{n}_{0}(u,v)-f_{0}(u,v)\right)^{2}dudv\right)^{1/2}\bar{f}_{1} \\
  &\leq \epsilon \left(\bar{f}^{n}_{0} + \bar{f}_{1}\right) \text{for $n > m(\epsilon)$ where $m(\epsilon)$ is from the hypothesis of Lemma 1}\\
 &\leq \varepsilon \text{ for any $n > m\left(\varepsilon\right)$ where $\varepsilon = \epsilon(\bar{f}^{n}_{0} + \bar{f}_{1})$}
\end{align*}
where $\bar{f}^{n}_{0} = \left(\int\int f^{n}_{0}(u,v)^{2}dudv\right)^{1/2}$ and $\bar{f}_{1} = \left(\int\int f_{1}(u,v)^{2}dudv\right)^{1/2}$, the first inequality is due to Cauchy-Schwarz and the triangle inequality, and the second is due to Lemma 1. \newline

To demonstrate Part ii, we bound $\left|\sum_{r \in [n]}\lambda^{n}_{\sigma_{n}(r)0}\lambda^{n}_{r1} - \sum_{r\in[n]}\lambda_{\sigma_{n}(r)0}\lambda_{r1}\right|$ where the sum $\sum_{r\in[n]}\lambda_{\sigma_{n}(r)0}\lambda_{r1}$ is a function of the $n$ largest eigenvalues of $f_{0}$ and $f_{1}$ in absolute value. The remainder $\left|\sum_{r\in[n]}\lambda_{\sigma_{n}(r)0}\lambda_{r1} - \sum_{r}\lambda_{\sigma(r)0}\lambda_{r1}\right|$ can be made arbitrarily small since $\sum_{r}\lambda_{\sigma(r)0}\lambda_{r1} := \lim_{n \to \infty}\sum_{r\in[n]}\lambda_{\sigma_{n}(r)0}\lambda_{r1}$. We write
\begin{align*}
&\left|\sum_{r \in [n]}\lambda^{n}_{\sigma_{n}(r)0}\lambda^{n}_{r1} - \sum_{r\in[n]}\lambda_{\sigma_{n}(r)0}\lambda_{r1}\right|
= \left|\sum_{r \in [n]}\left(\lambda^{n}_{\sigma_{n}(r)0}\lambda^{n}_{r1} - \lambda_{\sigma_{n}(r)0}\lambda_{r1}\right)\right| \\
&= \left|\sum_{r \in [n]}\left(\lambda^{n}_{\sigma_{n}(r)0}-\lambda_{\sigma_{n}(r)0}\right)\lambda^{n}_{r1} + \sum_{r \in [n]}\left(\lambda^{n}_{r1}-\lambda_{r1}\right)\lambda_{\sigma_{n}(r)0}\right| \\
&\leq  \left(\sum_{r \in [n]}\left(\lambda^{n}_{r0}-\lambda_{r0}\right)^{2}\right)^{1/2}\left(\sum_{r \in [n]}\left(\lambda^{n}_{r1}\right)^{2}\right)^{1/2} + \left(\sum_{r \in [n]}\left(\lambda^{n}_{r1}-\lambda_{r1}\right)^{2}\right)^{1/2}\left(\sum_{r \in [n]}\left(\lambda_{r0}\right)^{2}\right)^{1/2} \\
&=  \left(\sum_{r \in [n]}\left(\lambda^{n}_{r0}-\lambda_{r0}\right)^{2}\right)^{1/2}\bar{f}^{n}_{1} + \left(\sum_{r \in [n]}\left(\lambda^{n}_{r1}-\lambda_{r1}\right)^{2}\right)^{1/2}\bar{f}_{0}
\end{align*}
where the first inequality is due to Cauchy-Schwarz and the triangle inequality. Since $f_{t}^{n}$ and $f_{t}$ are  bounded functions then for every $\epsilon > 0$ there exists a $R,m' \in \mathbb{N}$ such that $\sum_{r \in [n]-[R]}\left(\lambda^{n}_{rt}\right)^{2} < \epsilon$ and $\sum_{r \in [n]-[R]}\left(\lambda_{rt}\right)^{2} < \epsilon$ for every $n > m'$ and $t \in \{0,1\}$. As a result, 

\begin{align*}
&\left(\sum_{r \in [n]}\left(\lambda^{n}_{r0}-\lambda_{r0}\right)^{2}\right)^{1/2}\bar{f}^{n}_{1} + \left(\sum_{r \in [n]}\left(\lambda^{n}_{r1}-\lambda_{r1}\right)^{2}\right)^{1/2}\bar{f}_{0}\\
&\leq \left(\sum_{r \in [R]}\left(\lambda^{n}_{r0}-\lambda_{r0}\right)^{2}\right)^{1/2}\bar{f}^{n}_{1} + \left(\sum_{r \in [R]}\left(\lambda^{n}_{r1}-\lambda_{r1}\right)^{2}\right)^{1/2}\bar{f}_{0} + 2\sqrt{\epsilon}(\bar{f}^{n}_{1}+\bar{f}_{0}) \text{ for $n > m'(\epsilon)$}  \\
&\leq \sqrt{R} \left(\int\int\left(f^{n}_{0}(u,v) - f_{0}(u,v)\right)^{2}dudv \right)^{1/2}\bar{f}^{n}_{1} + \sqrt{R} \left(\int\int\left(f^{n}_{1}(u,v) - f_{1}(u,v)\right)^{2}dudv \right)^{1/2}\bar{f}_{0}  \\
&\hspace{20mm}+ 2\sqrt{\epsilon}(\bar{f}^{n}_{1}+\bar{f}_{0}) \text{ for $n > m'(\epsilon)$} \\
&\leq (\sqrt{R}\tilde{\epsilon} + 2\sqrt{\epsilon}) (\bar{f}^{n}_{1} + \bar{f}_{0}) \text{ for $n > \max(m'(\epsilon),m(\tilde{\epsilon}))$ where $m(\tilde{\epsilon})$ is from the hypothesis of Lemma 1}\\
&\leq  \varepsilon/2 \text{ for $n > \max(m'(\varepsilon^{2}/(8\bar{f}^{n}_{1} + 8\bar{f}_{0})^{2}),m( \varepsilon /(4\sqrt{R}\bar{f}^{n}_{1} + 4\sqrt{R}\bar{f}_{0}) ))$}
\end{align*}
where the third inequality follows because the eigenvalues of compact Hermitian operators are Lipschitz continuous (see the corollary to Lemma B3 in Online Appendix Section B.1) and the last inequality follows if $\epsilon$, $R$, and $m'$ are chosen so that $\epsilon = \varepsilon^{2}/(8\bar{f}^{n}_{1} + 8\bar{f}_{0})^{2}$ and then $\tilde{\epsilon}$ and $m$ are chosen so that $\tilde{\epsilon} = \varepsilon /(4\sqrt{R}\bar{f}^{n}_{1} + 4\sqrt{R}\bar{f}_{0}) $. The claim follows.  $\square$

\end{flushleft}

\begin{flushleft}
\textbf{Lemma 4:}  If $f_{0}^{n}$ and $f_{1}^{n}$ take values in $\{0,1\}$ then $\max\left(\sum_{r \in [n]}\left((\lambda_{r0}^{n})^{2}+(\lambda_{r1}^{n})^{2}\right) - 1,0\right) \leq  \int\int f^{n}_{1}(u,v)f^{n}_{0}(u,v)dudv \leq \min\left(\sum_{r \in [n]}(\lambda_{r0}^{n})^{2},\sum_{r \in [n]}(\lambda_{r1}^{n})^{2}\right)$.
\end{flushleft}

\begin{flushleft}
\textbf{Proof of Lemma 4:} The upper bound follows  
\begin{align*}
\int\int f^{n}_{1}(u,v)f^{n}_{0}(u,v)dudv 
\leq \min_{t \in \{0,1\}}\int\int f^{n}_{t}(u,v)dudv
&=  \min_{t \in \{0,1\}}\int\int (f^{n}_{t}(u,v))^{2}dudv\\
   &= \min_{t \in \{0,1\}}\sum_{r \in [n]}(\lambda_{rt}^{n})^{2}.
\end{align*}
The lower bound follows
 \begin{align*}
 \int\int f^{n}_{1}(u,v)&f^{n}_{0}(u,v)dudv 
 = \int\int f^{n}_{1}(u,v)\left( 1 - \left(1 - f^{n}_{0}(u,v)\right)\right)dudv \\
 &\geq \int\int f^{n}_{1}(u,v)dudv - \min \left(\int\int f^{n}_{1}(u,v)dudv , \int\int\left(1 - f^{n}_{0}(u,v)\right)dudv\right)\\
 &= \max\left(0,\int\int f^{n}_{1}(u,v)dudv + \int\int f^{n}_{0}(u,v)dudv - 1\right) \\
  &= \max\left(0,\int\int (f^{n}_{1}(u,v))^{2}dudv + \int\int (f^{n}_{0}(u,v))^{2}dudv - 1\right) \\
 &= \max\left(\sum_{r \in [n]}\left((\lambda_{r0}^{n})^{2}+(\lambda_{r1}^{n})^{2}\right) - 1,0\right). 
 \end{align*}
 The claim follows. $\square$
\end{flushleft}

\subsection{Proposition 1}
Let $f_{t}(u,v) = \mathbbm{1}\{Y_{t}^{*}(u,v) \leq y_{t}\}$. For any $n\in \mathbb{N}$ let $S_{i} := \left(\frac{i-1}{n},\frac{i}{n}\right]$, $F_{t}^{n}$ be an $n\times n$ matrix with $F_{ij,t}^{n} \in \mathbb{R}$ as its $ij$th entry, and $f_{t}^{n}(u,v) = \sum_{ij}F_{ij,t}^{n}\mathbbm{1}\{u \in S_{i}, v \in S_{j}\}$ such that $\int\int\left(f_{t}(u,v)-f_{t}^{n}(u,v)\right)^{2}dudv \to 0$ as $n\to\infty$ as per Lemma 1. Let $\{\lambda_{rt}\}$ denote the eigenvalues of  $f_{t}$ and $\{\lambda^{n}_{rt}\}$ the eigenvalues of  $f_{t}^{n}$. \newline

For any $\epsilon > 0$ there exists an $m \in \mathbb{N}$ such that for every $n > m$
\begin{align*}
 \int\int f_{1}(u,v)f_{0}(u,v)dudv
 &< \int f_{1}^{n}(u,v)f_{0}^{n}(u,v)dudv + \epsilon \\
 &\leq  \min\left(\sum_{r}\lambda^{n}_{r1}\lambda^{n}_{r0},\sum_{r}(\lambda^{n}_{r1})^{2},\sum_{r}(\lambda^{n}_{r0})^{2}\right) + \epsilon  \\
& < \min\left(\sum_{r}\lambda_{r1}\lambda_{r0},\sum_{r}\lambda^{2}_{r1},\sum_{r}\lambda^{2}_{r0}\right) + 2\epsilon
\end{align*}
where the first inequality is due to Part i of Lemma 3, the second inequality is the intersections of the upper bounds in Lemmas 2 and 4, and the third inequality is due to Part ii of Lemma 3. Similarly, 
\begin{align*}
 \int\int f_{1}(u,v)f_{0}(u,v)dudv
 &> \int f_{1}^{n}(u,v)f_{0}^{n}(u,v)dudv - \epsilon \\
 &\geq \max\left(\sum_{r}\lambda^{n}_{r1}\lambda^{n}_{s(r)0},\sum_{r }\left((\lambda_{r0}^{n})^{2}+(\lambda_{r1}^{n})^{2}\right) - 1,0\right) - \epsilon \\
 &> \max\left(\sum_{r}\lambda_{r1}\lambda_{s(r)0},\sum_{r}\left(\lambda_{r0}^{2}+\lambda_{r1}^{2}\right) - 1,0\right) - 2\epsilon.
\end{align*}
Since $\epsilon > 0$ is arbitrary, the claim follows. $\square$

\subsection{Proposition 2}
We use the same notation and definitions as in the proof of Proposition 1 above. For any $y_{1},y_{0} \in \mathbb{R}$ such that $y_{1}-y_{0} = y$ we have
\begin{align*}
 \int\int \mathbbm{1}&\{Y_{1}^{*}(u,v)-Y_{0}^{*}(u,v) \leq y\} dudv
\geq  \int\int \mathbbm{1}\{Y_{1}^{*}(u,v) \leq y_{1}\}\mathbbm{1}\{-Y_{0}^{*}(u,v) < -y_{0}\}  dudv \\
&= \int\int \mathbbm{1}\{Y_{1}^{*}(u,v) \leq y_{1}\}dudv -  \int\int \mathbbm{1}\{Y_{1}^{*}(u,v) \leq y_{1}\}\mathbbm{1}\{Y_{0}^{*}(u,v) \leq y_{0}\}  dudv \\
&=\int\int f_{1}(u,v) dudv -  \int\int f_{1}(u,v)f_{0}(u,v)dudv \\
&\geq \sum_{r}\lambda_{r1}^{2} - \min\left(\sum_{r}\lambda_{r1}^{2},\sum_{r}\lambda_{r0}^{2},\sum_{r}\lambda_{r1}\lambda_{r0}\right) \\
&= \max\left(\sum_{r}(\lambda_{r1}^{2}-\lambda_{r0}^{2}),\sum_{r}(\lambda_{r1}^{2} - \lambda_{r1}\lambda_{r0}),0\right)
\end{align*}
and 
\begin{align*}
 \int\int \mathbbm{1}&\{Y_{1}^{*}(u,v)-Y_{0}^{*}(u,v) \leq y\} dudv
 \leq \int\int  \max\left(\mathbbm{1}\{ Y_{1}^{*}(u,v) \leq y_{1} \},\mathbbm{1}\{-Y_{0}^{*}(u,v) < -y_{0}\}\right) dudv\\
&= 1 +  \int\int \mathbbm{1}\{Y_{1}^{*}(u,v) \leq y_{1}\}\mathbbm{1}\{Y_{0}^{*}(u,v) \leq y_{0}\}  dudv - \int\int \mathbbm{1}\{Y_{0}^{*}(u,v) \leq y_{0}\}dudv \\
&\leq 1 + \min\left(\sum_{r}\lambda_{r1}^{2},\sum_{r}\lambda_{r0}^{2},\sum_{r}\lambda_{r1}\lambda_{r0}\right) - \sum_{r}\lambda_{r0}^{2} \\
&= 1 + \min\left(\sum_{r}(\lambda_{r1}^{2}-\lambda_{r0}^{2}),\sum_{r}(\lambda_{r1}\lambda_{r0} - \lambda_{r0}^{2}),0\right)
\end{align*}
where the the first inequality in both systems is due to the fact that for any $u,v \in [0,1]$, 
\begin{align*}
\mathbbm{1}\{Y_{1}^{*}(u,v) \leq y_{1}\}\mathbbm{1}\{-Y_{0}^{*}(u,v) < -y_{0}\} 
\leq \mathbbm{1}\{Y_{1}^{*}(u,v)-Y_{0}^{*}(u,v) \leq y\}\\
\leq \max\left(\mathbbm{1}\{ Y_{1}^{*}(u,v) \leq y_{1} \},\mathbbm{1}\{-Y_{0}^{*}(u,v) < -y_{0}\}\right)
\end{align*}
and the second inequality in both systems is due to the upper bound in Proposition 1. Since these inequalities hold for any $y_{1},y_{0} \in \mathbb{R}$ such that $y_{1} - y_{0} = y$, the claim follows. $\square$

\subsection{Proposition 3}
This result is an infinite dimensional analog of the Hoffman-Wielandt inequality (see Lemma B6 in Online Appendix Section B.1). Let $f_{t}(u,v) = Y_{t}^{*}(u,v)$. For any $n\in \mathbb{N}$ let $S_{i}^{n} := \left(\frac{i-1}{n},\frac{i}{n}\right]$, $F_{t}^{n}$ be an $n\times n$ matrix with $F_{ij,t}^{n} \in \mathbb{R}$ as its $ij$th entry, and $f_{t}^{n}(u,v) = \sum_{ij}F_{ij,t}^{n}\mathbbm{1}\{u \in S_{i}^{n}, v \in S_{j}^{n}\}$ such that $\int\int\left(f_{t}(u,v)-f_{t}^{n}(u,v)\right)^{2}dudv \to 0$ as $n\to\infty$ as per Lemma 1. Let $\{\sigma_{rt}\}$ denote the eigenvalues of $f_{t}$ and $\{\sigma_{rt}^{n}\}$ the eigenvalues of $f_{t}^{n}$. 

 For any $\epsilon > 0$ there exists an $m \in \mathbb{N}$ such that for every $n > m$
\begin{align*}
\int\int&(f_{1}(u,v)-f_{0}(u,v))^{2}dudv \\ 
&= \int\int f_{1}(u,v)^{2}dudv + \int\int f_{0}(u,v)^{2}dudv - 2\int\int f_{1}(u,v)f_{0}^{*}(u,v)dudv \\
&\geq \int\int f_{1}(u,v)^{2}dudv + \int\int f_{0}(u,v)^{2}dudv - 2\int\int f^{n}_{1}(u,v)f^{n}_{0}(u,v)dudv - \epsilon \\
&\geq \sum_{r}\sigma_{r1}^{2} + \sum_{r}\sigma_{r0}^{2} - 2\sum_{r}\sigma^{n}_{r1}\sigma^{n}_{r0} - \epsilon \\
&\geq \sum_{r}\sigma_{r1}^{2} + \sum_{r}\sigma_{r0}^{2} - 2\sum_{r}\sigma_{r1}\sigma_{r0} - 2\epsilon\\
&= \sum_{r}(\sigma_{r1}-\sigma_{r0})^{2} - 2\epsilon
\end{align*}
where the first inequality is due to Part i of Lemma 3, the second inequality is due to the upper bound of Lemma 2, and the third inequality is due to Part ii of Lemma 3.  \newline

The claim then follows from the fact that $\int\int STE(u,v;\phi)^{2}dudv =  \sum_{r}(\sigma_{r1}-\sigma_{r0})^{2}$ for any choice of orthogonal basis $\{\phi_{r}\}_{r\in \mathbb{N}}$. Specifically,  
\begin{align*}
\int\int STE(u,v;\phi)^{2}dudv 
&= \int\int\sum_{r,s}(\sigma_{r1}-\sigma_{r0})(\sigma_{s1}-\sigma_{s0})\phi_{r}(u)\phi_{r}(v)\phi_{s}(u)\phi_{s}(v)dudv \\
&= \sum_{r,s}(\sigma_{r1}-\sigma_{r0})(\sigma_{s1}-\sigma_{s0}) \left[\int\phi_{r}(u)\phi_{s}(u)du\right]^{2} \\
&=  \sum_{r}(\sigma_{r1}-\sigma_{r0})^{2}
\end{align*}
where the last equality is because  $\{\phi_{r}\}_{r\in \mathbb{N}}$ is orthogonal and so $\left[\int\phi_{r}(u)\phi_{s}(u)du\right]^{2} = \mathbbm{1}\{r=s\}$. $\square$

\subsection{Proposition 4}
Let $g: \mathbb{R} \to \mathbb{R}$ admit the series representation $g(x) = \sum_{s}c_{s}x^{s}$, $(\sigma_{rt},\phi^{*}_{rt})$ be the $r$th eigenvalue and eigenfunction pair of $Y_{t}^{*}$, and $(\sigma_{rt},\phi_{rt})$ be the $r$th eigenvalue and eigenfunction pair of $Y_{t}$ ordered so that the eigenvalues are decreasing. 
 Then 
\begin{align*}
Y_{1}^{*}(u,v) = g(Y_{0}^{*}(u,v)) = \sum_{s}c_{s}Y_{0}^{*}(u,v)^{s} 
= \sum_{r,s}c_{s}\sigma_{r0}^{s}\phi^{*}_{r0}(u)\phi^{*}_{r0}(v) =  \sum_{r}g(\sigma_{r0})\phi^{*}_{r0}(u)\phi^{*}_{r0}(v)
\end{align*}
where $Y_{0}^{*}(u,v)^{s}  = \int\int...\int Y_{0}^{*}(u,\tau_{1})Y_{0}^{*}(\tau_{1},\tau_{2})...Y_{0}^{*}(\tau_{s-1},v)d\tau_{1}d\tau_{2}...d\tau_{s-1}$ is the $s$th operator power of $Y_{0}^{*}$ evaluated at $(u,v)$ and the third equality follows from the fact that for any bounded symmetric measurable function $h$ with eigenvalue-eigenfunction pairs $\{(\rho_{r},\psi_{r})\}_{r \in \mathbb{N}}$ we have $h^{s}(u,v) = \sum_{r}\rho_{r}^{s}\psi_{r}(u)\psi_{r}(v)$. Since $Y_{1}^{*}(u,v) = \sum_{r}\sigma_{1r}\phi^{*}_{r1}(u)\phi^{*}_{r1}(v)$, it follows from  the assumption that $g$ is not decreasing that $\sigma_{r1} = g(\sigma_{r0})$ and $\phi^{*}_{r1} = \phi^{*}_{r0}$. As a result,
\begin{align*}
Y^{*}_{1}-Y_{0}^{*} = \sum_{r}(g(\sigma_{r0}) - \sigma_{r0})\phi^{*}_{r0}\phi^{*}_{r0}
 =  \sum_{r}\left(\sigma_{r1}-\sigma_{r0}\right)\phi^{*}_{r0}\phi^{*}_{r0}
 =  \sum_{r}\left(\sigma_{r1}-\sigma_{r0}\right)\phi^{*}_{r1}\phi^{*}_{r1}.
\end{align*}
Since $Y_{t}^{*}(u,v) = Y_{t}(\varphi_{t}(u),\varphi_{t}(v))$ we have  $\phi^{*}_{r1}(u) = \phi_{r1}(\varphi_{1}(u))$ and  $\phi^{*}_{r0}(u) = \phi_{r0}(\varphi_{0}(u))$. As a result, $STT(u,v) =  \sum_{r}\left(\sigma_{r1}-\sigma_{r0}\right)\phi_{r1}(u)\phi_{r1}(v)$ and $STU(u,v) =  \sum_{r}\left(\sigma_{r1}-\sigma_{r0}\right)\phi_{r0}(u)\phi_{r0}(v)$ imply
\begin{align*}
Y^{*}_{1}(u,v)-Y_{0}^{*}(u,v) = STT(\varphi_{1}(u),\varphi_{1}(v))= STU(\varphi_{0}(u),\varphi_{0}(v)).
\end{align*}
and so because $\varphi_{1},\varphi_{0} \in \mathcal{M}$,
\begin{align*}
\int\int\mathbbm{1}\left\{Y^{*}_{1}(u,v)-Y_{0}^{*}(u,v)  \leq y\right\} dudv
&= \int\int\mathbbm{1}\left\{STT(u,v) \leq y\right\}dudv \\
&= \int\int\mathbbm{1}\left\{STU(u,v) \leq y\right\} dudv
\end{align*}
as claimed. $\square$

\end{document}


\maketitle
%

\appendix
This is the online appendix for \cite{auerbach2022heterogeneous}. Section B states auxiliary lemmas used to prove Propositions 1-4 in the main text. Section C contains examples of experiments with rank invariant treatment effects and heterogeneous spillover effects. Section D discusses extensions to row and column heterogeneity, estimation, and inference.\looseness=-1 

\setcounter{section}{1}
\section{Auxiliary lemmas}
The following auxiliary lemmas are used to demonstrate Propositions 1-4 in the main text. \looseness=-1 
\subsection{Lemmas used in Appendix Section A}
\begin{flushleft}
Lemma B1 (Lusin): For every measurable $f: [0,1]^{2} \to \mathbb{R}$ and $\epsilon > 0$ there exists a compact $E_{\epsilon} \subseteq [0,1]^{2}$ with Lebesgue measure at least $1-\epsilon$ such that $f$ is continuous when restricted to $E_{\epsilon}$. See \cite{dudley2002real} Theorem 7.5.2. 
\end{flushleft}

\begin{flushleft}
Lemma B2 (Spectral):  Let $f :[0,1]^{2}\to \mathbb{R}$ be a bounded symmetric measurable function and $T_{f}: L_{2}[0,1] \to L_{2}[0,1]$ the associated integral operator $(T_{f}g)(u) = \int f(u,\tau)g(\tau)d\tau$.  $T_{f}$ admits the spectral decomposition $f(u,v) = \sum_{r=1}^{\infty}\lambda_{r}\phi_{r}(u),\phi_{r}(v)$ in the sense that $(T_{f}g)(u) = \int f(u,\tau)g(\tau)d\tau = \sum_{r=1}^{\infty}\lambda_{r}\phi_{r}(u) \int\phi_{r}(\tau)g(\tau)d\tau$ for any $g \in L_{2}[0,1]$. Each $(\lambda_{r},\phi_{r})$ pair satisfies $\int f(u,\tau)\phi_{r}(\tau)d\tau = \lambda_{r}\phi_{r}(u)$ where $\{\lambda_{r}\}_{r=1}^{\infty}$ is a multiset of bounded real numbers with $0$ as its only limit point and $\{\phi_{r}\}_{r=1}^{\infty}$ is an orthogonal basis of $L_{2}[0,1]$. See \cite{birman2012spectral} equation (5) preceding Theorem 4 in Chapter 9.2.
\newline

The spectral decomposition described in Lemma B2 is related to but distinct from a decomposition by the same name for matrices. Specifically, if $Y$ is an $N\times N$ dimensional symmetric real-valued matrix then it admits the spectral decomposition $Y_{ij} = \sum_{r=1}^{N}\lambda_{r}\phi_{ir}\phi_{jr}$. Each $(\lambda_{r},\phi_{r})$ pair satisfies $\sum_{j=1}^{N}Y_{ij}\phi_{jr} = \lambda_{r}\phi_{ir}$  where $\{\lambda_{r}\}_{r=1}^{N}$ is a multiset of real numbers and $\{\phi_{ir}\}_{i,r=1}^{N}$ is an orthogonal matrix with $r$th column denoted by the $N\times 1$ dimensional vector $\phi_{r}$. 
\end{flushleft}

\begin{flushleft}
Lemma B3 (Continuity): Let $f,g :[0,1]^{2}\to \mathbb{R}$ be bounded symmetric measurable functions with positive eigenvalues $\{\lambda^{+}_{r}(f),\lambda^{+}_{r}(g)\}_{r}$ and negative eigenvalues $\{\lambda^{-}_{r}(f),\lambda^{-}_{r}(g)\}_{r}$ both ordered to be decreasing in absolute value. Suppose $\left(\int\int \left(f(u,v)-g(u,v)\right)^{2}dudv\right)^{1/2} \leq \epsilon$. Then $|\lambda^{+}_{i}(U)-\lambda^{+}_{i}(W)| \leq \epsilon$ and $|\lambda_{i}^{-}(U)-\lambda_{i}^{-}(W)| \leq \epsilon$. See \cite{birman2012spectral}, equation (19) following Theorem 8 in Chapter 9.2. \newline 

In Lemma 3 of Appendix Section A.4 in the main text, we use the following corollary of this result and Theorem 368 of \cite{hardy1952inequalities} (Lemma B5 below). That is, $\left(\sum_{r \in [R]}\left(\lambda_{r}(f)-\lambda_{r}(g)\right)^{2}\right)^{1/2} \leq \sqrt{R}\left(\int\int \left(f(u,v)-g(u,v)\right)^{2}dudv\right)^{1/2}$ where $\{\lambda_{r}(f),\lambda_{r}(g)\}_{r \in [R]}$ are the $R$ largest (in absolute value) eigenvalues of $f$ and $g$ ordered to be decreasing. 
\end{flushleft}

\begin{flushleft}
Lemma B4 (Birkhoff): For every $M \in \mathcal{D}^{+}_{n}$ there exists an $m \in \mathbb{N}$, $\alpha_{1},...,\alpha_{m} > 0$, and $P_{1},...,P_{m} \in \mathcal{P}_{n}$ such that $\sum_{t=1}^{m}\alpha_{t} = 1$ and $M_{ij} = \sum_{t=1}^{m}\alpha_{t}P_{ij,t}$. See \cite{birkhoff1946three}. 
\end{flushleft}

\begin{flushleft}
Lemma B5 (Hardy-Littlewood-Polya Theorem 368): For any $m\in \mathbb{N}$ and $g,h \in \mathbb{R}^{m}$ we have $\sum_{r=1}^{m} g_{(r)}h_{(m-r+1)} \leq  \sum_{r=1}^{m} g_{r}h_{r}  \leq  \sum_{r=1}^{m} g_{(r)}h_{(r)}$ where $g_{(r)}$ is the $r$th order statistic of $g$. See \cite{hardy1952inequalities}, Section 10.2, Theorem 368.
\end{flushleft}

\begin{flushleft}
Lemma B6 (Hoffman-Wielandt): Let $\{\lambda_{r}(F)\}_{r \in [n]}$ and $\{\lambda_{r}(G)\}_{r \in [n]}$ be the eigenvalues of two $n\times n$ real symmetric matrices $F$ and $G$, ordered to be decreasing. Then $\sum_{r=1}^{n}\left( \lambda_{r}(F) - \lambda_{r}(G)\right)^{2} \leq \sum_{i=1}^{n}\sum_{j=1}^{n}\left(F_{ij}-G_{ij}\right)^{2}$. See \cite{hoffman1953hw}.
\end{flushleft}

\section{Examples}
\subsection{Examples of rank invariant treatment effects}
We provide four concrete examples of heterogeneous treatment effects with outcome matrices from the economics literature. We show that under certain conditions implied by economic theory the treatment effect is rank invariant in the sense of Definition 2 of Section 3.2.3 of the main text. As a result of our Proposition 4, the distributions of the STT and STU are both equal to the DTE. In these examples, we take the population of agents to be finite, although our results also apply to the infinite population case. \looseness=-1 

\subsubsection{Information diffusion}
This example follows \cite{banerjee2013diffusion,cruz2017politician,bramoulle2018diffusion}. $N$ agents are linked in a social network as described by an $N\times N$ symmetric binary adjacency matrix $G$. $G_{ij} = 1$ if agents $i$ and $j$ are linked and $G_{ij} = 0$ otherwise. Information about a new product or social program diffuses over the social network in $T$ discrete time periods. In an initial period $0$, one agent receives the relevant piece of information. For periods $t = 1,...,T$, agents who received the information in period $t-1$, transmit it to their neighbors in period $t$. Specifically, if an agent receives the information $M$ times from their neighbors in period $t-1$, the number of times they transmit the information to each of their neighbors in time period $t$ is the sum of $M$ independent Bernoulli($\alpha$) trials. The parameter $\alpha$ describes the probability that an agent will transmit information to their neighbors once they receive it. \looseness=-1 

The outcome of interest is the expected number of times agent $j$ receives the information in the $T$ time periods when agent $i$ is initially informed in period $0$. Proposition 1 of \cite{bramoulle2018diffusion} implies that it is given by
\begin{align*}
E[Y_{ij}] = \sum_{t=1}^{T}\alpha^{t}[G^{t}]_{ij}.
\end{align*}
where $[G^{t}]_{ij} = \sum_{s_{1}}\sum_{s_{2}}...\sum_{s_{t-1}}G_{is_{1}}G_{s_{1}s_{2}}...G_{s_{t-1}j}$ is the $ij$th entry of the $t$th operator power of $G$.  \looseness=-1 

Now consider an intervention that increases $\alpha$, the probability of information transmission between agents. For example, the intervention may be a new advertisement campaign. Let $\alpha(s)$ denote the transmission probability and $E[Y_{ij}](s)$ the resulting outcome matrix with and without the campaign as indexed by $s \in \{0,1\}$. Then
\begin{align*}
E[Y_{ij}](s) = \sum_{r=1}^{n}\left(\sum_{t=1}^{T}\alpha(s)^{t}\lambda_{r}^{t}\right)\phi_{ir}\phi_{jr}
\end{align*} 
where $(\lambda_{r},\phi_{r})$ are the eigenvalue and eigenvector pairs of $G$. Under certain conditions on $T$ and $\alpha(1),\alpha(0)$, the treatment effect is rank invariant in the sense of Definition 2 of Section 3.2.3 that $E[Y_{ij}](1) = g(E[Y_{ij}](0))$ where $g$ is the matrix lift of a nondecreasing function. For example, if $\alpha(1) > \alpha(0)$ and $T$ is taken to infinity then $g(x) = \frac{\alpha(1)x}{\alpha(0) - (\alpha(1) - \alpha(0))x}$ which is increasing in $x$. More generally, if $\alpha(0) < \alpha (1) < \left(\frac{1}{T}\right)^{1/(T-2)}$ (for example, $\alpha(1),\alpha(0) < .5$ and $T > 3$) then one can show that such a $g$ exists and is nondecreasing, although to our knowledge it does not have a tractable analytical representation. 
\looseness=-1 

\subsubsection{Factor model}
See generally \cite{bai2008large,stock2016dynamic}. We focus on the setting of a ``classic factor model'' to simplify the exposition. One can extend the example to more general settings, but we leave this to future work. \looseness=-1 

The returns on $N$ assets in $T$ time periods are described by the factor model
\begin{align*}
X_{it} = \sum_{r=1}^{R}\lambda_{ir}F_{tr} + e_{it}
\end{align*}
where $R \leq \min(N,T)$, $F_{tr}$ describes the return of factor $r$ in time period $t$, $\lambda_{ir}$ describes the exposure of asset $i$ to factor $r$, and $e_{it}$ describes the idiosyncratic return. The classic normalizations are $\sum_{t=1}^{T}F_{tr}F_{ts} = \mathbbm{1}\{r=s\}$, $\sum_{i=1}^{N}\lambda_{ir}\lambda_{is} = \nu_{r}\mathbbm{1}\{r=s\}$, and $E\left[e_{it}e_{jt}\right] = \sigma_{it}^{2}\mathbbm{1}\{i=j\}$. $F_{tr}$ and $\lambda_{ir}$ are fixed; $e_{it}$ is stochastic with mean zero.\looseness=-1 

The outcome of interest is the covariance matrix of asset returns
\begin{align*}
E\left[Y_{ij}\right]  = E\left[\sum_{t=1}^{T}X_{it}X_{jt}\right] = \sum_{r=1}^{R}\lambda_{ir}\lambda_{jr} + \sum_{t=1}^{T}\sigma_{it}^{2}\mathbbm{1}\{i=j\}.
\end{align*} \looseness=-1 

Now consider an intervention that scales $\sigma_{it}^{2}$, the variance of the idiosyncratic error. For example, the intervention may be the announcement of an economic policy to be implemented in the future. The effects of the policy are uncertain. Let $\sigma_{it}^{2}(s)$ denote the variance of $e_{it}$, $\sigma_{i}^{2}(s) = \sum_{t=1}^{T}\sigma_{it}^{2}(s)$, and $E\left[Y_{ij}\right](s)$ the resulting outcome matrix before and after the announcement as indexed by $s \in \{0,1\}$. Then 
\begin{align*}
E\left[Y_{ij}\right](s)  = \sum_{r=1}^{R}\lambda_{ir}\lambda_{jr} + \sum_{t=1}^{T}\sigma_{it}^{2}(s)\mathbbm{1}\{i=j\}.
\end{align*}
and by the normalization on $\lambda_{ir}$
\begin{align*}
E\left[\tilde{Y}_{ab}\right](s) := \sum_{i=1}^{N}\sum_{j=1}^{N}E\left[Y_{ij}\right](s)\lambda_{ia}\lambda_{jb} - \nu_{a}^{2}\mathbbm{1}\{a = b\}  = \sum_{i=1}^{N}\sigma_{i}^{2}(s)\lambda_{ia}\lambda_{ib} 
\end{align*}
for $a,b = 1,...,R$. Then the treatment effect is rank invariant for the rotated and shifted outcome matrix $E\left[\tilde{Y}_{ab}\right]$ if $\sigma_{i}^{2}(1) = g(\sigma_{i}^{2}(0))$ for some nondecreasing function. For example, if $\sigma_{i}^{2}(s) = \rho(s)\sigma_{i}^{2}$ for some scalars $\rho(1) > \rho(0)$. This may be the case if the announcement scales the idiosyncratic variation in returns for all assets, regardless of their factor structure. 

Practically, this example suggests an identification strategy where the researcher first identifies $\lambda_{ir}$ and $\nu_{r}$ from $Y$, constructs $\tilde{Y}$, uses rank invariance to identify the distribution of treatment effects for the outcome matrix $\tilde{Y}$, and then uses $\lambda_{ir}$ and $\nu_{r}$ to identify the distribution of treatment effects for $Y$. 

\subsubsection{Social interaction}
This example follows \cite{ballester2006s,calvo2009peer}. $N$ agents are linked in a social network as described by an $N\times N$ symmetric binary adjacency matrix $G$. $G_{ij} = 1$ if agents $i$ and $j$ are linked and $G_{ij} = 0$ otherwise. Agents take $K$ real-valued actions. The $k$th action of agent $i$ is described by $A_{ik}$. It may describe, for example, how much agent $i$ smokes or invests in a risky venture. The utility agent $i$ receives from choosing action $A_{ik}$ depends on the total amount of the action taken by their peers. Specifically, 
\begin{align*}
U_{i}(A_{k}) = \eta_{ik}A_{ik} - \frac{1}{2}A_{ik}^{2} + \beta \sum_{j=1}^{N}G_{ij}A_{jk}A_{ik}  
\end{align*}
where $\eta_{ik} \sim_{iid}(0,\sigma^{2}_{k})$ is an idiosyncratic shock. The parameter $\beta$ describes the size of the peer effect. That is, how much agents are influenced by their peers. Under the assumption that $I - \beta G$ is invertible, there exists a unique Nash equilibrium 
\begin{align*}
A_{k} = \left(I - \beta G\right)^{-1}\eta_{k}
\end{align*}
where $A_{k} = (A_{1k},...,A_{Nk})$. 

The outcome of interest is the correlation of actions between agent pairs. It is given by 
\begin{align*}
E[Y_{ij}] = E\left[\sum_{k=1}^{K}A_{ik}A_{jk}\right] = \left[\left(I - \beta G\right)^{-1}\Sigma \left(I - \beta G\right)^{-1}\right]_{ij}
\end{align*}
where $\Sigma_{ij} = E[\sum_{k=1}^{K}\eta_{ik}\eta_{jk}]$ is a diagonal matrix with $\Sigma_{ii} = \sum_{k=1}^{K}\sigma_{k}^{2} := \sigma^{2}$. 

Now consider an intervention that increases $\beta$, the peer effect size parameter. For example, the intervention may be a school program that better informs students about their peers' actions. Let $\beta(s)$ denote the peer effects parameter and $E[Y_{ij}](s)$ the resulting outcome matrix with and without the program as indexed by $s \in \{0,1\}$. Then 
\begin{align*}
E[Y_{ij}](s) = \sum_{r=1}^{N}\left(\frac{\sigma^{2}}{(1-\beta(s)\lambda_{r})^{2}}\right)\phi_{ir}\phi_{jr}
\end{align*}
where $(\lambda_{r},\phi_{r})$ are the eigenvalue and eigenvector pairs of $G$. If $\beta(1) > \beta(0) > 0$ then the treatment effect is rank invariant in the sense of Definition 2 of Section 3.2.3 that $E[Y_{ij}](1) = g(E[Y_{ij}](0))$ where $g$ is the matrix lift of a nondecreasing function. Specifically, one can verify that $g(x) = \sigma^{2}\left( 1 - \frac{\beta(1)}{\beta(0)}\left(1-\left(\frac{\sigma^{2}}{x}\right)\right)^{1/2}\right)^{-2}$ which is well defined and nondecreasing for $x$ in the range of outcome matrices $Y_{0}$ that satisfy the condition that $I - \beta(0) G$ is invertible.   

\subsubsection{Link formation}
This example follows \cite{graham2017econometric}. $N$ agents are linked in a social network as described by an $N\times N$ symmetric binary adjacency matrix $G$. $G_{ij} = 1$ if $i$ and $j$ are linked and $G_{ij} = 0$ otherwise. The marginal transferable utility agents $i$ and $j$ receive from forming a link depends on homophily. That is, their proximity in a $K$-dimensional social characteristic space. Specifically,
\begin{align*}
U_{ij}(G_{ij} = 1) - U_{ij}(G_{ij} = 0) = \alpha_{i} + \alpha_{j} - \beta \sum_{k=1}^{K}(x_{ik} - x_{jk})^{2}  + \eta_{ij}
\end{align*} 
where the fixed effect $\alpha_{i}$ describes agent $i$'s degree heterogeneity or popularity, $x_{ik}$ describes the $k$th social characteristic of agent $i$, and the idiosyncratic error $\eta_{ij}$ is iid logistic. The parameter $\beta$ describes the size of the homophily effect. That is, how much link formation is influenced by agent proximity in the social characteristic space.  

The conditional probability that utility-maximizing agents $i$ and $j$ form a link is given by
\begin{align*}
E[G_{ij}] = \Lambda(\alpha_{i} + \alpha_{j} - \beta \sum_{k=1}^{K}(x_{ik} - x_{jk})^{2}).  
\end{align*}
where $\Lambda$ is the standard logistic distribution function. We define the conditional logit function
\begin{align*}
Y_{ij} = \Lambda^{-1}(E[G_{ij}]) = (\alpha_{i}+\beta\sum_{k}x_{ik}^{2}) + (\alpha_{j}+\beta\sum_{k}x_{jk}^{2}) +2\beta\sum_{k}x_{ik}x_{jk}.  
\end{align*}

To simplify our exposition, we take as our outcome of interest the demeaned conditional logit function 
\begin{align*}
E\left[\tilde{Y}_{ij}\right] = 2 \beta \tilde{X}_{ij} 
\end{align*}
where $\tilde{Y}_{ij} = Y_{ij} - \frac{1}{N}\sum_{i=1}^{N}Y_{ij}  - \frac{1}{N}\sum_{j=1}^{N}Y_{ij}$ and $\tilde{X}_{ij} = \sum_{k}x_{ik}x_{jk} - \frac{1}{N}\sum_{i,k=1}^{N}x_{ik}x_{jk} - \frac{1}{N}\sum_{j,k=1}^{N}x_{ik}x_{jk}$. It is straightforward to also demonstrate rank invariance for the undemeaned conditional logit function, following the logic of Section D.1 below. 

Now consider an intervention that increases $\beta$, the homophily size parameter. For example, the intervention may be a technology that decreases the costs of communication between locations.  Let $\beta(s)$ denote the homophily parameter and $E\left[\tilde{Y}_{ij}\right](s)$ the resulting outcome matrix with and without the communication technology as indexed by $s \in \{0,1\}$. Then  
\begin{align*}
\tilde{Y}_{ij}(s)= \sum_{r=1}^{N}\left(2\beta(s)\lambda_{r}\right)\phi_{ir}\phi_{jr}
\end{align*}
where $(\lambda_{r},\phi_{r})$ are the eigenvalue and eigenvector pairs of $\tilde{X}$. If $\beta(1) > \beta(0) > 0$ then the treatment effect is rank invariant in the sense of Definition 2 of Section 3.2.3 that $E[Y_{ij}](1) = g(E[Y_{ij}](0))$ where $g$ is the matrix lift of a nondecreasing function. Specifically, one can verify that $g(x) = \frac{\beta(1)}{\beta(0)}x$ which is increasing. 

\subsection{Examples of heterogeneous spillover effects}
We provide two additional concrete examples of settings with heterogeneous spillover effects. 

\subsubsection{Treatment spillovers}
This example follows \cite{bajari2021multiple}. Consider a buyer-seller experiment where pairs of buyers and sellers are assigned to an information treatment. For example, the setting may be an online marketplace where a buyer-seller pair is treated if the online platform explicitly recommends the seller's product to the buyer. The outcome of interest $Y_{ij}$ is the size of the transaction between buyer $i$ and seller $j$. Let $X_{ij} = 1$ if buyer $i$ and seller $j$ are assigned to treatment and $X_{ij} = 0$ otherwise. 

Following \cite{bajari2021multiple}, we assume local interference  (their Assumption 5.4). That is, the outcome $Y_{ij}$ between buyer $i$ and seller $j$ depends on whether $i$ and $j$ are treated, the number of sellers $l$ for which $i$ and $l$ are treated, and the number of buyers $k$ for which $k$ and $j$ are treated. For example, buyer $i$ may be more likely to buy from seller $j$ if the platform recommends one of seller $j$'s products, all else equal. Buyer $i$ may be less likely to buy from seller $j$ if the platform recommends products from seller $j$'s competitors, all else equal. 

Formally, we model
\begin{align*}
Y_{ij} = f_{ij}\left(X_{ij}, \sum_{l}X_{il},\sum_{k}X_{kj}\right).
\end{align*}
Let $f_{ij}\left(x_{b},x_{s}\right)$ be the expected outcome for agent pair $ij$ when $\sum_{l}X_{il} = x_{b}$ and $\sum_{k}X_{kj} = x_{s}$ for some $x_{b},x_{s} \in \mathbb{Z}_{+}$, i.e. $\sum_{t \in \{0,1\}}f_{ij}\left(t, x_{b},x_{s}\right)P(X_{ij} =t)$. In this example, our parameter of interest is the distribution of treatment spillover effects 
\begin{align*}
\frac{1}{NM}\sum_{i \in [N],j \in [M]}\mathbbm{1}\{ f_{ij}\left(x^{1}_{b},x^{1}_{s}\right) - f_{ij}\left(x^{0}_{b},x^{0}_{s}\right) \leq y\}.
\end{align*}
In words, it is the fraction of buyer-seller pairs whose change in outcome, after altering the number of relevant treated agent pairs for $i$ and $j$ from $(x^{0}_{b},x^{0}_{s})$ to $(x^{1}_{b},x^{1}_{s})$ is less than $y$. \looseness=-1

To identify the distribution of treatment spillover effects, we propose the following experiment. First randomly assign the treatment to pairs of buyers and sellers. Then form two groups. The first group collects all of the buyers that belong to exactly $x^{1}_{b}$ treated buyer-seller pairs and all of the sellers that belong to exactly $x^{1}_{s}$ treated buyer-seller pairs. The second group similarly collects all of the buyers and sellers that belong to $x^{0}_{b}$ and $x^{0}_{s}$ treated pairs. The next step is to use the matrix of outcomes associated with each group $Y_{t} := \{f_{ij}\left(X_{ij}, x^{t}_{b},x^{t}_{s}\right)\}_{i,j \in \text{ group } t}$ to compute $\bar{Y}_{t} := \{f_{ij}\left(x^{t}_{b},x^{t}_{s}\right)\}_{i,j \in \text{ group } t}$ in the case that $P(X_{ij} = t)$ is fixed by the researcher and estimate $\bar{Y}_{t}$ in the case that  $P(X_{ij} = t)$ is to be recovered from data (see Section D.2 below). Finally, after symmetrization as in Section 5.1, the distribution of spillover effects can be characterized exactly as in Section 3. \looseness=-1

\subsubsection{Market externalities}
Consider the setting of a market economy with $N$ agents and $L$ goods. For a fixed price $p \in \mathbb{R}^{L-1}$, agent $i$'s demand for the $l$th good is given by the function $Q_{il}(p)$ with $Q_{i}(p) = \{Q_{i1}(p),...,Q_{iL}(p)\} \in \mathbb{R}^{L}$. $Q_{il}$ may be negative in which case $i$ is a supplier of good $l$. An equilibrium market price $p^{*}(0)$ is assumed to satisfy the market clearing condition $\sum_{i=1}^{N}Q_{i}^{*}(0) = 0$ where  $Q_{i}^{*}(0) = Q_{i}(p^{*}(0))$ is agent $i$'s equilibrium demand. Absent any market intervention, the equilibrium price and quantity $(p^{*}(0),Q_{1}^{*}(0),...,Q_{N}^{*}(0))$ is realized.  \looseness=-1

We are interested in understanding the impact of a market intervention such as a price floor on the equilibrium demand matrix between agents and goods. An equilibrium market price $p^{*}(1)$ is assumed to satisfy the market clearing condition $\sum_{i=1}^{N}Q_{i}^{*}(1) = 0$ and restriction $p^{*}(1) \geq c$ where $Q_{i}^{*}(1) = Q_{i}(p^{*}(1))$ is agent $i$'s equilibrium demand under the price floor and $c \in \mathbb{R}^{L-1}$ is chosen by the policy maker. Under the price floor intervention, the equilibrium price and quantity $(p^{*}(1),Q_{1}^{*}(1),...,Q_{N}^{*}(1))$ is realized. \looseness=-1

Our interest is in the distribution of treatment effects
\begin{align*}
\frac{1}{NL}\sum_{i \in [N], l \in [L]}\mathbbm{1}\{Q^{*}_{il}(1) - Q^{*}_{il}(0) \leq y\}.
\end{align*}  
In words, it is the fraction of agent and good pairs whose difference in equilibrium demand with and without the price ceiling is less than $y$. There are market externalities in this example because while price floor may only nominally restrict one agent or item, the equilibrium condition implies that implementing the policy may result in changes in the equilibrium demand matrix for any agent and item. \looseness=-1

To identify the distribution treatment effects with market externalities, we suppose that the researcher is given data from the following natural experiment. They observe a matrix of equilibrium demand for a population of agents and goods from a region without the price floor. They observe another matrix of equilibrium demand for a different population of agents and goods from a region with the price floor. The two regions may not have any agents or goods in common, but they are assumed to be comparable in the sense that the distribution of potential outcomes (here equilibrium demand) is the same for both regions. For example, the regions may be located nearby each other but under different political jurisdictions. After symmetrization as in Section 5.1, the two outcome matrices matrices can be used to characterize the distribution of equilibrium treatment effects exactly as in Section 3.  \looseness=-1

\section{Extensions}
We provide additional details about some of the extensions described in Section 5 of the main text. 
\subsection{Row and column heterogeneity}
While the bounds from Section 3 are valid for any symmetric outcome matrices, they may be uninformative when there is substantial variation in the row and column variances. In such cases, we recommend allowing for row and column heterogeneity following ideas of \cite{finke1987quadratic}.

\subsubsection{Bounds on the DPO and DTE}
In Section 5.2 of the main text we write the DPO as 
\begin{align*}
F(y_{1},y_{0}) = \int\int \prod_{t \in \{0,1\}}\left(\alpha_{t}(\varphi_{t}(u)) + \alpha_{t}(\varphi_{t}(v))\right) dudv + \int\int \prod_{t \in \{0,1\}}\epsilon_{t}(\varphi_{t}(u),\varphi_{t}(v))dudv
\end{align*}
where $\mathbbm{1}\{Y^{*}_{t}(u,v) \leq y_{t}\} = \alpha_{t}(u) + \alpha_{t}(v) + \epsilon_{t}(u,v)$ and $\int\epsilon_{t}(\tau,v)d\tau = \int\epsilon_{t}(u,\tau)d\tau = 0$ for every $u,v \in [0,1]$. We bound the two summands separately. Specifically, the upper bound is constructed by
\begin{align*}
F(y_{1},y_{0}) \leq \max_{\varphi_{1},\varphi_{0} \in \mathcal{M}}\left[\int\int \prod_{t \in \{0,1\}}\left(\alpha_{t}(\varphi_{t}(u)) + \alpha_{t}(\varphi_{t}(v))\right) dudv + \int\int \prod_{t \in \{0,1\}}\epsilon_{t}(\varphi_{t}(u),\varphi_{t}(v))dudv\right] \\
\leq  \max_{\varphi_{1},\varphi_{0} \in \mathcal{M}}\left[\int\int \prod_{t \in \{0,1\}}\left(\alpha_{t}(\varphi_{t}(u)) + \alpha_{t}(\varphi_{t}(v))\right) dudv\right] + \max_{\varphi_{1},\varphi_{0} \in \mathcal{M}}\left[\int\int \prod_{t \in \{0,1\}}\epsilon_{t}(\varphi_{t}(u),\varphi_{t}(v))dudv\right]. 
\end{align*}
The first summand is bounded from above by 
\begin{align*}
2\max_{\varphi_{1},\varphi_{0} \in \mathcal{M}}\left[\int \alpha_{1}(\varphi_{1}(u))\alpha_{0}(\varphi_{0}(u)) du\right]  
+ 2\alpha_{1}\alpha_{0}
\leq 2\int\alpha_{1}^{+}(u)\alpha_{0}^{+}(u)du  + 2\alpha_{1}\alpha_{0}
\end{align*}
where $\alpha_{t} = \int \alpha_{t}(u)du$ and $\alpha_{t}^{+}$ is the ``increasing rearrangement'' or quantile function of $\alpha_{t}$ (i.e. $\alpha_{t}^{+}$ is nondecreasing and equal to $\alpha_{t}$ up to a measure preserving transformation). See Theorem 378 of \cite{hardy1952inequalities} or the second proof of Theorems 2.1 and 2.5 of \cite{whitt1976bivariate} for details. Following Proposition 1 of Section 3, the second summand is bounded from above by 
\begin{align*}
\min\left(\sum_{r}\lambda_{r1}^{2},\sum_{r}\lambda_{r0}^{2},\sum_{r}\lambda_{r1}\lambda_{r0}\right)
\end{align*}
where $\lambda_{rt}$ refers to the eigenvalues of $\epsilon_{t}$ and the sums are defined as in Section 3.2.1 of the main text. Together, the two bounds imply that 
\begin{align*}
F(y_{1},y_{0}) \leq 2\int\alpha_{1}^{+}(u)\alpha_{0}^{+}(u)du + 2\alpha_{1}\alpha_{0} + \min\left(\sum_{r}\lambda_{r1}^{2},\sum_{r}\lambda_{r0}^{2},\sum_{r}\lambda_{r1}\lambda_{r0}\right).
\end{align*}

By the same logic, the lower bound on the DPO is
\begin{align*}
F(y_{1},y_{0}) \geq 2\int\alpha_{1}^{+}(u)\alpha_{0}^{+}(1-u)du + 2\alpha_{1}\alpha_{0} + \max\left(\sum_{r}\left(\lambda_{r1}^{2} + \lambda_{r0}^{2}\right) - 1,\sum_{r}\lambda_{r1}\lambda_{s(r)0},0\right).
\end{align*}

Bounds on the DTE can be constructed from those on the DPO following exactly the logic of Proposition 2 in Section 3 of the main text.

\subsubsection{Spectral treatment effects}
Suppose the rank invariance assumption that $\alpha_{1}(\varphi_{1}(u)) = g_{\alpha}(\alpha_{0}(\varphi_{0}(u))$ and $\epsilon_{1}(\varphi_{1}(u),\varphi_{1}(v)) = g_{\epsilon}(\epsilon_{0}(\varphi_{0}(u),\varphi_{0}(v))$ for every $u,v \in [0,1]$ where $g_{\alpha}$ is a nondecreasing function and $g_{\epsilon}$ is the matrix lift of a nondecreasing function as in Definition 2 of Section 3.2.3. Define the spectral treatment effect with row and column heterogeneity to be
\begin{align*}
STE(u,v;\phi) = \left(\alpha_{1}^{+}(u) - \alpha_{0}^{+}(u)\right) + \left(\alpha_{1}^{+}(v) - \alpha_{0}^{+}(v)\right) + \sum_{r}(\sigma_{r1} - \sigma_{r0})\phi_{r}(u)\phi_{r}(v)
\end{align*}
where $\{\phi_{r}\}$ is any orthogonal basis in $L^{2}([0,1])$ and $\{\sigma_{rt}\}$ are the eigenvalues of $\epsilon_{t}$. Similarly define $STT(u,v) = STE(u,v;\phi_{1})$ and $STU(u,v) = STE(u,v;\phi_{0})$ where $\phi_{1}$ and $\phi_{0}$ refer to the eigenfunctions of $\epsilon_{1}$ and $\epsilon_{0}$ respectively. 

Then by the logic of Standard Result 4 in Section 2 and Proposition 4 in Section 3 
\begin{align*}
Y^{*}_{1}(u,v) - Y^{*}_{0}(u,v) &= (\alpha_{1}(\varphi_{1}(u)) - \alpha_{0}(\varphi_{0}(u))) + (\alpha_{1}(\varphi_{1}(v)) - \alpha_{0}(\varphi_{0}(v))) 
\\ &\hspace{10mm}+ (\epsilon_{1}(\varphi_{1}(u),\varphi_{1}(v)) - \epsilon_{0}(\varphi_{0}(u),\varphi_{0}(v))) \\
&= (g_{\alpha}(\alpha_{0}(\varphi_{0}(u))) - \alpha_{0}(\varphi_{0}(u))) + (g_{\alpha}(\alpha_{0}(\varphi_{0}(v))) - \alpha_{0}(\varphi_{0}(v)) 
\\ &\hspace{10mm}+ (g_{\epsilon}(\epsilon_{0}(\varphi_{0}(u),\varphi_{0}(v))) - \epsilon_{0}(\varphi_{0}(u),\varphi_{0}(v))) \\
&= (g_{\alpha}(\alpha_{0}(\varphi_{0}(u))) - \alpha_{0}(\varphi_{0}(u))) + (g_{\alpha}(\alpha_{0}(\varphi_{0}(v))) - \alpha_{0}(\varphi_{0}(v)) 
\\ &\hspace{10mm}+ \sum_{r}(g_{\epsilon}(\sigma_{r0}) - \sigma_{r0})\phi_{r0}(\varphi_{0}(u)) \phi_{r0}(\varphi_{0}(v))
\\ &= STE(\varphi_{0}(u),\varphi_{0}(v);\phi_{0}))
\end{align*}

Since $\varepsilon_{1}$ and $\varepsilon_{0}$ are rank invariant, they have the same eigenfunctions (see the proof of Proposition 4 in Section A.5), and so $ STE(\varphi_{0}(u),\varphi_{0}(v);\phi_{0})) =  STE(\varphi_{0}(u),\varphi_{0}(v);\phi_{1}))$. It follows that under the above rank invariance assumption $Y^{*}_{1}-Y_{0}^{*}$, $STT$, and $STU$ all have the same distribution. 

\subsection{Randomization inference}
Our paper is mainly about identification, but for completeness we also sketch how one can conduct randomization-based inference about the impact of the treatment in three of the motivating examples from Section 1.1 of the main text. We focus on the global point null of no treatment effect, take the populations to be finite and the potential outcomes to be fixed, and do not explicitly consider network interference. One can, however, similarly consider infinite populations, test other such hypotheses, or invert the tests to construct point estimates and confidence intervals in the sense of \cite{hodges2012estimates,rosenbaum2002observational}, see for instance \cite{athey2018exact,basse2019randomizationT,basse2019randomization}. Whereas  \cite{auerbach2022testing} compares outcome matrices defined on the same set of agents, this section compares outcome matrices defined on two (or more) different sets of agents. \looseness=-1 

\subsubsection{Matched pair design}
This example is based on  \cite{azoulay2010superstar}, see Example 2 in Section 1.1 of the main text. The data consists of a collection of $V$ matched pairs of research groups. One group from each pair is affected by the death of a superstar ($t=1$), the other group is not $(t=0$). We assume that the treatment is (as good as) randomly assigned so that its ex-ante probability is the same for each group in a given pair. The treatment assignment is also independent across pairs of groups. \looseness=-1 

Let $Y_{ij,sv,t}$ be the potential research productivity of researchers $i$ and $j$ in group $s \in \{1,2\}$ of pair $v \in [V]$ under treatment $t \in \{0,1\}$. $\tilde{Y}_{ij,v,t}$ is the observed research productivity for $i$ and $j$ in the group that is actually assigned treatment $t$ in pair $v$. That is,  $\tilde{Y}_{ij,v,1} = Y_{ij,1v,1}$ if the first group in pair $v$ is treated, $\tilde{Y}_{ij,v,1} = Y_{ij,2v,1}$ if the second group in pair $v$ is treated, etc.  $N_{sv}$ is the number of researchers in group $s $ of pair $v$ and $\tilde{N}_{tv}$ is the number of researchers in the group that is actually assigned treatment $t$ in pair $v$.   \looseness=-1 

The null hypothesis is that the research productivity between pairs of researchers in each group are completely unaffected by the death of the superstar. That is, \looseness=-1 
\begin{align*}
H_{0}: Y_{ij,sv,t} = Y_{ij,sv,t'} \text{ for every } i,j \in [N_{sv}], s \in \{1,2\}, v \in [V] \text{ and } t,t' \in \{0,1\}.
\end{align*}

For a test statistic, we propose the difference in the eigenvalues of the thresholded outcome matrices associated with the treated and untreated groups in each pair
\begin{align*}
T = \max_{v \in [V]}\sup_{y \in \mathbb{R}}\sum_{r}\left(\lambda_{r,v,1}(y) - \lambda_{r,v,0}(y)\right)^{2}
\end{align*}
where $\lambda_{r,v,t}(y)$ is the $r$th eigenvalue of the thresholded outcome matrix $\mathbbm{1}\{\tilde{Y}_{v,t} \leq y\}/\tilde{N}_{tv}$ and $\tilde{Y}_{v,t}$ is a $\tilde{N}_{tv}\times \tilde{N}_{tv}$ matrix with $ij$th entry equal to $\tilde{Y}_{ij,v,t}$. \looseness=-1 

For a reference distribution, we propose re-randomizing the treatment assignments within each pair. For $A \in \mathbb{N}$ and $a \in [A]$, let  $\rho_{v,a}$ be a collection of independent Bernoulli ($1/2$) random variables, 
\begin{align*}
\tilde{Y}_{v,1}^{a} = \tilde{Y}_{v,1}\rho_{v,a} + \tilde{Y}_{v,0}(1-\rho_{v,a}), \\
\tilde{Y}_{v,0}^{a} = \tilde{Y}_{v,1}(1-\rho_{v,a}) + \tilde{Y}_{v,0}\rho_{v,a},
\end{align*}
and 
\begin{align*}
T^{a} =   \max_{v \in [V]}\sup_{y \in \mathbb{R}}\sum_{r}\left(\lambda_{r,v,1}^{a}(y) - \lambda_{r,v,0}^{a}(y)\right)^{2}
\end{align*}
where $\lambda^{a}_{r,v,t}(y)$ is the $r$th eigenvalue of the thresholded outcome matrix $\mathbbm{1}\{\tilde{Y}^{a}_{v,t} \leq y\}/\tilde{N}^{a}_{tv}$, $\tilde{N}^{a}_{1v} = \tilde{N}_{1v}\rho_{v,a} + \tilde{N}_{0v}(1-\rho_{v,a})$, and $\tilde{N}^{a}_{0v} = \tilde{N}_{1v}(1-\rho_{v,a}) + \tilde{N}_{0v}\rho_{v,a}$. \looseness=-1 

By \cite{lehmann2006testing} Theorem 15.2.1, the test that rejects $H_{0}$ whenever
\begin{align*}
(A+1)^{-1}\left(1 + \sum_{a \in [A]}\mathbbm{1}\{T^{a} \geq T\} \right) \leq \alpha
\end{align*}
is level $\alpha$. It is powered to detect deviations in the eigenvalues of the thresholded outcome matrices associated with each treatment. That such deviations detect a large class of heterogeneous treatment effects follows Proposition 2 in the main text.  \looseness=-1 

\subsubsection{Double randomization with uncensored outcomes}
This example is based on the conjunctive simple multiple randomization design of \cite{bajari2021multiple}, see Example 4 in Section 1.1 of the main text. A group of $B$ buyers and $S$ sellers are independently randomized to one of two groups. The probability that any buyer or seller is assigned to group $1$ is $\pi \in (0,1)$. Every buyer-seller pair where both the buyer and the seller of that pair are assigned to group $1$ is given an information treatment. \looseness=-1 

$Y_{ij,st}$ records the potential transaction between buyer $i$ and seller $j$ in the event that $i$ is assigned to group $s \in \{1,2\}$ and $j$ is assigned to group $t \in \{1,2\}$. $\tilde{Y}_{ij}$ is the observed transaction for buyer $i$ and seller $j$ under their realized group assignments. We call this example uncensored because the researcher observes transactions for every pair of agents. To simplify arguments, we assume that the potential transactions for a buyer-seller pair do not depend on exactly which other buyers or sellers are assigned to groups 1 and 2. This may be the case when the number of buyers and sellers assigned to each group is large. \looseness=-1 

The null hypothesis is that the group assignments have no effect on the potential transactions between buyers and sellers in the marketplace. That is, \looseness=-1 
\begin{align*}
H_{0}: Y_{ij,st} = Y_{ij,s't'} \text{ for every } i \in [B], j \in [S], s,s',t,t' \in \{1,2\}.
\end{align*}

For a test statistic, we propose the difference in the thresholded eigenvalues of the outcome matrices associated with each treatment
\begin{align*}
T = \max_{s,s',t,t' \in \{1,2\}}\sup_{y \in \mathbb{R}}\sum_{r}\left(\lambda_{r,st}(y) - \lambda_{r,s't'}(y)\right)^{2}.
\end{align*}
where $\lambda_{r,st}(y)$ is the $r$th eigenvalue of the symmetrized (see Section 5.4 of the main text) thresholded outcome matrix $\{\mathbbm{1}\{\tilde{Y}^{\dagger}_{ij} \leq y\}/N_{st}\}_{i, j \in B(s) \cup S(t)}$, $B(s) = \{i \in B: i \text{ is assigned to group } s \}$ and $S(t) = \{j \in S: i \text{ is assigned to group } t \}$, and $N_{st} = |B(s)| + |S(t)|$. \looseness=-1

For a reference distribution, we propose re-randomizing the individual treatment assignments. For $A \in \mathbb{N}$ and $a \in [A]$, let $\rho_{i,a}^{B}$ and $\rho_{j,a}^{S}$ be a collection of independent Bernoulli($\pi$) random variables, $B^{a}(s) = \{i \in B : \rho_{i,a}^{B}+1 = s\}$ and $S^{a}(t) = \{j \in S : \rho_{j,a}^{S}+1 = t\}$ be the set of buyers and sellers re-randomized to group $s$ and $t$ respectively,
\begin{align*}
\tilde{Y}_{st}^{a} = \{\tilde{Y}_{ij}\}_{i \in B^{a}(s), j \in S^{a}(t)}
\end{align*} 
and 
\begin{align*}
T^{a} = \max_{s,s',t,t' \in \{1,2\}}\sup_{y \in \mathbb{R}}\sum_{r}\left(\lambda^{a}_{r,st}(y) - \lambda^{a}_{r,s't'}(y)\right)^{2}
\end{align*}
where $\lambda^{a}_{r,st}(y)$ is the $r$th eigenvalue of the symmetrized thresholded outcome matrix $\mathbbm{1}\{\tilde{Y}^{a\dagger}_{st} \leq y\}/\tilde{N}^{a}_{st}$ and $\tilde{N}^{a}_{st} = |B^{a}(s)|+|S^{a}(t)|$.\looseness=-1 

By \cite{lehmann2006testing} Theorem 15.2.1, the test that rejects $H_{0}$ whenever
\begin{align*}
(A+1)^{-1}\left(1 + \sum_{a \in [A]}\mathbbm{1}\{T^{a} \geq T\} \right) \leq \alpha
\end{align*}
 is level $\alpha$. It is powered to detect deviations in the eigenvalues of the thresholded outcome matrices associated with each treatment. That such deviations detect a large class of heterogeneous treatment effects follows Proposition 2 in the main text.

\subsubsection{Double randomization with censored outcomes}
This example is based on \cite{comola2021treatment}, see Example 1 in Section 1.1 of the main text. A collection of households are randomly chosen to participate in a savings program. Each household is assigned to participate independently with probability $\pi \in (0,1)$. \looseness=-1 

$Y_{ij,t}$ records the potential risk sharing link for households $i$ and $j$ when both ($t=1$) or neither ($t=0$) participate in the program. $\tilde{Y}_{ij,t}$ records the observed risk sharing link for households $i$ and $j$ that were actually assigned to treatment $t$. $\tilde{N}_{t}$ is the number of households actually assigned to treatment $t$. We call this example censored because the researcher only observes the potential risk sharing links for pairs of agents assigned to the same treatment. To simplify arguments, we assume that the number of participants is small relative to the number of non-participants (i.e. $\pi \approx 0$). \looseness=-1 

The null hypothesis is that participation in the savings program has no effect on the potential risk sharing links between pairs of households. That is, \looseness=-1 
\begin{align*}
H_{0}: Y_{ij,t} = Y_{ij,t'} \text{ for every } t,t' \in \{0,1\}.
\end{align*}

For a test statistic, we propose the difference in the eigenvalues of the thresholded outcome matrices associated with the treated and untreated household pairs
\begin{align*}
T = \sup_{y \in \mathbb{R}}\sum_{r}\left(\lambda_{r,1}(y) - \lambda_{r,0}(y)\right)^{2}
\end{align*}
where $\lambda_{r,t}(y)$ is the $r$th eigenvalue of the thresholded outcome matrix $\mathbbm{1}\{\tilde{Y}_{t} \leq y\}/\tilde{N}_{t}$.\looseness=-1 

For a reference distribution, we propose re-randomizing the individual treatment assignments. For $A \in \mathbb{N}$ and $a \in [A]$, let $\rho_{i,a}$ be a collection of independent Bernoulli($\pi$) random variables, 
\begin{align*}
\tilde{Y}_{1}^{a} = \{Y_{ij,0}\}_{\rho_{i,a} = 1, \rho_{j,a} = 1},
\end{align*}
and 
\begin{align*}
T^{a} =  \sup_{y \in \mathbb{R}}\sum_{r}\left(\lambda^{a}_{r,1}(y) - \lambda_{r,0}(y)\right)^{2}
\end{align*}
where $\lambda^{a}_{r,1}$ is the $r$th eigenvalue of the thresholded outcome matrix $\mathbbm{1}\{\tilde{Y}^{a}_{1} \leq y\}/\tilde{N}^{a}_{1}$ where $\tilde{N}_{1}$ is the number of households re-randomized to treatment $1$. \looseness=-1 

By \cite{lehmann2006testing} Theorem 15.2.1, the test that rejects $H_{0}$ whenever
\begin{align*}
(A+1)^{-1}\left(1 + \sum_{a \in [A]}\mathbbm{1}\{T^{a} \geq T\} \right) \leq \alpha
\end{align*}
is level $\alpha$. It is powered to detect deviations in the eigenvalues of the thresholded outcome matrices associated with each treatment. That such deviations detect a large class of heterogeneous treatment effects follows Proposition 2 in the main text.  \looseness=-1

\subsection{Estimation and large sample inference}
Our paper is mainly about identification, but for completeness we also sketch how one can estimate and conduct inference about the bounds on the DTE, DPO, and the distribution of STE using sampled, mismeasured, or missing data. The data is assumed to be drawn from an infinite population. Our sketch relies on kernel density smoothing along the lines of \cite{horowitz1992smoothed}. Alternative strategies may potentially lead to more accurate inferences in practice, but we leave their study to future work. \looseness=-1  

\subsubsection{Assumptions about the data generating process}
The researcher does not observe the outcome functions $Y_{t}: [0,1]^{2} \to \mathbb{R}$ for $t \in \{0,1\}$. They have instead a stochastic approximation $\hat{Y}_{t} : [0,1]^{2} \to \mathbb{R}$ whose accuracy depends on a sample size $N$. We give more information about the relationship between $\hat{Y}_{t}$ and $Y_{t}$ below.  \looseness=-1 

For example, the researcher may observe the $N\times N$ matrix $M_{t}$ with $ij$th entry  $M_{ij,t} = \left(Y_{t}(w_{i,t},w_{j,t}) + \epsilon_{ij,t}\right)\eta_{ij,t}$. The agent types are sampled from a population as described by $w_{i,t} \sim_{iid} F_{w}$. The outcomes are observed with measurement error as described by $\epsilon_{ij,t} \sim_{iid} F_{\epsilon}$.  Some outcomes are missing as described by $\eta_{ij,t} \sim_{iid} Bernoulli(p_{t})$ with $p_{t} \in (0,1)$. To construct $\hat{Y}_{t}$, the researcher first estimates the entries of $Y_{t}(w_{i,t},w_{j,t})$ conditional on $\{w_{i,t}\}_{i, \in [N]}$. This may be done by local averaging, k-means clustering, linear regression, principal components analysis, spectral thresholding, etc. See for instance \cite{bai2008large,bonhomme2015grouped,chatterjee2015matrix,stock2016dynamic,jochmans2019fixed,graham2020network}. $\hat{Y}_{t}$ is then the function embedding (see Appendix Section A.1.1) of this matrix of estimates, reweighted by the inverse density of $w_{i,t}$, see \cite{hsieh2018non}. \looseness=-1 

To demonstrate consistency of our estimators (specified below), our main assumption is that $\hat{Y}_{t}$ is consistent in MSE. That is, 
\begin{flushleft}
Assumption D1: \begin{align*}
MSE(\hat{Y}) := \max_{t \in \{0,1\}}\int\int\left(\hat{Y}_{t}(u,v) - Y_{t}(u,v)\right)^{2}dudv \to_{p} 0 \text{ as } N \to \infty.
\end{align*}
\end{flushleft}
Nearly all of the methods proposed in the literature, including those referenced above, satisfy this property under certain regularity conditions. We show that under Assumption 1 and additional regularity conditions below, the rate of convergence of our estimators decreases with the MSE of $\hat{Y}_{t}$. As a result, Assumption D1 implies that our estimators are consistent. \looseness=-1 

To construct a distribution for large sample inference, our main assumption is that $Y_{t}$ is determined by a linear model. Specifically, 
\begin{flushleft}
Assumption D2: For $t \in \{0,1\}$, $Y_{t}(u,v) = X_{t}(u,v)\beta_{t}$  and $\hat{Y}_{t}(u,v) = X_{t}(u,v)\hat{\beta}_{t}$ with $\beta_{t},\hat{\beta}_{t} \in \mathbb{R}^{K}$ for some $K \in \mathbb{N}$,  $\left(\hat{\beta}_{t}-\beta_{t}\right) \to_{d} \mathcal{N}\left(0,V_{t}\right)$ as $N \to \infty$ for some covariates $X_{t}(u,v)$ observed up to a measure preserving transformation (see Section 3.1.4.), $V_{t}$ can be consistently estimated, and $\hat{\beta}_{1}$ and $\hat{\beta}_{0}$ have independent entries. 
\end{flushleft}
Linear models are common in economics, see for instance \cite{bonhomme2015grouped,jochmans2019fixed,auerbach2022identification}. We believe the arguments of this section can also be applied to other classes of models, but at the cost of a more complicated characterization. We leave this to future work.  \looseness=-1

\subsubsection{Additional regularity conditions}
We rely on the following regularity conditions, see relatedly Assumptions K1, K2, 6, and 9 of \cite{horowitz1992smoothed}. They are modified to fit our setting. 

\begin{flushleft}
Assumption D3: 
\begin{enumerate}
\item[i] $K: \mathbb{R} \to \mathbb{R}$ is  everywhere twice differentiable with $|K|$, $|K'|$, and $|K''|$ uniformly bounded, $\lim_{u \to \infty}K(u) = 0$ and $\lim_{u \to -\infty}K(u) = 1$. $\int \left[K'(u)\right]^{4}du$, $\int \left[K''(u)\right]^{2}du$, and $\int \left[u^{2}K''(u)\right]^{4}du$ are finite. For some $P \in \mathbb{N}$, $P \geq 2$, and $p \in [P]$ let $\int |u^{p}K'(u)|du$ be finite with $\int u^{p}K'(u)du = 0$.  
\item[ii] $h({N})$ is a bandwidth sequence such that $h \to 0$, $h^{p-P-1}\int_{|hu| > \eta}|u^{p}K'(u)|du \to 0$, $h^{-1}\int_{|hu| > \eta}|K''(u)|du \to 0$, and $h^{-P}MSE(\hat{Y}) \to \infty$ as $N \to \infty$ for any $\eta > 0$ and $p \in [P]$.  
\item[iii] $F_{STE}(y;\phi) = \int\int\mathbbm{1}\{STE(u,v;\phi) \leq y\}dudv$ is everywhere smooth with uniformly bounded derivatives.
\end{enumerate}
\end{flushleft}

\begin{flushleft}
Assumption D4: 
\begin{enumerate}
\item[i] $K: \mathbb{R} \to \mathbb{R}$ is  everywhere twice differentiable with $|K|$, $|K'|$, and $|K''|$ uniformly bounded, $\lim_{u \to \infty}K(u) = 0$,  $K(0) = 1$, and $\lim_{u \to -\infty}K(u) = 1$. $\int \left[K'(u)\right]^{4}du$, $\int \left[K''(u)\right]^{2}du$, and $\int \left[u^{2}K''(u)\right]^{4}du$ are finite. For some $P \in \mathbb{N}$, $P \geq 2$ and $p \in [P]$ let $\int |u^{p}K'(u)|du$ be finite with $\int_{u \leq 0} u^{p}K'(u)du = 0 = \int u^{p}K(u)K'(u)du = 0$ for $p \in [P]$.  

\item[ii] $h({N})$ is a bandwidth sequence such that $h \to 0$, $h^{p-P-1}\int_{|hu| > \eta}|u^{p}K'(u)|du \to 0$, $h^{-1}\int_{|hu| > \eta}|K''(u)|du \to 0$, and $h^{-P}MSE(\hat{Y})^{1/2}\to \infty$ as $N \to \infty$ for any $\eta > 0$ and $p \in [P]$. 

\item[iii] $F_{t}(y_{t}) = \int\int\mathbbm{1}\{Y_{t}(u,v) \leq y_{t}\}dudv$ is everywhere smooth with uniformly bounded derivatives.
\end{enumerate}
\end{flushleft}

A consequence of Assumption D4(i) is that $2 \int K(u)K'(u)du = -1$ and $\int_{u \leq 0} K'(u) = 0$. We use these restrictions in our proofs below. 

\subsubsection{Consistent estimation of the bounds on the DPO and DTE}
Let $\{\lambda_{rt}\}_{r \in [R]}$ and $\{\hat{\lambda}_{rt}\}_{r \in [R]}$ denote the $R$ largest eigenvalues of the functions $\mathbbm{1}\{Y_{t} \leq y_{t}\}$ and $K\left(\left(\hat{Y}_{t} - y_{t}\right)/h\right)$ respectively in absolute value, ordered to be decreasing. To estimate the bounds on the DPO and DTE, we propose using $\sum_{r \in \mathbb{N}}\hat{\lambda}_{rt}\hat{\lambda}_{rt'}$ to estimate $ \sum_{r \in\mathbb{N}}\lambda_{rt}\lambda_{rt'}$ for $t,t' \in \{0,1\}$. We show that

\begin{flushleft}
Proposition D1: Under Assumptions D1 and D4  
\begin{align*}
\sup_{y_{t},y_{t'} \in \mathbb{R}}\left|\sum_{r \in \mathbb{N}}\hat{\lambda}_{rt}\hat{\lambda}_{rt'} - \sum_{r \in\mathbb{N}}\lambda_{rt}\lambda_{rt'}\right| = O_{p}\left(MSE\left(\hat{Y}\right)\right). 
\end{align*}
\end{flushleft}

\begin{flushleft}
Sketch of proof of Proposition D1: Write
\begin{align*}
\left|\sum_{r \in \mathbb{N}}\hat{\lambda}_{rt}\hat{\lambda}_{rt'} - \sum_{r \in\mathbb{N}}\lambda_{rt}\lambda_{rt'}\right| 
\leq \left|\sum_{r \in \mathbb{N}}\left(\hat{\lambda}_{rt} - \lambda_{rt}\right)\lambda_{rt'}\right|+\left|\sum_{r \in \mathbb{N}}\left(\hat{\lambda}_{rt'} - \lambda_{rt'}\right)\lambda_{rt}\right| + r_{N} 
\end{align*}
where $r_{N} = \left|\sum_{r \in \mathbb{N}}\left(\hat{\lambda}_{rt} - \lambda_{rt}\right)\left(\hat{\lambda}_{rt'} - \lambda_{rt'}\right)\right|$ is asymptotically negligible relative to the first two terms. The first two summands are bounded
\begin{align*}
 \left|\sum_{r}\left(\hat{\lambda}_{rt} - \lambda_{rt}\right)\lambda_{rt'}\right|
  \leq \left[\sum_{r}\left(\hat{\lambda}_{rt} - \lambda_{rt}\right)^{2}\right]^{1/2}F_{t'}(y_{t'})^{1/2}
\end{align*}
by Cauchy-Schwarz since $\left(\sum_{r}\lambda_{rt}^{2}\right)^{1/2} = F_{t}(y_{t})^{1/2}$. The term $ \left[\sum_{r}\left(\hat{\lambda}_{rt} - \lambda_{rt}\right)^{2}\right]^{1/2}$ is further bounded 
\begin{align*}
\left[\sum_{r}\left(\hat{\lambda}_{rt} - \lambda_{rt}\right)^{2}\right]^{1/2} 
&\leq \left[\int\int \left( K\left(\frac{\hat{Y}_{t}(u,v) - y_{t}}{h}\right) - \mathbbm{1}\{Y_{t}(u,v) \leq y_{t}\}\right)^{2}dudv\right]^{1/2} \\
 &\leq \left[\int\int\left(K\left(\frac{\hat{Y}_{t}(u,v)- y_{t}}{h}\right) - K\left(\frac{Y_{t}(u,v)}{h}\right)\right)^{2}dudv\right]^{1/2} \\
 &\hspace{20mm}+  \left[\int\int \left( K\left(\frac{Y_{t}(u,v)- y_{t}}{h}\right) - \mathbbm{1}\{Y_{t}(u,v) \leq y_{t}\}\right)^{2}dudv\right]^{1/2}.
\end{align*}

We show that Assumption D4 implies that the second summand is $o_{p}\left(h^{\frac{P}{2}}\right)$ in Section D.3.6, Lemma D1 below, see also \cite{horowitz1992smoothed}, Lemma 5. The first term in the block of equations is $O_{p}\left(MSE\left(\hat{Y}\right)\right)$ because
\begin{align*}
\left[\int\int\left(K\left(\frac{\hat{Y}_{t}(u,v)- y_{t}}{h}\right) - K\left(\frac{Y_{t}(u,v)- y_{t}}{h}\right)\right)^{2}dudv\right]^{1/2} = \\
\left[\int\int \left( K'\left(\frac{Y_{t}(u,v)-y_{t}}{h}\right)\left[\frac{\hat{Y}_{t}(u,v) - Y_{t}(u,v)}{h}\right]\right)^{2}dudv\right]^{1/2} + s_{N}
\end{align*}
where $s_{N}$ is asymptotically negligible since $K$ is smooth. The claim follows since $p$ is chosen in Assumption D4 so that $h^{P/2}$ is $o_{p}\left(MSE\left(\hat{Y}\right)\right)$.\hspace{2mm} $\square$
\end{flushleft}

\subsubsection{Inference on the bounds on the DPO and DTE}
\begin{flushleft}
Proposition D2: Under Assumptions D1,D2, and D4
\begin{align*}
P\left(\left|\sum_{r \in \mathbb{N}}\hat{\lambda}_{rt}\hat{\lambda}_{rt'} - \sum_{r \in\mathbb{N}}\lambda_{rt}\lambda_{rt'}\right| > \epsilon \right)  \leq P\left(\left(\xi_{t}'A_{t}\xi_{t}\right)^{1/2}F_{t'}(y_{t'})^{1/2}+\left(\xi_{t'}'A_{t'}\xi_{t'}\right)^{1/2}F_{t}(y_{t})^{1/2}  > \epsilon\right) + o_{p}(1)
\end{align*}
where $A_{t} = V_{t}^{1/2}\Omega_{t}V_{t}^{1/2}$, $\Omega_{t} = \int\int \frac{1}{h^{2}}K'\left(\frac{Y_{t}(u,v)-y_{t}}{h}\right)^{2}X_{t}(u,v)'X_{t}(u,v)dudv$, and  $\left(\xi_{1},\xi_{0}\right) \sim \mathcal{N}\left(0,I_{dim(\beta_{1})+dim(\beta_{0})}\right)$. 
\end{flushleft}

\begin{flushleft}
Sketch of proof of Proposition D2: From the proof sketch of Proposition D1 in Section D.3.3, Assumptions D1 and D3 imply that 
\begin{align*}
\left|\sum_{r \in \mathbb{N}}\hat{\lambda}_{rt}\hat{\lambda}_{rt'} - \sum_{r \in\mathbb{N}}\lambda_{rt}\lambda_{rt'}\right| \leq \left[\int\int \left( K'\left(\frac{Y_{t}(u,v)-y_{t}}{h}\right)\left[\frac{\hat{Y}_{t}(u,v) - Y_{t}(u,v)}{h}\right]\right)^{2}dudv\right]^{1/2}F_{t'}(y_{t'})^{1/2} \\
+ \left[\int\int \left( K'\left(\frac{Y_{t'}(u,v)-y_{t'}}{h}\right)\left[\frac{\hat{Y}_{t'}(u,v) - Y_{t'}(u,v)}{h}\right]\right)^{2}dudv\right]^{1/2}F_{t}(y_{t})^{1/2} + o_{p}(1).
\end{align*}
The claim then follows from Assumption D2, since
\begin{align*}
\int\int \left( K'\left(\frac{Y_{t}(u,v)-y_{t}}{h}\right)\left[\frac{\hat{Y}_{t}(u,v) - Y_{t}(u,v)}{h}\right]\right)^{2}dudv \\
= \left(\hat{\beta}_{t} - \beta_{t}\right)'\left[\int\int \frac{1}{h^{2}}K'\left(\frac{Y_{t}(u,v)-y_{t}}{h}\right)^{2}X_{t}(u,v)'X_{t}(u,v)dudv\right] \left(\hat{\beta}_{t} - \beta_{t}\right) 
\to_{d} \xi_{t}' A_{t} \xi_{t}. \hspace{2mm} \square
\end{align*}
\end{flushleft}

Once can make inferences about $\sum_{r \in \mathbb{N}}\lambda_{rt}\lambda_{rt'}$ and the bounds on the DPO and DTE in practice by replacing $A_{t}$ with the estimator $\hat{A} =  \hat{V}_{t}^{1/2 '}\hat{\Omega}_{t}\hat{V}_{t}^{1/2}$ where $\hat{V}_{t}$ is a consistent estimator of $V_{t}$, $\hat{\Omega}_{t} = \int\int \frac{1}{h^2}K'\left(\frac{\hat{Y}_{t}(u,v)-y_{t}}{h}\right)^{2}X_{t}(u,v)'X_{t}(u,v)dudv$, and $\hat{F}_{t}(y_{t}) = \int\int \mathbbm{1}\{\hat{Y}_{t}(u,v) \leq y_{t}\}dudv$. One can use the right-hand side to construct critical values or confidence intervals in the usual way. Replacing $\frac{1}{h^{2}}K'\left(\frac{\hat{Y}_{t}(u,v)-y_{t}}{h}\right)^{2}$ with $\sup_{u}|\frac{1}{h^{2}}K'\left(\frac{u}{h}\right)^{2}|$ and $\hat{F}_{t'}(y_{t'})$ with $1$ allows for uniformly valid inferences over $y_{t},y_{t'} \in \mathbb{R}$.

\subsubsection{Consistent estimation of the distribution of spectral treatment effects}
Let $\{\sigma_{rt}\}_{r \in [R]}$ and $\{\hat{\sigma}_{rt}\}_{r \in [R]}$ denote the $R$ largest eigenvalues of the functions $Y_{t}$ and $\hat{Y}_{t}$ respectively in absolute value, ordered to be decreasing. We propose the estimator $\hat{STE}(u,v;\phi) = \sum_{r\in [R]}\left(\hat{\sigma}_{r1}-\hat{\sigma}_{r0}\right)\phi_{r}(u)\phi_{r}(v)$ for the parameter $STE(u,v;\phi) = \sum_{r\in [R]}\left(\sigma_{r1}-\sigma_{r0}\right)\phi_{r}(u)\phi_{r}(v)$ and the estimator  $\int\int K\left(\frac{\hat{STE}(u,v;\phi)-y}{h}\right)dudv$ for the distribution of STE, $\int\int \mathbbm{1}\{STE(u,v;\phi) \leq y\}$. 

\begin{flushleft}
Proposition D3: Under Assumptions D1 and D3
 \begin{align*}
\sup_{y \in \mathbb{R}}\left|\int\int K\left(\frac{\hat{STE}(u,v;\phi)-y}{h}\right)dudv - \int\int \mathbbm{1}\{STE(u,v;\phi) \leq y\}dudv\right| = o_{p}\left(1\right).
\end{align*}
\end{flushleft}

\begin{flushleft}
Sketch of proof of Proposition D3: Write
\begin{align*}
&\left|\int\int K\left(\frac{\hat{STE}(u,v;\phi)-y}{h}\right)dudv - \int\int \mathbbm{1}\{STE(u,v;\phi) \leq y\}dudv\right| \\
&\leq \left| \int\int K\left(\frac{\hat{STE}(u,v;\phi)-y}{h}\right)dudv -\int\int K\left(\frac{STE(u,v;\phi)-y}{h}\right)dudv  \right| \\
&\hspace{20mm}+ \left|\int\int K\left(\frac{STE(u,v;\phi)-y}{h}\right)dudv - \int\int \mathbbm{1}\{STE(u,v;\phi) \leq y\}dudv\right|.
\end{align*}
We show that Assumption D3 implies that the second summand is $o_{p}\left(h^{P}\right)$ in Section D.3.6 below, see also \cite{horowitz1992smoothed}'s Lemma 5. The first summand is $o_{p}\left(1\right)$ because 
\begin{align*}
\left|\int\int K'\left(\frac{STE(u,v;\phi)-y}{h}\right)\left[\frac{\hat{STE}(u,v;\phi)-STE(u,v;\phi)}{h}\right]dudv\right|  \\
= \left| \sum_{r\in[R]}\left((\hat{\sigma}_{r1}-\hat{\sigma}_{r0})-(\sigma_{r1} - \sigma_{r0})\right)W_{r}\right| 
\leq \left[\sum_{s \in \{0,1\}}||\hat{Y}_{s} - Y_{s}||_{2}\right]\left(\sum_{r \in [R]}W_{r}^{2}\right)^{1/2} 
\end{align*}
plus an asymptotically negligible term, where $W_{r} = \int\int \frac{1}{h}K'\left(\frac{STE(u,v;\phi)-y}{h}\right)\phi_{r}(u)\phi_{r}(v)dudv$ and $\sum_{r}W_{r}^{2}$ is finite because $K'$ is square integrable by assumption. The claim follows since $P$ is chosen in Assumption D3 so that  $h^{P}$ is $o_{p}\left(MSE\left(\hat{Y}\right)\right)$. \hspace{1mm} $\square$
\end{flushleft}

\subsubsection{Inference on the distribution of spectral treatment effects}
\begin{flushleft}
Proposition D4: Under Assumptions D1-D3
\begin{align*}
P\left(\left|\int\int K\left(\frac{\hat{STE}(u,v;\phi)-y}{h}\right)dudv - \int\int \mathbbm{1}\{STE(u,v;\phi) \leq y\}dudv\right| > \epsilon \right) \\
 \leq P\left(\left[\left(\xi_{1}'B_{1}\xi_{1}\right)^{1/2}+\left(\xi_{0}'B_{0}\xi_{0}\right)^{1/2}\right]\left(\sum_{r}W_{r}^{2}\right)^{1/2}  > \epsilon\right)
\end{align*}
where $W_{r} = \int\int \frac{1}{h}K'\left(\frac{STE(u,v;\phi)-y}{h}\right)\phi_{r}(u)\phi_{r}(v)dudv$, $B_{t} = V_{t}^{1/2}\left[\int\int X_{t}(u,v)'X_{t}(u,v)dudv\right]V_{t}^{1/2}$, and  $(\xi_{1}\hspace{1mm} \xi_{0}) \to \mathcal{N}\left(0,I_{dim(\beta_{1})+dim(\beta_{0})}\right)$.
\end{flushleft} 

\begin{flushleft}
Sketch of proof of Proposition D4: From the proof sketch of Proposition D3 in Section D.3.5, Assumptions D1 and D3 imply that
\begin{align*}
\left|\int\int K\left(\frac{\hat{STE}(u,v;\phi)-y}{h}\right)dudv - \int\int \mathbbm{1}\{STE(u,v;\phi) \leq y\}dudv\right|  \\
\leq  \left[\sum_{s \in \{0,1\}}||\hat{Y}_{s} - Y_{s}||_{2}\right]\left(\sum_{r \in [R]}W_{r}^{2}\right)^{1/2} + o_{p}\left(MSE\left(\hat{Y}\right)\right)
\end{align*}
The claim then follows from Assumption D2, since 
\begin{align*}
\int\int\left(\hat{Y}_{t}(u,v)-Y_{t}(u,v)\right)^{2}dudv = \left(\hat{\beta}_{t}-\beta_{t} \right)'\left[\int\int X_{t}(u,v)'X_{t}(u,v)\right]\left(\hat{\beta}_{t}-\beta_{t}\right) \to_{d} \xi'_{t}B_{t}\xi_{t}. \hspace{1mm} \square
\end{align*}
\end{flushleft}

One can make inferences about $\int\int \mathbbm{1}\{STE(u,v;\phi) \leq y\}dudv$ in practice by replacing $W_{r}$ with  the estimator $\hat{W}_{r} =  \frac{1}{h}\int\int K'\left(\frac{\hat{STE}(u,v;\phi)-y}{h}\right)\phi_{r}(u)\phi_{r}(v)dudv$ and $B_{t}$ with $\hat{B}_{t} = \hat{V}_{t}^{1/2}\left[\int\int X_{t}(u,v)'X_{t}(u,v)dudv\right]\hat{V}_{t}^{1/2}$. One can use the right-hand side to construct critical values or confidence intervals in the usual way. 

\subsubsection{Additional details about some of the calculations}
\begin{flushleft}
Lemma D1: Under Assumption D4,
\begin{align*}
 \left[\int\int \left( K\left(\frac{Y_{t}(u,v)- y_{t}}{h}\right) - \mathbbm{1}\{Y_{t}(u,v) \leq y_{t}\}\right)^{2}dudv\right] = o_{p}\left(h^{P}\right).
 \end{align*} 
\end{flushleft}

\begin{flushleft}
Sketch of proof of Lemma D1: Write 
\begin{align*}
\int\int \left( K\left(\frac{Y_{t}(u,v)- y_{t}}{h}\right) - \mathbbm{1}\{Y_{t}(u,v) \leq y_{t}\}\right)^{2}dudv\\ 
 = \int\int  K\left(\frac{Y_{t}(u,v)- y_{t}}{h}\right)^{2}dudv +F_{t}(y_{t}) - 2 \int\int  K\left(\frac{Y_{t}(u,v)- y_{t}}{h}\right)\mathbbm{1}\{Y_{t}(u,v) \leq y_{t}\}dudv. 
\end{align*}
The first summand 
\begin{align*}
\int\int &K\left(\frac{Y_{t}(u,v)- y_{t}}{h}\right)^{2}dudv 
= -2 \int K(\tau)K'(\tau)F_{t}(y_{t} + h\tau)d\tau \\
&= -2 \int K(\tau)K'(\tau) \left[F_{t}(y_{t}) + f_{t}(y_{t})h\tau + ... + \frac{1}{P!}f_{t}^{P}(y_{t})h^{P}\tau^{P} + o_{p}\left(h^{P}\right)\right]d\tau \\
&= F_{t}(y_{t}) + o_{p}\left(h^{P}\right)
\end{align*}
where the first equality is due to a change in variables and integration by parts, the second equality is Taylor's Theorem, and the last equality is by the choice of $K$ in Assumption D4: $2 \int K(\tau)K'(\tau)d\tau = -1$ and $\int K(\tau)K'(\tau)\tau^{p}d\tau = 0$ for $p \in [P]$. \newline

Similarly, the third summand
\begin{align*}
 \int\int &K\left(\frac{Y_{t}(u,v)- y_{t}}{h}\right)\mathbbm{1}\{Y_{t}(u,v) \leq y_{t}\}dudv
= \int K'(\tau)\mathbbm{1}\{\tau \leq 0\}F_{t}(y_{t} + h\tau)d\tau \\
&=\int K'(\tau)\mathbbm{1}\{\tau \leq 0\} \left[F_{t}(y_{t}) + f_{t}(y_{t})h\tau + ... + \frac{1}{P!}f_{t}^{P}(y_{t})h^{P}\tau^{P} + o_{p}\left(h^{P}\right)\right]d\tau \\
&= o_{p}\left(h^{P}\right)
\end{align*}
where the last equality is also by the choice of $K$ in Assumption D4: $\int_{\tau \leq 0}K'(\tau)\tau^{P}d\tau = 0$ for $p = 0$ and $p  \in [P]$. The claim follows. \hspace{1mm} $\square$
\end{flushleft}

\begin{flushleft}
Lemma D2: Under Assumption D3,
\begin{align*}
\left|\int\int K\left(\frac{STE(u,v;\phi)-y}{h}\right)dudv - \int\int \mathbbm{1}\{STE(u,v;\phi) \leq y\}dudv\right|
 = o_{p}\left(h^{P}\right).
 \end{align*} 
\end{flushleft}

\begin{flushleft}
Sketch of proof of Lemma D2: Write 
\begin{align*}
\int\int &K\left(\frac{STE(u,v;\phi)-y}{h}\right)dudv 
= \int K'(\tau)F_{t}(y_{t} + h\tau)d\tau \\
&= \int K'(\tau) \left[F_{t}(y_{t}) + f_{t}(y_{t})h\tau + ... + \frac{1}{P!}f_{t}^{P}(y_{t})h^{P}\tau^{P} + o_{p}\left(h^{P}\right)\right]d\tau \\
&= F_{t}(y_{t}) + o_{p}\left(h^{P}\right)
\end{align*}
where the first equality is due to a change in variables and integration by parts, the second equality is Taylor's Theorem, and the last equality is by the choice of $K$ in Assumption D3: $\int K'(\tau)d\tau = -1$ and $\int K'(\tau)\tau^{p}d\tau = 0$ for $p \in [P]$. \hspace{1mm} $\square$
\end{flushleft}

\subsection{Restricting the weights of the STT to be non-extrapolative}
In Section 3.2.2 we interpret the the STT as the difference between $Y_{1}$ and a counterfactual formed by a weighted average of $Y_{0}$. The weights are potentially extrapolative in the sense that they may be negative and not necessarily integrate to $1$ (although they necessarily do satisfy these properties under our matrix generalization of rank invariance in Definition 2 of the main text). In this section, we describe a modified version of the STT that is not extrapolative in the sense that the weights are necessarily nonnegative and integrate to $1$, even when the rank invariance condition does not hold.  \looseness=-1

To do this, we first show that the STT is arbitrarily well approximated by the solution to a quadratic programming problem where the weights are optimized over a set of orthogonal matrices.  We then propose a modified version of the problem where the weights are instead optimized over the set of doubly stochastic matrices. By construction, these weights are nonnegative and sum to one. \looseness=-1

\subsubsection{The STT is well-approximated by the solution to a quadratic programming problem}
By definition of the STT in Section 3.2.2, we have that for $W_{R}(u,s) = \sum_{r=1}^{R}\phi_{r1}(u)\phi_{r0}(s)$,
\begin{align*}
STT(u,v) =  Y_{1}(u,v) - \lim_{R \to \infty}\int\int Y_{0}(s,t)W_{R}(u,s)W_{R}(v,t)dsdt.
\end{align*}

To motivate our modification of the STT, we approximate $Y_{1}$ and $Y_{0}$ with the function embeddings of their matrix approximations $Y_{1}^{N}$ and $Y_{0}^{N}$ (see Lemma 1 of Appendix Section A.1.3 for a definition). Specifically, we define 
\begin{align*}
STT^{N}(u,v) = Y^{N}_{\lceil Nu\rceil \lceil Nv \rceil, 1} - \int\int Y_{\lceil Ns\rceil \lceil Nt \rceil,0}^{N}W_{\lceil Nu\rceil \lceil Ns \rceil}^{N}W_{\lceil Nv\rceil \lceil Nt \rceil}^{N}dsdt 
\end{align*}
where for $i,j \in [N]$, $Y^{N}_{ij,t}$ is the $ij$th entry of the $N \times N$ matrix approximation to $Y_{t}$, $\{\sigma_{rt}^{N},\phi_{rt}^{N}\}_{r \in [N]}$ are the eigenvalue-eigenvector pairs of $Y_{t}^{N}$, and $W_{ik}^{N} = \sum_{r \in [N]}\phi^{N}_{ir1}\phi^{N}_{kr0}$. Following Lemma 1 in Appendix A.1.3 of the main text, 
\begin{align*}
\int\int\left(STT(u,v) - STT^{N}(u,v)\right)^{2}dudv \to 0 \text{ as } N \to \infty.
\end{align*}

The weights $W^{N}$ can be derived as the solution to a quadratic programming problem. Specifically, 
\begin{align}\label{orthoRes}
W^{N} = \text{argmin}_{O \in \mathcal{O}_{N}} \sum_{i,j \in [N]}\left(Y^{N}_{ij,1} - \sum_{k,l \in [N]}Y^{N}_{kl,0}O_{ik}O_{jl}\right)^{2}
\end{align}
since 
\begin{align*}
\sum_{i,j \in [N]}\left(Y^{N}_{ij,1} - \sum_{k,l \in [N]}Y^{N}_{kl,0}O_{ik}O_{jl}\right)^{2}
 &= \sum_{i,j \in [N]}\left(Y^{N}_{ij,1}\right)^{2} + \sum_{k,l \in [N]}\left(Y^{N}_{kl,0}\right)^{2} - 2\sum_{i,j,k,l \in [N]}Y_{ij,1}^{N}Y_{kl,0}^{N}O_{ik}O_{jl} \\
 &= \sum_{r \in [N]}\left(\sigma^{N}_{r1}\right)^{2} + \sum_{s \in [N]}\left(\sigma^{N}_{s0}\right)^{2} - 2 \sum_{r,s \in [N]}\sigma^{N}_{r1}\sigma^{N}_{s0}\left[O_{ik}\phi^{N}_{ir1}\phi^{N}_{ks0}\right]^{2} \\
 &\geq \sum \left( \sigma_{r1}^{N} - \sigma_{r0}^{N}\right)^{2}
\end{align*}
where the inequality follows from the fact that $ \sum_{r,s \in [N]}\sigma^{N}_{r1}\sigma^{N}_{s0}\left[O_{ik}\phi^{N}_{ir1}\phi^{N}_{ks0}\right]^{2} \leq \sum_{r\in[N]}\sigma^{N}_{r1}\sigma^{N}_{r0}$ (see Section 4.3.2). The inequality holds with equality when $O = W^{N}$, which demonstrates (\ref{orthoRes}). \looseness=-1

\subsubsection{An alternative non-extrapolative treatment effects parameter}
In the quadratic programming problem derivation of $W^{N}$, the weight matrix is only restricted to be orthogonal, so it generally may have entries that are negative and do not sum to one. One can impose these restrictions by altering the space of weight matrices to be $\mathcal{D}^{+}_{N}$ in (\ref{orthoRes}) instead of $\mathcal{O}_{N}$. That is,
\begin{align}\label{altRes}
\tilde{X}^{N} = \text{argmin}_{D \in \mathcal{D}^{+}_{N}}\sum_{i,j \in [N]}\left(Y^{N}_{ij,1} - \sum_{k,l \in [N]}Y^{N}_{kl,0}D_{ik}D_{jl}\right)^{2}
\end{align}
and consider the alternative treatment effects parameter
\begin{align*}
Y^{N}_{\lceil Nu\rceil \lceil Nv \rceil, 1} - \int\int Y_{\lceil Ns\rceil \lceil Nt \rceil,0}^{N}X_{\lceil Nu\rceil \lceil Ns \rceil}^{N}X_{\lceil Nv\rceil \lceil Nt \rceil}^{N}dsdt 
\end{align*}
While this problem does not, to our knowledge, have a closed form analytical solution, it is a convex programming problem and so straightforward to solve using standard tools. \looseness=-1

\bibliographystyle{aer}
\bibliography{literature}